# Performance Provisioning and Energy Efficiency in Cloud and Distributed Computing Systems

(The initial title of this thesis is: Statistical performance Prediction and Evaluation of Distributed Computing Systems)

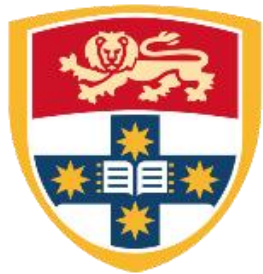

A thesis submitted in fulfilment of the requirements for the
degree of Doctor of Philosophy in the School of Information Technologies at
The University of Sydney

Nikzad Babaii Rizvandi

January 2013

# Abstract

In recent years, the issue of energy consumption in high performance computing (HPC) systems has attracted a great deal of attention. In response to this, many energy-aware algorithms have been developed in different layers of HPC systems, including the hardware layer, service layer and system layer. These algorithms are of two types: first, algorithms which directly try to improve the energy by tweaking frequency operation or scheduling algorithms; and second, algorithms which focus on improving the performance of the system, with the assumption that efficient running of a system may indirectly save more energy.

In this thesis, we develop algorithms in both layers. First, we introduce three algorithms to directly improve the energy of scheduled tasks at the hardware level by using Dynamic Voltage Frequency Scaling (DVFS). Second, we propose two algorithms for modelling and resource provisioning of MapReduce applications (a well-known parametric distributed framework currently used by Google, Yahoo, Facebook and LinkedIn) based on its configuration parameters. Certainly, estimating the performance (e.g., execution time or CPU clock ticks) of a MapReduce application can be later used for smart scheduling of such applications in clouds or clusters.

To evaluate the algorithms, we have conducted extensive simulation and real experiments on a 5-node physical cluster with up to 25 virtual nodes, using both synthetic and real world applications. Also, the proposed new algorithms are compared with existing algorithms by experimentation, and the experimental results reveal new information on the performance of these algorithms, as well as on the properties of MapReduce and DVFS. In the end, three open problems are revealed by the experimental observations, and their importance is explained.

# Acknowledgements

I first thank my supervisors Prof. Albert Y. Zomaya, Dr. Javid Taheri, Dr. Young Choon Lee, and my co-supervisors at National ICT Australia (NICTA), Dr. Joachim Gudmundsson and Prof. Aruna Seneviratne, for their full support and valuable advice during my PhD study. Without their help, it would have been impossible to complete the PhD journey.

I also thank my spouse and my parents for their selfless care and warm encouragement. I would also like to thank my little boy who was born during my PhD and brings great inspiration and happiness to my life.

Finally, I thank my colleagues and friends in the both school of IT in USyd and Network group at NICTA for the happy time they have shared with me. We have enjoyed discussions of research problems, BBQs and dinners, various sports, and other fun activities together. Especially, I'd like to thank Dr Roksana Boreli and Mr Thava Iyer at NICTA for their advice and help when I was working at Trusted networking group as a software engineer.

# List of Figures

# List of Tables

# Table of Contents

# Chapter 1. Background and Thesis Overview

Recently, primarily due to operational and environmental factors, such as significant increases in energy price, and carbon emissions associated with energy (electricity) generation and transportation, the issue of energy consumption has extended to a much broader range of systems, including High Performance Computing Systems (HPCS; i.e., multicores, grids, and clouds) and servers (on a smaller scale).

The energy consumption of these systems is associated with various monetary, environmental and system performance issues. For example, Earth Simulator and Petaflop are two HPCS systems that use 12 and 100 megawatts of peak power, respectively. With an approximate price of 100 dollars per megawatt, their energy costs during peak operation times are 1,200 and 10,000 dollars per hour, respectively; this is beyond the acceptable budget of many (potential) HPCS operators. In addition to the cost of power, cooling is another issue that must be addressed due to the negative effects of high temperature on electronic components. The rising temperature of a circuit not only derails the circuit from its normal activity but also decreases the lifetime of its components. A formula based on the Arrhenius Law indicates that the life expectancy of components decreases 50% for every 10 $^\circ$C increase in temperature. In the opposite direction, the lifetime of components will be doubled for each 10 $^\circ$C decrease [3].

The energy consumption issue with respect to different parts of HPCS systems, and existing techniques to decrease energy use in such systems, are described below.

## 1.1 Energy Consumption of Unused Servers

To run a server, a company has to spend a huge amount of money every year on hardware, updating software licences, operational support and energy. However, the company should consider whether the installed server is beneficial for the company, and if it is necessary to have it working all the time. Some analysts estimate that, on average, around one-sixth (15%) of full-time servers in a company are not actively used on a daily basis. This indicates that of the 44 million servers in the world, around 4.7 million servers are not doing any useful work. The potential savings by turning off these servers are large: globally, $3.8 billion in energy costs alone and $24.7 billion in the total cost of running non-productive servers, according to a study by 1E Company in partnership with the Alliance to Save Energy (ASE) [4]. From an environmental perspective, this amount of wasted energy results in 11.8 million tons of carbon dioxide per year, which is equivalent to the CO2 pollution of 2.1 million cars. In the U.S. alone, the figure is 3.17 million tons of carbon dioxide, or 580,678 cars.

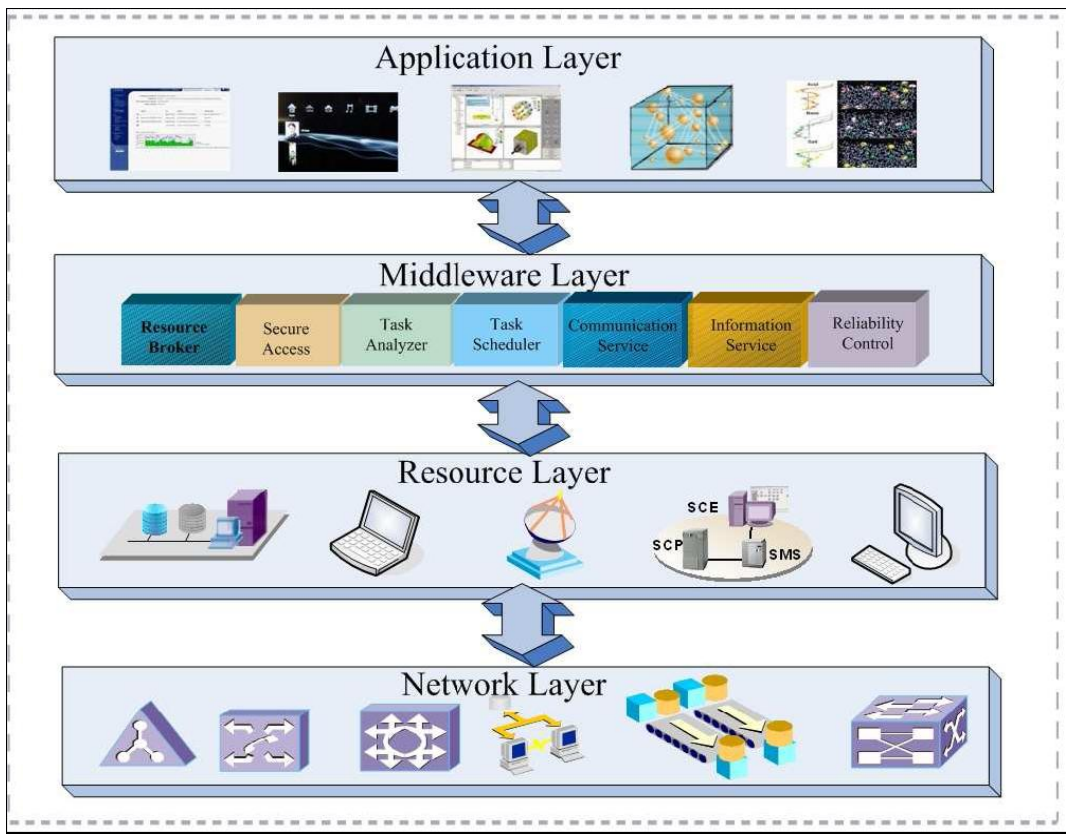

**Figure 1-1. High Performance computing platforms architecture [2].**

## 1.2 Reducing Energy in Active Servers

In addition to reducing the number of unused servers, it is necessary to apply appropriate techniques to decrease the energy use of active servers in HPCS with negligible influence on their performance. According to Figure 1-1, the power management issue in HPCS platforms can be categorised into four layers: the application layer, middleware layer, resource layer and network layer [2].

### 1.2.1 Application/service layer

Until now, all user applications on HPCS, such as applications in science, engineering, business, and finance, are optimised to increase the performance of the application, which is defined as its speed, accuracy or stability. In introducing energy-aware applications, the challenge is to design sophisticated multilevel and multi-domain energy management applications with a minor impact on performance. The first step is to explore the relationship between performance and energy in HPCS. Indeed, the energy consumption of an application is strongly dependent on the number of instructions needed to execute the application, and the number of transactions with the storage unit (or memory). These two items are also involved in the application's completion time. Moreover, instructions with a fewer number of clock cycles generally consume less energy. However, a complex model is necessary to optimise the relationship between performance and energy. For example, the order of certain operations and bit representation of data can alter energy consumption without necessarily changing the completion time.

### 1.2.2 Middleware layer

Applications in a HPC platform such as a grid or cloud interact indirectly with the computing nodes and storage units in the resource layer by means of the middleware layer. In other words, the middleware layer acts as a bridge between the application layer and the resource layer. Therefore, this layer has to provide several intelligent units, such as resource broker, security access, job analyser, job scheduler, communication service, information service and reliability control, to optimise the relationship between the application and resource layers [2]. Due to the intelligent nature of these units, this layer is too susceptible for applying energy-efficient techniques, particularly in task scheduling. Until recently, scheduling has been optimised in order to minimise a cost function, generally the makespan – the whole execution time of a set of tasks. Now, a new cost function involving both makespan and energy is necessary.

### 1.2.3 Resource layer

The resource layer consists of a wide range of resources, in particular, computing nodes and storage units. This layer generally interacts with hardware devices and also the operating system (OS), and therefore is responsible for controlling all distributed resources in high performance computing systems. In the recent past, several mechanisms have been developed for better/more efficient power management of hardware and operating systems. The majority of these are hardware approaches, particularly for processors. Dynamic power management (DPM) and dynamic voltage-frequency scaling (DVFS) are perhaps the most appealing methods incorporated into much of the recent hardware. In DPM, hardware devices such as CPU have the capability to switch from an idle mode to one or more lower-power modes [5]. In DVFS, energy saving is achieved by taking advantage of the fact that the power consumption in CMOS circuits has a direct relationship with frequency and the square of voltage supply. In this case, the execution time and power consumption are controllable by switching between different frequencies and voltages. Figure 1-2 shows the principle of the DVFS method. As can be seen, in DVFS a device uses the slack time (idle time) of a task to execute the task in a lower voltage-frequency. The relation between energy and voltage-frequency in CMOS circuits is:

$$\begin{cases} E = C_{eff} f v^2 t \\ f \propto \dfrac{(v - v_t)^2}{v_t} \end{cases}$$

Where $C_{eff}$ and $v_t$ are circuit switching capacity and circuit threshold voltage, respectively. $t$ is the execution time of the task in the current frequency ($f$). Therefore, by reducing voltage and frequency the energy consumption of the device can be reduced. However, both the DPM and DVFS techniques may have some negative effects on power consumption of a device in both active and idle modes, and create a transition overload for switching between states or voltage/frequencies. Transition overload is especially important in the DPM technique: if the transition latencies between lower-power modes are assumed to be negligible, then energy can be saved by simply switching between these modes. However, this assumption is rarely correct and therefore switching between low-power modes affects performance. To minimise this effect, workload batching techniques are utilised to alter the workload so that most of the small idle times are merged into fewer large ones [5].

   Even though DPM and DVFS techniques are widely used in the resource layer to save power in computing nodes and storage units, they may have a negative influence on performance in complicated ways. This is because the overall system computation rate depends on both speed and cooperation between multiple elements inside the system. For instance, increasing CPU speed may not increase the computation rate when the memory cannot send/receive data any faster. So, in order to construct a good model to study energy-efficient techniques, other devices of the system and their coordination should be addressed [5].

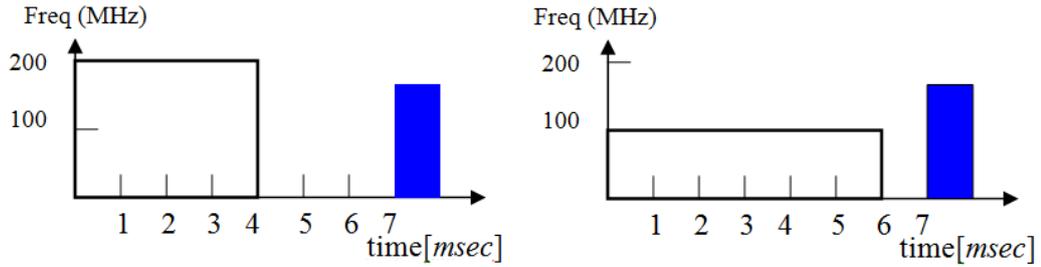

**Figure 1-2. Dynamic voltage-frequency scaling technique: (left) original task (right) voltage-frequency scaled task. This slack time is created by scheduler between this task and the next task (indicated as Blue).**

Another important issue in the resource layer is storage units. As computing nodes typically produce and process a massive amount of data, and it is necessary to store this data, the storage units have a massive number of interactions with the computing nodes in HPCS. This huge number of interactions means that storage units are constantly active, concurrent with computing nodes. As a result, after computing nodes, the storage units are the largest energy consumer in HPCS, such that the storage devices spend almost one-third (27%) of the total consumed energy in a data centre. Even worse, this percentage is rising by 60% annually due to the increasing need for storage equipment. Thus, new technologies to design storage devices and new algorithms to deal with them are highly desirable to achieve energy-efficient HPCS. Some of these algorithms include dynamic power management schemes, power-aware cache management strategies, power-aware prefetching schemes, software-directed power management techniques, and multi-speed settings [2].

### 1.2.4 Network layer

Routing and transferring packets and enabling network services to the resource layer are the major responsibilities of the network layer in HPCS. The major challenge in building energy-efficient networks is to measure and predict network behaviour, and optimise the balance between energy consumption and performance. The key to understanding and predicting network behaviours is to truly model the network. Such models are useful to help researchers to deeply understand the interaction between energy and performance in networks. Some of the major challenges in the design of energy-efficient networks are the following[5]:

- The developed models, first, must be applicable to the real world. Then these models should represent the networks comprehensively, in that they should give a full understanding of interactions between time, space and energy.
- New energy-efficient routing algorithms need to be developed. The current routing algorithms are mostly concerned with optimising the packet paths and also the reliability of paths.
- As the existing protocols have not been designed for energy efficiency, new energy-efficient protocols should be developed. The new protocols firstly must be strong against attack. Then, these protocols should have the capability to balance the trade-off between network latency and energy consumption. This capability is important as it addresses a key question in cloud computing: which scenario is efficient? Sending data to the cloud, doing computation in the cloud and then receiving the result, or doing computation locally in the host?
- Development of new methods for network traffic engineering is another issue in energy-efficient network design. For example, it is important to determine whether using half the processing or network capacity for a longer time is more energy efficient than using the full network capacity for half the time. The same question should be asked in many other situations, such as using network links, multicore processors, and datacentre servers. Answering this question – which depends on several items including the relationship between capacity and energy consumption – can produce a guide map to obtain the optimal relationship between energy and network capacity and also an energy-proportional (i.e., no power consumption when there is no traffic) design plan to create network devices like IP routers and Ethernet switches.

## 1.3 Thesis Outline and Main Results

The research presented in this thesis consists of two main parts. The research problem and main results of each part are described below.

In the first part, this author was interested in energy efficient task scheduling in the resource layer of distributed computing systems using the Dynamic Voltage-Frequency Scaling (DVFS) technique. To start the study of this topic, we completed an in-depth survey of the existing techniques based on DVFS. Then, two algorithms were proposed. Unlike most of the research in this area which uses only one processor's frequency to execute a task in a task-graph, the basic idea of these algorithms was to first deploy the multiple frequencies available in a DVFS-enabled processor to execute tasks, and then optimise the whole task graph; this is a promising way to reclaim slack times between tasks in a scheduled task-graph and thus achieve energy efficiency. During our experiments, we observed that a task reaches its optimal energy consumption when it is executed by, at most, two frequencies of DVFS-enabled processor. Surprisingly, these two frequencies are adjacent when a simplified model of DVFS is applied. To this end, we prove this observation mathematically and propose another algorithm to pick these frequencies. Refer to a technical report written by researchers at UC-Berkerly [6], the most energy-efficient way of scheduling a computation is to put all hardware into the highest-performance state and race to complete as quickly as possible. Then drop the hardware to low power modes. Therefore n the second part of the present study, we focus on deep understanding and modeling of MapReduce performance– as a famous distributed/network-based computing service in application/service layer of public clouds such as Amazon Elastic MapReduce. These understanding and modeling of MapReduce performance give valuable tools for smart job scheduling in the middleware layer of a cloud system to execute jobs in an efficient completion time; thus performance prediction and automatic tuning of MapReduce is addressed by proposing two statistical-based algorithms as follows. In the first algorithm, statistical pattern matching analysis is used to find similar CPU utilisation time patterns between two MapReduce applications. The idea is that if two applications are considered 'computationally similar' for short data files, they will be fairly 'similar' for large data sizes too. The aim is to find an optimal number of map tasks and number of reduce tasks for running a new unknown MapReduce application by first categorising it based on its CPU utilisation patterns, and then estimate its optimal running parameters based on similar applications in the same category/class. In the second algorithm, statistical regression is used to model dependency between the MapReduce network traffic load of applications and two main MapReduce configuration parameters (i.e., number of map tasks, and number of reduce tasks). These statistical techniques can form the basis for maximizing the performance of MapReduce framework by automatic tuning of workloads.

The rest of this thesis is organised as follows. Chapter 2 highlights the related work in energy efficient task scheduling in DVFS-enabled processors. Chapter 3 describes our first algorithm for deploying the highest and lowest frequencies of a DVFS-enabled processor to execute tasks. In Chapter 4, multiple frequency selection of all frequencies in such processors is used to execute task graphs considering energy consumption minimisation. The experiments in this chapter lead to some observations on using multiple frequencies in DVFS for energy minimisation; these observations are mathematically proved in Chapter 5. The second part of the thesis starts with Chapter 6, in which an in-depth survey of the state-of-the-art resource provision techniques in distributed computing systems, including MapReduce, is presented. Chapter 7 reports on our new uncertain pattern matching algorithm for finding applications with a similar CPU time pattern. Chapter 8 describes the dependency analysis and the analytical approach we have used to profile, model and predict the network load of MapReduce applications, followed by the conclusions of the study and future research directions in Chapter 9. For the reader's convenience, Chapters 3, 4, and 5 from the first part and Chapters 7 and 8 from the second part are written in a self-contained fashion.

# Part I- Energy Efficient Distributed Computing Using Dynamic Voltage-Frequency Scaling

# Chapter 2. Background and Literature Review on Energy Efficiency in Distributed Computing Systems

## 2.1 Introduction

Research on low power systems has received a great amount of attention in recent years since the sustainability of current technologies and practices has become a serious issue. A few example systems where lowering power usage is critical are the following:

- *Wireless sensors:* several sensors extract data from the environment concurrently, transmit this data to a processing unit and receive processed data accompanied by appropriate commands from the processing unit [7, 8]. The sensors and their receiver/transmitter are generally powered by battery and/or solar cells.

- *Satellite circuits:* Satellites typically involve a massive number of complex circuits that must work in low power. These circuits are supplied by solar cells, the only available power supply in satellites.

- *Robots and surveillance devices:* these devices are heavily used in the army, mine extraction and in difficult or unsafe environments for humans.

- *Cell phones and laptops:* these devices are powered by batteries which are expected to work for a long time.

Stiff increases in energy price and the environmental impact of carbon dioxide emissions associated with energy generation and transportation have forced the issue of reducing energy consumption to be extended to a broader range of systems including High Performance Computing Systems (HPCS).

Various issues such as resource management at both the software and hardware levels must be addressed to reduce energy consumption in HPCS. An important issue in hardware resource management is how to reduce power usage in processors. In the recent past, many hardware-based approaches have been made to efficiently reduce energy consumption, particularly for processors. Dynamic voltage-frequency scaling (DVFS) is perhaps the most appealing method incorporated into many recent processors. Energy saving with this method is based on the fact that the power consumption in CMOS circuits has a direct relationship with frequency and the square of voltage supply. Thus, the execution time and power consumption can be controlled by switching between a processor's frequencies and voltages. Although this approach was initially designed for single processor task scheduling [9], it has recently received much attention in multiprocessor systems as well [3, 10].

DVFS technique and task scheduling can be combined in two ways: (1) schedule generation, and (2) slack time reclamation. In schedule generation, tasks graph are (re)scheduled on DVFS-enabled processors in a global cost function including both energy saving and makespan to meet both energy and time constraints at the same time [11, 12]. In slack time reclamation, which works as a post-processing procedure on the output of scheduling algorithms, the DVFS technique is used to minimise the energy consumption of tasks in a schedule generated by a separate scheduler. However, the existing methods based on the DVFS technique have two major shortcomings: (1) most of them focus on schedule generation and do not adequately take the slack time reclamation approaches into account to save more energy; and (2) the existing slack time reclamation methods use only one frequency for each task among the discrete set of a processor's frequencies. Using one frequency usually results in uncovered slack time where the processor and other devices only waste energy.

In this chapter we focus on slack time reclamation and propose a new slack time reclamation technique, Multiple Frequency Selection DVFS (MFS-DVFS). The key idea is to execute each task with a linear combination of more than one frequency, such that the combination results in using the lowest energy by covering the whole slack time of the task. We have tested our algorithm with both random and real-world application task graphs and compared it with the results of previous research in [10]. The experimental results show that our approach can achieve energy usage almost identical to the optimum energy saving.

## 2.2 Energy Efficiency in HPCS

Many electronic systems in our life, such as satellite systems, cell-phones, game instruments and so on, use rechargeable batteries as their power supply. Although the battery capacity has grown significantly in recent years (battery capacity increases 5% per year), battery life is still the major drawback for most electronic systems. In addition to power-aware battery-based systems, the issue of energy consumption has recently attracted a great amount of attention in high performance computing systems (HPCS). The energy consumption issue in such systems can be classified into three groups: (1) system-level resource allocation, (2) service-level energy-load distribution, and (3) task-scheduling level (Figure 2-1).

At the system-level, the problem is how to distribute computational resources (e.g. CPU, network, memory and I/O) between large scale data storage and processing centres (such as supercomputers and data centres). Fairly distributing resources among applications (or services) not only requires obtaining individual adaptation of resources but also requires an understanding of the interaction between individual resources when they work as a system. Therefore, the big challenge here is to find both the relationships among system resources and their trade-off, which may result in an optimal balance between performance, QoS (quality of service) and energy consumption [13]. Among the different technologies at system-level for managing resources between workloads, virtualisation becomes a key technology in data centres. Virtualisation allows the computational resources to be shared between different workloads. Many of the incoming workloads to data centres are medium size workloads which often require a small fraction of the computational resources. The servers typically spend around 70% of their maximum power consumption even in low utilisation. With virtualisation, such workloads can be run within a virtual machine (VM), causing a significant saving in overall energy usage. The associated VMs may require fewer resources and therefore they can be run on a single hardware unit. It is obvious that less hardware is used in overall, so less energy is wasted for both working on and cooling of the servers.

At the service-level, energy reduction concerns load balancing, scheduling and mapping workloads. The main challenge is to utilise appropriate algorithms to both multiplex/demultiplex workloads in order to save energy and make a trade-off between performance and service cost reduction due to energy savings. Also, to avoid hotspots in data centres due to high-loaded nodes, services can be moved from nodes with high load and high temperature to nodes with a smaller load and lower temperature. Generally, this movement of services should happen when the destination nodes can operate the services in an energy efficient way [13].

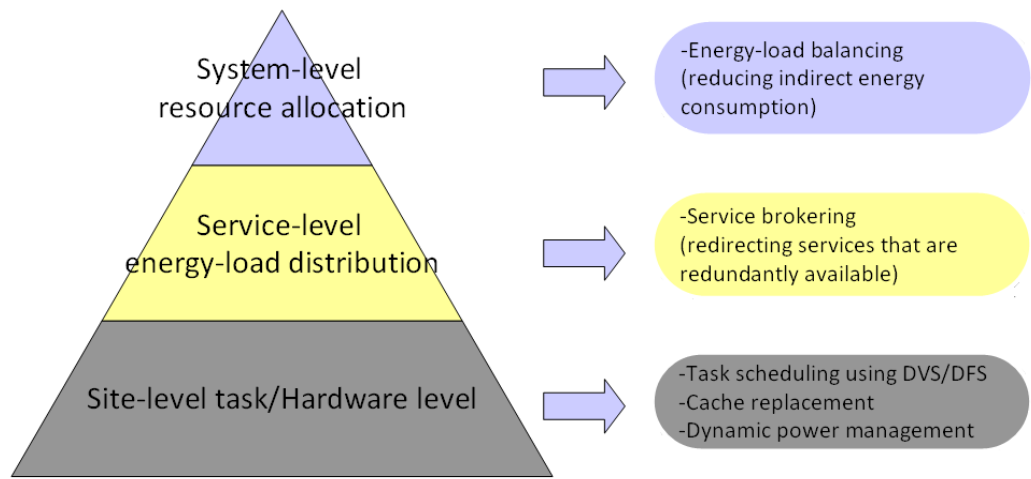

Figure 2-1. Energy consumption levels in HPCS.

At the site level/hardware level, the focus of this chapter, the operating system (OS) and hardware configuration, such as dynamic power management, micro-architecture techniques and dynamic voltage scaling, are used to decrease power. Here, the typical question could be: what is a suitable OS/hardware configuration to process tasks in the shortest possible time and with minimum energy?

## 2.3 Dynamic Power Management (DPM)

Dynamic power management (DPM) is an operating system-level mechanism to dynamically configure the hardware. This mechanism tries to use the minimum number of active components for the requested services and performance [14]. Generally, computing systems and their components are utilised non-uniformly by workloads during operation time. Therefore by dividing system operation time into different working states, the next state of workload can be predicted regarding the previous states of workload in combination with the current state.

DPM is a set of computational techniques used to save power by selectively turning on and off system components or moving them to low-power states in idle time. Typically, there is a power manager (PM) to control the components' working states. The PM makes decisions based on observing how the workload behaves during runtime and making some assumption about the workload. The PM then follows some policies to manage power. For example, one simple policy which is already used in laptops is turning off the system after spending pre-defined seconds of time in idle. In the literature, several studies on DPM-based task scheduling can be found. In [15], the authors present an online greedy scheduling algorithm for independent tasks. The algorithm attempts to reduce energy consumption by ordering task execution so that devices can have continuous idle periods. Moreover, reducing the number of device on/off events shrinks transition delays between device operating states. Unlike most real-time DPM techniques which are CPU-centric, the authors in [16] focus on I/O devices by proposing an offline branch-and-bound algorithm that aims to find the optimal energy schedule for a given task set and meet all real-time deadlines. As the operating system (OS) observes the relationship between processes and hardware devices, in [17], an approach to reduce energy consumption of I/O devices in interactive systems is proposed. In this approach, after estimating the utilisation of a device from each running process on the system by the OS, a device goes into a low-power state if it is not running any process. Also, it is proved that even when no timing dependency is considered, solving optimal low-power task scheduling for DPM on multiple devices is categorised as an NP-hard problem.

### 2.3.1 Simple shutdown

It is important to note that clock gating, a power-saving technique utilised in many synchronous circuits, does not decrease the loss of power [14]. First, power loss still exists in a clock circuit when there is local clock gating or the clock generator is active. Second, leakage currents in components always waste power even if all clocks are off. Therefore, clock gating may be an inefficient way to save power in battery-based devices. A simple but useful policy for battery-based devices is simply turning off them in the idle time. The main benefit of this policy is that it can be used in almost all electronic systems, both digital and analog, sensors and transducers. However, the major drawback is the time of activating the device (as the components must be initialised), which typically is more than the clock gating.

### 2.3.2 OSPM

Microsoft OnNow is a power management tool at the operating system level which supports OSPM (Operating System-directed configuration and Power Management) in personal computers with the capability to move to different power states. This method uses the hardware capability of computers to move to sleep mode instead of turning off [14]. The general policy in OSPM is that the operating system automatically sends the computer to sleep mode when the computer is in idle for some time or when the user press a button on the front of the computer to indicate that the session is completed. In sleep mode, processor and peripherals do not work, but some components are still working to capture events. If an event happens in either software or hardware, the system wakes up and goes to the active state.

With OnNow, the operating system can control the global power-state transitions. Depending on the global power policy, the system may move to sleep depending on the reports coming from the hardware. Also, the current running applications can be configured to wake up the system. As each component of the system has its own power states, it can go to sleep independent from other components even when some parts of system are working. Therefore, another policy of the system is to adapt the special requirements of each application with both system capabilities and operating system information in order to save energy without a harmful influence on the work that the user is doing.

### 2.3.3 Advanced Configuration and Power Interface (ACPI)

Advanced Configuration and Power Interface (ACPI), supported by Intel, Microsoft and Toshiba, is a standard to simplify OSDM by introducing a standard interface to control system resources [14]. This standard, however, does not provide a unique and proper procedure to control all systems efficiently. For example, ACPI-related policies are different between desktop computers and laptops. Therefore it is necessary to apply different policies on a system and examine which combination of policies can help the system work most energy efficiently. A study to find the best policies for desktops and laptops has been published [18]. Among existing operating systems, Microsoft Windows 98 is the only one that fully supports ACPI, while Windows XP/Vista/200, Linux and the desktop version of SunOS partially support ACPI [14].

## 2.4 Micro-Architecture Techniques

The increase of circuit temperature due to the sharp increase in the number of transistors per cm2 has led circuit designers to enhance and modify the architecture and design of microprocessors (such as on-chip system integrations, multi-clock frequency implementation, and multiple voltages). As in CMOS circuits, there is a direct relationship between the clock frequency and power consumption and consequently circuit temperature; thus, the issue of power consumption in a circuit can easily limit the maximum clock frequency. In other words, in order to face this problem, power/performance and power/cost optimisation must be considered when designing a new microprocessor. Two major micro-architecture techniques to power reduction are adaptive architecture and cache management [14].

### 2.4.1 Adaptive architecture

Since there is strong dependency between the performance of DPM policies and workload statistics, a static prediction of workload becomes ineffective when the workload is either unknown or non-stationary. Therefore, DPM requires some form of adaptation. The idea of adaptation is based on predicting workload timer duration, or sometimes using several timeouts for non-stationary workloads. Krishnan et al in [19] use a set of timeout values where each timeout is associated with an index to show how well the timeout performs. The timeout which performs best among the set of available ones is chosen as the adaptation policy. Another study by Helmbold et al in [20] keeps a list of candidate timeouts and assigns a weight to each timeout. The actual timeout is the weighted average of all candidates. To find more information about prediction policies, refer to Langen and Juurlink in [21].

### 2.4.2 Cache management

Memory is one of the most power-hungry components in most computer systems. Based on a report from IBM eServer machine [22], memory uses about 40% of the total energy, which ranks it as the second largest power consumer in computing systems after the processor. In addition to computing systems, memory is still one of the biggest power consumers in PDAs and laptops. The reduction of memory power consumption is important, as in most computing systems memory works and consumes power continuously.

Power-Aware Virtual Memory (PAVM), proposed by Huang et al. in [23], is a recent and powerful scheme to reduce memory power consumption by allocating active pages to the same memory. This scheme is used for the main memory and at operating system level. The idea behind this schema is that it turns on the active ranks of process $p$ and turns off inactive ranks while $p$ is being scheduled. A rank is defined as an active rank of process $p$ if, and only if, at least one page from the rank is mapped into the address space of p. The main drawback of this schema is the amount of buffer cache it uses. Buffer caches are used as a bridge between the hard disk and main memory. Increasing total memory size results in increasing the amount of buffer cache and therefore occupying a bigger part of the memory. To find more information about different cache management techniques based on the PAVM scheme refer to [22-24].

## 2.5 Dynamic Voltage-Frequency Scaling (DVFS)

Dynamic voltage-frequency scaling is a modern technique in computer architecture to reduce energy consumption of microprocessors or control the amount of heat generated by circuits. This technique is commonly utilised in battery-based devices such as laptops and cell phones where decreasing the energy usage of the battery is necessary. In addition, DVFS is used in high-computing nodes not only to decrease the power of the nodes, but also to save more energy to cool down the nodes. An approximation model shows that the dynamic power in CMOS circuits is a linear function of both switching frequency and voltage square as: $C.f.V^2$, where C is the effective switching capacity per clock cycle. Therefore, a workload (or task) can save more energy when it is executed in lower voltage and frequency. In general, a computing node executes several tasks with inter-task relationships (e.g., precedence constraints) simultaneously. These inter-task relationships typically incur slack time (idle time) between tasks which can be used by DVFS to reduce energy usage. Specifically, the slack time associated with a task is utilised to execute the task in a lower voltage-frequency; this in turn results in energy reduction.

In recent years, there has been a significant amount of work on low-power task scheduling using the DVFS technique, especially in real-time embedded systems and HPCS. The main idea of most existing algorithms is that the processor slack time should be filled by switching the processor operating frequency to the lowest possible frequency; this changes the DVFS problem to that of estimating the processors' slack time and the tasks' timing information (e.g. task deadline, task release time and task execution time). Based on the tasks' timing information availability, energy-aware task scheduling in embedded systems is categorised into two groups: real-time scheduling and non-real-time scheduling.

In real-time scheduling, task release time (i.e., arrival time) and total number of CPU cycles needed to complete each task (i.e., task execution time) are unknown *a priori*. There are many studies that apply DVFS in real-time scheduling scenarios. In [25] and [26], low-power scheduling of multi tasks while meeting given hard timing constraints in the operating system (OS) was studied. The main assumption in these coarse-grained DVFS approaches – which is difficult to achieve in practice – is that the total number of CPU cycles required to execute each task is fixed and available *a priori*. In [27], an energy-aware scheduling algorithm is formulated into a mathematical programming formulation of the scheduling and voltage/frequency assignment problem. To speed up the algorithm, a schedule table at design time is constructed to provide multiple scheduling options for each task; these multiple scheduling options may use complex algorithms to build the schedule table at design time.

A stochastic allocation, non-linear integer programming-based scheduling algorithm is proposed in [28] for multiprocessor platforms. In addition to exact analytic formulation of the stochastic objective function based on the task graph analysis for scheduling, the authors also extend the timetable constraint for conditional activities to improve their stochastic resource allocation. In [29], a probabilistic distribution of all tasks' execution time is used to better partition the workload and consequently reduce more energy consumption. Then a polynomial-time heuristic method is proposed to convert the problem of scheduling, which is NP-hard, into a probability-based load balancing problem; this problem then is solved with worst-fit decreasing bin-packing heuristic. The authors in [30] consider uncertainty in execution times of tasks, and model each varied execution time as a random variable. Then two algorithms are proposed for both the uniprocessor and multiprocessor using a probabilistic approach; these algorithms are designed to both minimise total energy consumption of the system and satisfy timing constraints with a guaranteed confidence probability. However, their algorithms for multiprocessor platforms suffer from exponential complexity.

In energy harvesting systems – embedded systems working in an environment without access to energy power – low-power scheduling using DVFS increases their lifetime. In such systems, tasks are periodic and generally simple; also their schedulers try to find an optimal trade-off between energy consumption and deadline miss rate of tasks – which means time constraints are not hard in such systems. In [31], Adaptive Scheduling DVFS (AS-DVFS) is proposed, which simplifies the original scheduling problem by decoupling/separating timing and energy constraints; then, AS-DVFS adjusts the processor/processors operating frequency: (1) based on workload information, and (2) under the timing and energy constraints, towards achieving the whole system energy efficiency. A scheduling algorithm in [32] reduces energy consumption: (1) by assigning tasks to the processing element with lower operating frequency, and (2) migrating these tasks among processors. The authors in [33] propose a Utilisation Based (UTB) algorithm that combines energy harvesting awareness, DVFS, and task slack time management to reduce energy consumption of periodic tasks on multiprocessor systems. The proposed low-complexity task scheduling algorithm is based on the concept of task CPU utilisation – defined as the worst-case task execution time divided by its period.

In [34], the authors propose an intra-task voltage scheduling algorithm based on a static timing analysis of an application: at first, a given task is partitioned into several segments, and then appropriate supply voltage (resulting from static timing analysis of previous segments and based on the worse-case execution time of the task) is assigned for each segment. The authors claim that this scheduling algorithm has a high energy reduction ratio by fully exploiting all the slack times and choosing a suitable voltage/frequency to fill these slack times as much as possible. A software feedback loop-based method is proposed in [35], where for each time slot a deadline is provided. Then, the processor operating frequency of the current slot is calculated based on the following: (1) the slack time generated in the previous slot, and (2) the worst-case execution time of the current slot. The main assumption in both mentioned DVFS methods is that the worse-case execution time of each segment/slot of a task is known, which is difficult to reach in many applications; for instance, in MPEG decoding it is too hard to precisely estimate/calculate the worse-case execution time of each frame. Although calculating a single unique worse-case execution time for all frames in a specific video can be an option – and also easy – it cannot significantly use the potential of DVFS-based techniques to reduce energy consumption. The authors in [36] focus on energy-efficient MPEG decoding by proposing an effective DVFS algorithm based on future workload prediction; frame-based history is used to predict the computational workload of an incoming frame. This prediction gives this opportunity to pick a suitable voltage/frequency for the processor, and provide the exact amount of computing power to decode the frame.

Therefore, the decoding time of a frame is divided into two parts: the frame-dependent part and frame-independent part. The statistical variation in the frame-dependent part is compensated by the frame-independent part; this allows a considerable energy saving with less degradation of QoS.

In non-real-time scheduling, the timing information of tasks is available in compile time. Having this information allows schedulers (such as the list scheduler) to be developed by maximising processor utilisation to meet all deadlines [9, 11]. This type of scheduling is used for most large-scale computational problems such as bioinformatics, chemistry and object recognition in machine vision applications [37]. In [38], tasks are first scheduled using a list-scheduling algorithm where the mutual of the slack time is used as the tasks' priorities. Then, these slack times are distributed among tasks on each critical path; thus, processors can move to lower frequency/voltage and consume less energy. In [39], scheduled tasks are re-scheduled by an extended list-scheduling algorithm. At each time step, the algorithm first calculates the energy saving of a task when it is scheduled at the current step and the next step. The difference between the energy of these two steps is represented as the task's energy saving. A task with a higher energy saving and lower slack time gets a higher priority to be scheduled. A two-phase solution is proposed in [40] where a version of an early-deadline-first scheduling algorithm is utilised in the first phase for assigning a task to a best-fit processor with regard to the task ready time and the processor free time; in the second phase, the proper voltage/frequency of the processor to run each task is then is solved by an Integer Linear Programming (ILP).

There are many algorithms in the literature for energy-efficient real-time and non-real-time scheduling embedded systems. These algorithms are suitable for platforms with a small number of processors, and mostly assume the tasks (periodic or aperiodic) are independent; however, a few algorithms have been developed for reduction of power consumption in HPCS using DVFS and DPM. In [41], the authors present a theoretical framework for energy-efficient e-business datacentres and introduce an automatic power and performance management methodology; their methodology uses a mathematically-rigorous optimisation technique to optimise performance/watt with regard to meeting performance constraints and minimising energy consumption. There are a few optimisation techniques for online scheduling algorithms, which generally work by deploying popular heuristic algorithms like MET, Min-Min, Max-Min, OLB, or fast greedy [42, 43]. Moreover, in [42, 44, 45], meta-heuristic algorithms such as Simulated Annealing and Tabu have been proposed.

The authors in [46] study the effect of virtualisation for power-aware consolidation and propose a dynamic configuration technique/algorithm to effectively optimise the power in virtualised server clusters, and dynamically manage it. In the case of consolidation of multiple services/applications on such virtualised clusters, a dynamic configuration was developed based on a mixed integer programming (MIP) formulation [47]. This power-efficient approach also takes into account the cost of servers frequently switched on/off. However, this approach is too slow to make a proper decision for an online scheduler. In [48], Virtual Machine (VM) is used for executing HPC applications to reduce energy consumption of virtualised datacentres by supporting VM migration and VM placement optimisation with less human interaction; in addition, the overhead of virtualisation is taken into account. In [49], the workload is distributed to different datacentres at various locations in order to minimise power consumption and guarantee SLA. Another dynamic job-scheduling policy is proposed in [50] for power-aware resource allocation in a virtualised datacentre. In addition to considering the overhead of virtualisation, the technique tries to gather workloads from separate machines into a smaller number of nodes, turns off more servers and thus reduces the overall datacentre power consumption.

In [51], the Just-in-time DVFS technique was presented to fill slack time in MPI programs. A system called Jitter was utilised to reduce the frequency on nodes with more slack times and fewer computations. The goal of Jitter was to be sure that the tasks came just in time without increasing overall execution time. Ge et al. in [3] applied the DVS technique on processors that did not work at peak performance during execution of a parallel application. The best processor frequency of each task was selected by analyzing computation and communication power profiles collected before execution. A method to reduce power consumption was presented in [12] by adaptively activating and deactivating hardware resources, especially memory for intensive HPC applications. Cache missing when accessing the main memory also plays a great role in adjusting and triggering processor slack times.

Lee and Zomaya in [12] presented a DVFS-based algorithm to minimise both completion time and energy consumption of precedence-constrained parallel jobs on HPC systems. This method tried to minimise a summation of two cost functions: completion time and energy. Therefore, the final result was a trade-off between the quality of scheduling and energy consumption. Ding et al. in [52] introduced the concept of energy scalability in formal terms. In addition to studying the energy efficiency/iso-efficiency concept, they extended an analytical model to study tradeoffs between performance and energy saving in HPCS. In [53], the slack times in real-time applications were classified into static, work and shared slack time groups for multiple dependent tasks on multiple DVFS-enabled processors. Then a dynamic dependency-aware task scheduling was proposed to adjust voltage/frequency of each processor regarding the tasks' real time deadlines. Hotta [54] presented a profiled-based power-performance optimisation method utilising DVFS in HPCS. Here, the execution of a program was divided into several regions. In trial steps, profile information of each region (including power and execution profiles) was extracted and then utilised to find the best combination of processors' voltages and frequencies.

The authors in [55] present their power-aware scheduling algorithms for bag-of-tasks applications with deadline constraints on DVS-enabled cluster systems; these scheduling algorithms try to both minimise power consumption and meet the deadlines specified by application users. In [56], an upper limit for system energy usage was chosen externally. Then, a combination of performance modeling and performance prediction was used to reduce execution times with respect to their predefined energy usage upper limit. After creating models for both execution time and energy consumption, key parameters of the models were estimated by executing a program a small number of times and then regressing the estimated parameters. Thus, for better estimation of parameters, the following steps were iterated until a proper schedule is achieved: (1) using models to predict each possible scheduling of tasks, (2) executing the program a few times with the best predicted schedule, and (3) updating estimated key parameters. Rountree et al in [57] proposed an energy-aware schedule generation algorithm for DVFS-enabled processors where a combination of all processor frequencies is involved in an overall linear programming optimisation. An energy reduction algorithm was proposed by Kimura et al in [10] for a power-scalable high performance cluster supporting DVFS. In a simplified version of this algorithm, a suitable frequency among a discrete set of the processor's frequencies is chosen for each task with regard to each task's slack time.

### 2.5.1 DVFS and DPM combination techniques

It is worth noting that there are several studies in the literature that investigated reducing power consumption by combining both DVFS and DPM. A Markovian decision processes-based DPM model is presented by the authors in [58], which is a uniform modelling framework for both DVS and DPM. Another stochastic approach is proposed in [59] where the authors combine the DVS and renewal theory-based DPM approach. Despite their effectiveness, these two stochastic approaches cannot handle tasks with hard deadline constraints or dependency; the hard deadline constraint is addressed in [60].

# Chapter 3. Linear Combinations of DVFS-enabled Processor Frequencies to Modify the Energy-Aware Scheduling Algorithms (first algorithm)

## 3.1 Introduction

To the best of our knowledge, most DVFS-based methods use only one frequency to process a task in graph. In this chapter, a new model (called Maximum-Minimum-Frequency DVFS algorithm or MMF-DVFS) is described. Unlike slack time reclamation in most existing DVFS-based algorithms which use only one frequency to process a task in graph, MMF-DVFS uses a linear combination of the highest and lowest frequencies of a processor to approach better energy consumption. As part of the present study, an extensive set of experiments and comparisons are conducted. Before describing the algorithm, some preliminaries about system and application models and the energy model of processors are explained. These preliminaries also apply to Chapters 4 and 5.

## 3.2 Preliminaries

### 3.2.1 System and application models

A parallel computing system is comprised of $N$ homogeneous processors with individual memories. In such systems, switching time between frequencies can be safely ignored in processors because the time taken to switch from one frequency to another ($30-150\mu\sec$ [61]) is significantly smaller than execution time of tasks (at least *1 ms*).

A set of dependent tasks $\left\{A^{(1)}, A^{(2)}, ..., A^{(M)}\right\}$ represented by a directed acyclic task graph (DAG) is also assumed to be executed in the modelled HPC system. Here, the $k^{\text{th}}$-task ($A^{(k)}$) has the following five parameters: $T^{(k)}$ is the whole of the available time a processor can assign to the task – a summation of the task's execution and slack time (Figure 3-1a); $t_i^{(k)}$ is the task execution time when frequency $f_i$ is used; $f_{ideal}^{(k)}$ is the ideal continuous frequency based on [25], which results in the optimum energy consumption (Figure 3-1c); $K^{(k)}$ is the required number of clock ticks (i.e. clock cycles) the task needs for its execution; and $t_{OS}^{(k)}$ is the time the processor spends on executing the task in the original scheduling (Figure 3-1a).

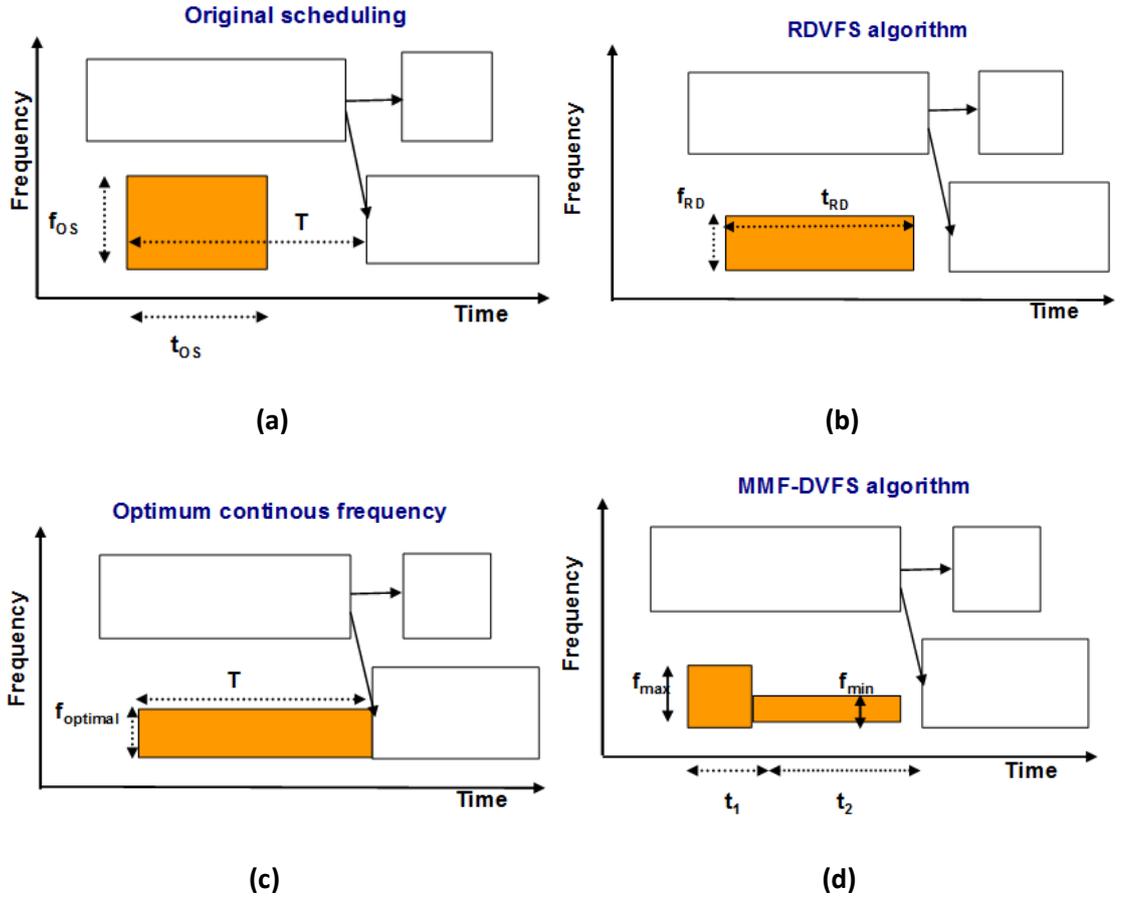

**Figure 3-1. Time representation of MMF-DVFS and other algorithms: (a) Original scheduling, (b) RDVFS algorithm, (c) optimum continuous frequency, and (d) MMF-DVFS algorithm.**

### 3.2.2 Energy model

DVFS−enabled processors can execute a task by using a discrete set of voltage-frequency pairs, $(f_i, v_i)$, in which $\{v_1 < v_2 < \cdots < v_N\}$ and $\left\{ \underbrace{f_1}_{f_{min}} < f_2 < \cdots < \underbrace{f_N}_{f_{max}} \right\}$. In CMOS-based processors, the power consumption of a processor consists of two parts: (1) a dynamic part that is mainly related to CMOS circuit switching energy, and (2) a static part that addresses the CMOS circuit leakage power. The whole power consumption ( $P_d$ ) is estimated as [61]:

$$\begin{cases} P_d = \lambda f v^2 + \mu v \\ f \propto \dfrac{(v - v_t)^2}{v_t} \end{cases} \qquad (3-1)$$

Where $f$, $\lambda$ and $v$ represent the processor's working frequency, the effective capacitance, and the processor's working voltage, respectively. Note that $v_t$ is a threshold voltage usually provided by a manufacturer. The general relationship between voltage, frequency and power is:

$$If \ (f_i, v_i) < (f_j, v_j) \ \Rightarrow P_d(f_i, v_i) < P_d(f_j, v_j) \qquad (3-2)$$

The overall energy consumption of $k^{\text{th}}$-task ( $A^{(k)}$ ) in a DAG is calculated as:

$$E^{(k)} = P_d t_i^{(k)} + P_l \left( T^{(k)} - t_i^{(k)} \right) \qquad (3-3)$$

where $P_l$ is the energy a processor consumes when it is in idle. With the almost-always true assumption that $\mu v$ is constant, and $v \gg v_t$, the relationship between frequency and voltage becomes proportional, i.e., $f \propto v$; therefore, Eqn. 3-1 is simplified as:

$$P_d = \lambda f^3 + \gamma \qquad (3-4)$$

### 3.2.3 Optimum Continuous Frequency

The optimal approach to remove slack time and, as a result, reduce the energy consumption of a processor, is for the processor to perform a task using a continuous frequency (Figure 3-1c). Before moving further, it is necessary to prove the following theorems:

**Theorem 1:** If $f_1$ and $f_2 (> f_1)$ execute a task in $t_1$ and $t_2$, respectively. Then,

$E^{(k)}(f_1, t_1) < E^{(k)}(f_2, t_2)$.

**Proof:**

**Figure 3-2. RDVFS algorithm**

$$E^{(k)}(f_2, t_2) - E^{(k)}(f_1, t_1)$$
$$= (\alpha f_2^3 + \gamma)t_2 + P_{Idle}(T^{(k)} - t_2) - (\alpha f_1^3 + \gamma)t_1 + P_{Idle}(T^{(k)} - t_1)$$
$$= \cdots = (f_2 - f_1)[\alpha f_1 f_2(f_2 + f_1) - \gamma + P_{Idle}] \geq 0$$

As generally $P_{Idle} > \gamma$, therefore the theorem 1 is proved.

**Theorem 2:** If processor frequency is continues (unrealistic assumption), the optimum energy for $k^{\text{th}}$-task is obtained when the task covers the whole task's slack time $(T^{(k)})$.

**Proof:** the result in theorem 1 shows that when a frequency covers the whole slack time it gives the optimum power consumption. Note that this frequency may not exist unless the frequency set is continuous.

Referring to theorem 2, for $k^{\text{th}}$-task $(A^{(k)})$, the optimum continuous frequency and its related energy are defined as $f^{(k)}_{Opt-cont.}$ and $E^{(k)}_{Opt-cont.}$ and are calculated as:

$$\begin{cases} f_{Opt-cont.}^{(k)} = f_N \dfrac{t_{OS}^{(k)}}{T^{(k)}} \\ E_{Opt-cont.}^{(k)} = \left(\alpha\left(f_{Opt-cont.}^{(k)}\right)^3 + \gamma\right) T^{(k)} \end{cases} \qquad (3-5)$$

In actual systems, however, frequencies must be chosen from a discrete set of frequencies. Also, finishing a task by its deadline may require choosing a frequency that is faster than the optimal frequency. Therefore, the underline{optimal discrete frequency} of $k^{\text{th}}$-task is the first frequency in the discrete set larger than $f_{Opt-cont.}^{(k)}$. This discrete frequency and its associated time are $f_{RD}^{(k)}$ and $t_{RD}^{(k)}$, respectively. The algorithm calculating this frequency is referred to as RDVFS for our comparison [10].

### 3.2.4 Reference Dynamic Voltage-Frequency Scaling (RDVFS)

RDVFS is a simplified version of the algorithm introduced by Kimura et al in [10] for power-scalable high performance clusters supporting DVFS. It reduces the energy consumption of processors by selecting the smallest available processor frequency ($f_{RDVFS}$) capable of finishing a task in a given time frame (Figure 3-1b). The details of RDVFS algorithm are shown in Figure 3-2.

For each task assigned to a processor, $f_{RDVFS}^{(k)}$, which is the first frequency larger than optimal frequency ($f_{Opt-cont.}^{(k)}$) calculated from Eqn. 3-5, is likely to be the best discrete frequency candidate to execute the task within the given time frame and covering its related slack time. As mentioned before, a major limitation of the RDVFS technique is the usage of only one frequency to execute the task.

## 3.3 Maximum-Minimum-Frequency DVFS algorithm (MMF-DVFS)

Using the simplified energy consumption model in Eqn. 3-4, the power consumption of $k^{\text{th}}$-task in MMF-DVFS algorithm is formulated as:

$$\begin{cases} Minimize: E\left(t_1^{(k)}, t_2^{(k)}\right) = f_{max}^3 t_1^{(k)} + f_{min}^3 t_2^{(k)} + P_{Idle}\left(T^{(k)} - t_1^{(k)} - t_2^{(k)}\right) \\ subject\ to \\ \qquad 1. f_{max} t_1^{(k)} + f_{min} t_2^{(k)} = f_{RDVFS}^{(k)} t_{RDVFS}^{(k)} \\ \qquad 2. 0 \leq t_1^{(k)} + t_2^{(k)} \leq T^{(k)} \\ \qquad 3. t_1^{(k)} \geq 0, t_2^{(k)} \geq 0 \end{cases} \qquad (3-6)$$

and also

$$\begin{cases} E\left(t_1^{(k)}, t_2^{(k)}\right) < E_{RDVFS}^{(k)} \\ E_{RDVFS}^{(k)} = \left(f_{RDVFS}^{(k)}\right)^3 t_{RDVFS}^{(k)} + P_{Idle}\left(T^{(k)} - t_{RDVFS}^{(k)}\right) \end{cases} \qquad (3-7)$$

Eqn. 3-6 is an optimisation problem and represents the power consumption problem: how to choose $t_1^{(k)}$ and $t_2^{(k)}$, with regard to constraints, so that the consumed energy of $k^{\text{th}}$-task ($T^{(k)}$) is minimised. Eqn. 3-7 constrains the outcome of Eqn. 3-6 to be less than $E_{RDVFS}^{(k)}$. For processing $k^{\text{th}}$-task, the processor has to spend the same number of cycles in both RDVFS and MMF-DVFS algorithms, as addressed in constraint 1 of Eqn. 3-6.

### 3.3.1 Algorithm

In this section, it is proved that the optimisation problem for finding $t_1^{(k)}$ and $t_2^{(k)}$ in Eqn. 3-6 can be solved directly.

**Theorem 3:** For $k^{\text{th}}$-task ($T^{(k)}$), the MMF-DVFS algorithm approaches a better solution than RDVFS algorithm when

$$t_{RDVFS}^{(k)} \leq \frac{\left[f_{max}^2 + f_{max}f_{min} + \frac{P_{Idle}}{f_{min}}\right]\frac{f_{max}}{f_{RDVFS}^{(k)}}T^{(k)}}{\left[f_{max}^2 + f_{max}f_{min} + f_{min}^{(k)} - f_{RDVFS}^{(k)}\right] + \frac{P_{Idle}}{f_{RDVFS}^{(k)}}} \qquad (3-8)$$

**Proof:** obtaining $t_2^{(k)}$ in constraint 1 of Eqn. 3-6 and replacing it in constraint 2 results in:

$$\frac{f_{RDVFS}^{(k)}t_{RDVFS}^{(k)} - T^{(k)}f_{min}}{f_{max} - f_{min}} \leq t_1^{(k)} \leq \frac{f_{RDVFS}^{(k)}t_{RDVFS}^{(k)}}{f_{max} - f_{min}} \qquad (3-9)$$

Replacing $t_2^{(k)}$ from constraint 1 into the optimisation problem in Eqn. 3-6 gives:

$$E\left(t_1^{(k)}\right) = t_1^{(k)}\left[f_{max}^3 - P_{Idle} - \frac{f_{max}}{f_{min}}(f_{min}^3 - P_{Idle})\right] + t_{RDVFS}^{(k)}\frac{f_{RDVFS}^{(k)}}{f_{min}}[f_{min}^3 - P_{Idle}]$$
$$+ P_{Idle}T^{(k)} = t_1^{(k)}\alpha + \beta \qquad (3-10)$$

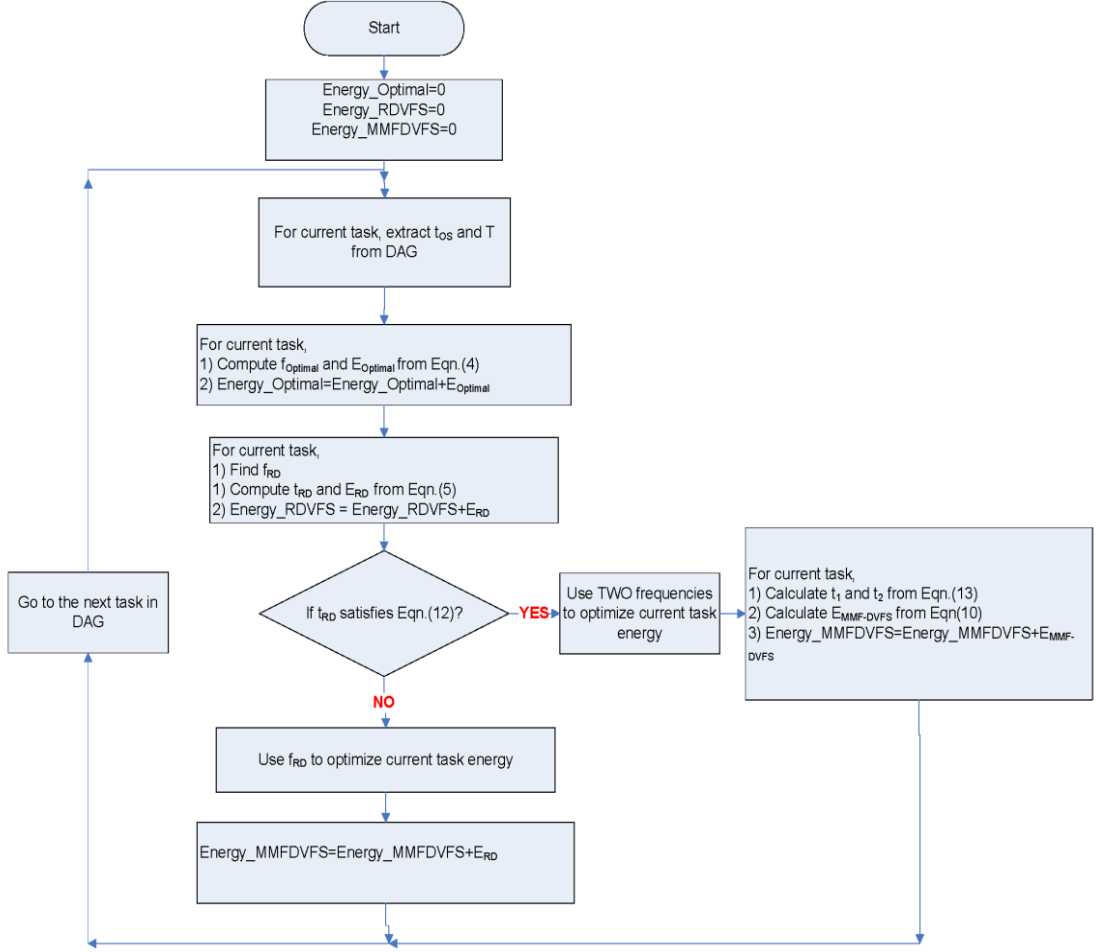

**Figure 3-3. MMF-DVFS flowchart**

By adding Eqn. 3-10 to Eqn. 3-7, new criteria for $t_1^{(k)}$ is now calculated as:

$$t_1^{(k)} \leq \frac{\left[\left(f_{RDVFS}^{(k)}\right)^2 - f_{min}^2 + \frac{P_{Idle}}{f_{min}}\right] f_{RDVFS}^{(k)} t_{RDVFS}^{(k)} - P_{Idle} t_{RDVFS}^{(k)}}{f_{max}[f_{max}^2 - f_{min}^2] + \frac{P_{Idle}}{f_{min}}[f_{max} - f_{min}]} \qquad (3-11)$$

Merging the left side of Eqn. 3-9 and Eqn. 3-11 results in:

$$t_{RDVFS}^{(k)} \leq \frac{\left[f_{max}^2 + f_{max}f_{min} + \frac{P_{Idle}}{f_{min}}\right] \frac{f_{min}}{f_{RDVFS}^{(k)}} T^{(k)}}{\left[f_{max}^2 + f_{max}f_{min} + f_{min}^2 - \left(f_{RDVFS}^{(k)}\right)^2\right] + \frac{P_{Idle}}{f_{RDVFS}^{(k)}}} \qquad (3-12)$$

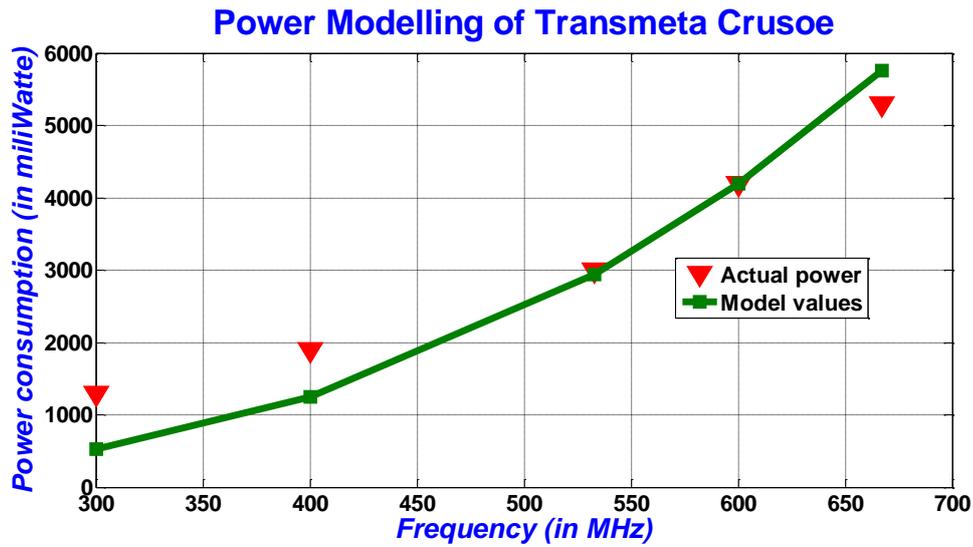

**(a)**

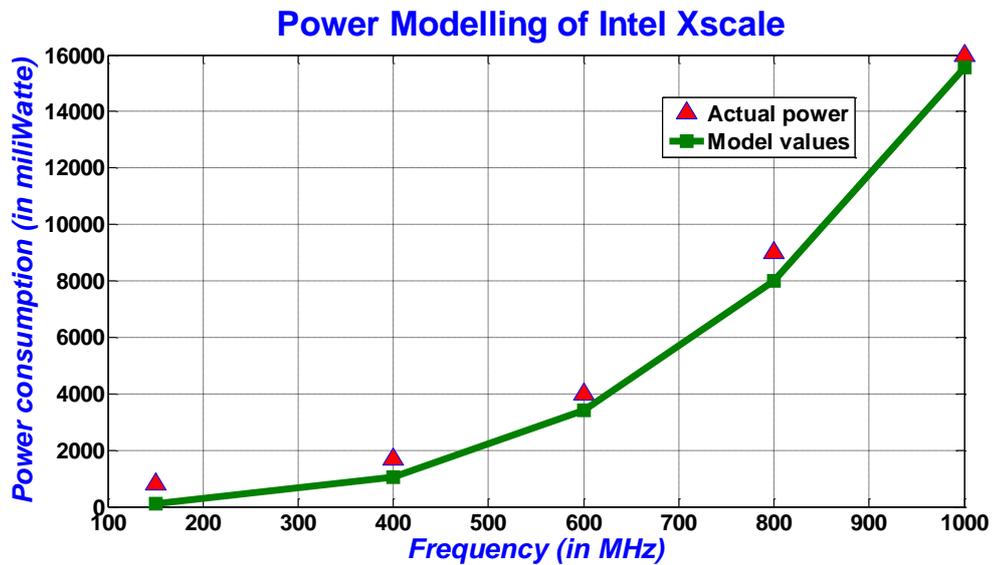

**(b)**

Figure 3-4. The least-square modelling of (a) Transmeta Crusoe, and (b) Intel Xscale processors.

The non-equality in Eqn. 3-12 in fact shows that rescheduling $k^{\text{th}}$-task using MMF-DVFS will lead to lower energy consumption than RDVFS for this task. Therefore, depending on whether the non-equality in Eqn. 3-12 is satisfied for this task or not, MMF-DVFS or RDVFS is used to determine how the task should be executed, respectively. Based on Eqn. 3-10, consumed energy $E\left(t_1^{(k)}\right)$ is an ascent function of $t_1^{(k)}$; therefore, the energy of the task is minimised when $t_1^{(k)}$ has its minimum possible value. Comparing Eqn. 3-9 and constraint 1 of Eqn. 3-6 gives the final value for $t_1^{(k)}$ and $t_1^{(k)}$ as:

$$\begin{cases} t_1^{(k)} = \dfrac{f_{RDVFS}^{(k)} t_{RDVFS}^{(k)} - T^{(k)} f_{min}}{f_{max} - f_{min}} \\ t_2^{(k)} = \dfrac{T^{(k)} f_{max} - f_{RDVFS}^{(k)} t_{RDVFS}^{(k)}}{f_{max} - f_{min}} \end{cases} \qquad (3-13)$$

Figure 3-3 shows the MMF-DVFS flowchart. The flowchart shows that MMF-DVFS works the same as RDVFS when a task cannot satisfy Eqn. 3-12.

## 3.4 Evaluation and Experimental Results

In this section we present the results on energy consumption obtained from simulating our MMF-DVFS algorithm in comparison with optimum continuous frequency, and RDVFS. To compare the algorithms, the following schedulers were used with different numbers of processors: (1) list scheduling, (2) list scheduling with Longest Processing Time first (LPT), and (3) list scheduling with Shortest Processing Time first (SPT). These schedulers are chosen as examples and any other scheduling algorithms can be used instead as our algorithm applies on slacks remained after scheduling and is independent on scheduler itself. The simulations were carried out using the simulator we developed as a part of this study.

### 3.4.1 Simulation settings

The voltage/frequency setting of two real processors are used in the simulations: Transmeta Crusoe [10] and Intel Xscale [62]. Table 3-1 shows the voltage/frequency and related power consumption of these processors along with the convex models of each processor. These models use least-square curve fitting to fit a convex function ($\alpha f^3 + \gamma$) to the frequency-power of two real processors, as shown in Figure 3-4.

The performance of the algorithms is evaluated with two sets of task graphs: randomly generated and real-word parallel applications. The two real-world applications used in our experiments were LU decomposition and Gauss-Jordan with DAGs extracted from [59]. A large number of variations in the number of processors and tasks are used for each application in the simulations. The random task graph set consisted of 1500 graphs with five graph sizes of 100, 200, 300, 400 and 500 nodes,

*Table 3-1. The voltage/frequency setting of two real processors in the experiments with their power consumption and convex models.*

| | Transmeta Crusoe | | |
|---|---|---|---|
| *Level* | *Frequency (MHz)* | *Voltage (V)* | *Power (W)* |
| **0** | 667 | 1.6 | 5.3 |
| **1** | 600 | 1.5 | 4.2 |
| **2** | 533 | 1.35 | 3.0 |
| **3** | 400 | 1.225 | 1.9 |
| **4** | 300 | 1.2 | 1.3 |
| *Convex model* | $P = 1.94 \times 10^{-5} \left(\dfrac{f}{10^6}\right)^3 + 4.44 \text{ mW}$ | | |
| | Intel Xscale | | |
| *Level* | *Frequency (MHz)* | *Voltage (V)* | *Power (W)* |
| **0** | 1000 | 1.8 | 1.6 |
| **1** | 800 | 1.6 | 0.9 |
| **2** | 600 | 1.3 | 0.4 |
| **3** | 400 | 1 | 0.17 |
| **4** | 150 | 0.75 | 0.08 |
| *Convex model* | $P = 1.55 \times 10^{-5} \left(\dfrac{f}{10^6}\right)^3 + 60 \text{ mW}$ | | |

together with three different schedulers on five sets of 2, 4, 8, 16 and 32 processors (Table 3-2).

These task graphs have different numbers of tasks, task distributions, communication costs and task dependencies. The execution cycle of these randomly generated tasks varied from 5–10 million cycles from a uniform distribution. Moreover, 150 real-world application task graphs based on the LU decomposition algorithm are utilised in experiments. For the real-world application graph, the same number of task graphs

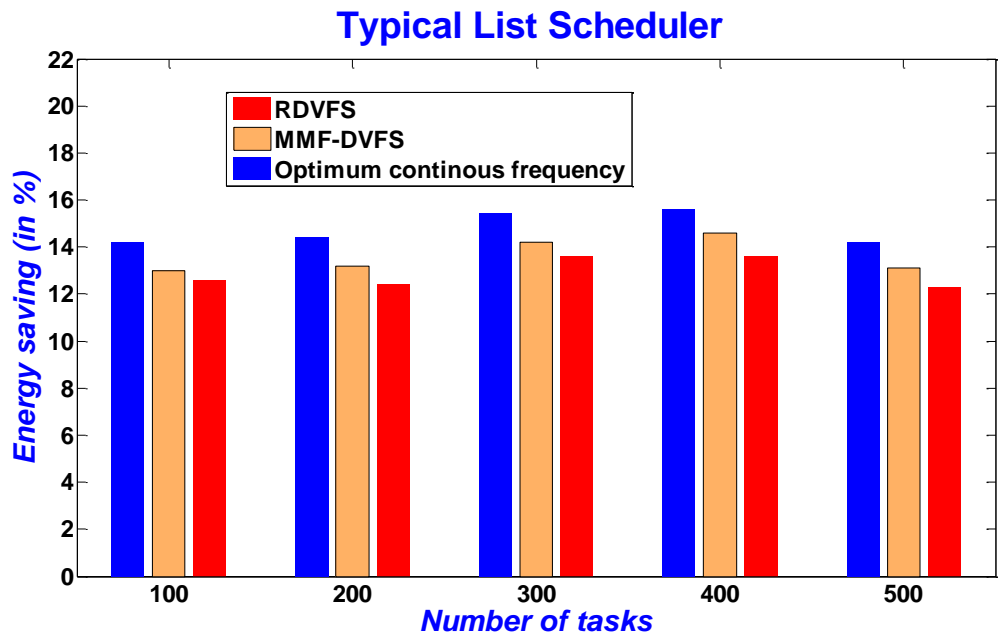

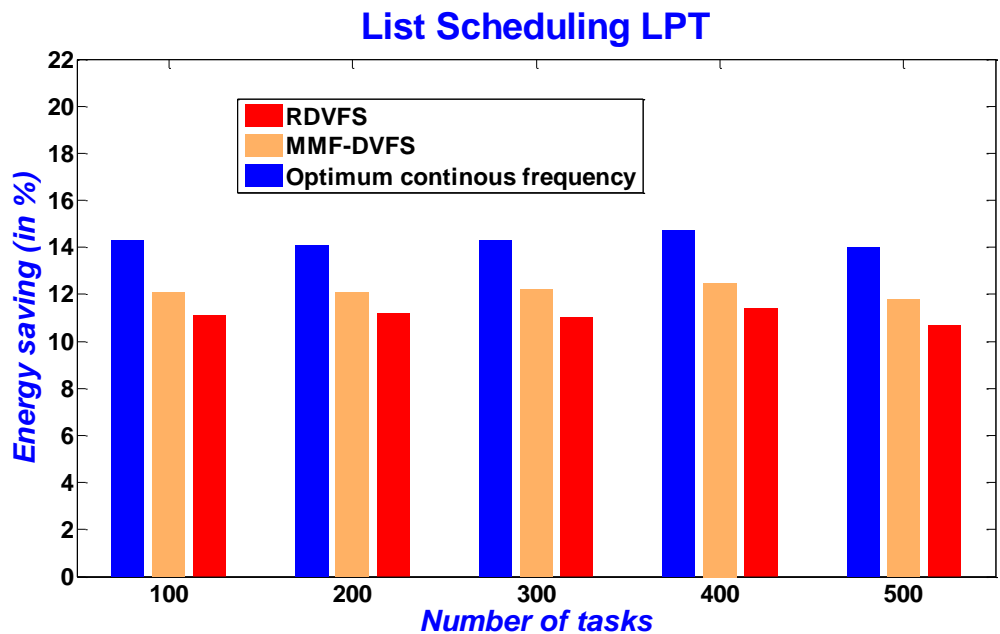

**Figure 3-5. The normalized energy consumption of MMF-DVFS on the number of tasks: The typical list scheduler, list scheduler with Longest Processing Time first (LPT), and list scheduler with Shortest Processing Time first (SPT).**

– ranging from 100 to 500 tasks – was investigated, with three schedulers and on five sets of processors.

### 3.4.2 Results

The simulation results of normalised energy consumption for all DAGs (Figures. 3-5 and 3-6) are shown in Table 3-3. This table clearly denotes the superior performance of MMF-DVFS scheduling compared to the other approaches in all cases. Figure 3-6 shows that although the efficiency of all algorithms, including MMF-DVFS, in saving energy in LU decomposition is significant, these algorithms have poorer performance on Gauss-Jordan tasks. For a deeper examination of this behaviour, a sample three level Gauss-Jordan application job scheduling on three processors is shown in Figure 3-7. Since there is no idle time among tasks in Gauss-Jordan graphs applications, none of these algorithms can efficiently reduce energy consumption.

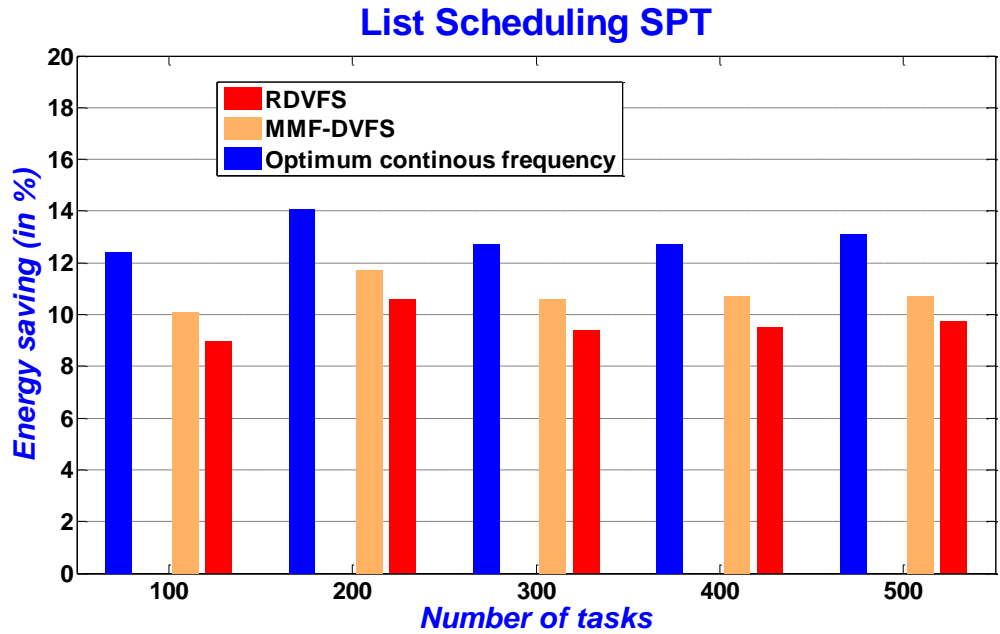

**Figure 3-5. (*Continued*)**

An interesting issue for further investigation is the relationship between energy consumption and the number of processors in our experiments. Increasing the number of processors expedites the processing time and consequently reduces the makespan; however, as a drawback, it also increases the system slack time. Figure 3-8 addresses this issue and illustrates the percentage of overall energy saving of the system due to the number of processors for random and LU decomposition task graphs. The graphs in this figure reveal that increasing the number of processors results in saving more energy. The major limitation on most DVFS-based algorithms working with one frequency (such as the RDVFS algorithm) is that the frequency combinations are fixed. Those algorithms work better when the processor can run at any arbitrary set of frequencies. However, due to technological issues, the number of valid frequencies is limited so that these algorithms have to choose the most appropriate frequency among a set of frequencies defined by DVFS. According to the fixed number of tick cycles for a task (constraint 1 in Eqn. 3-6) the relation among $t_{RD}^{(k)}$, $f_{RD}^{(k)}$, $f_N$ and $t_{OS}^{(k)}$ for $k^{\text{th}}$-task $\left(A^{(k)}\right)$ is:

$$t_{RD}^{(k)} = \frac{f_{RD}^{(k)}}{f_N} t_{OS}^{(k)}$$

| Parameter | Value |
|---|---|
| ***The number of tasks*** | [100, 200, 300, 400, 500] |
| ***The number of processors in clusters*** | [2, 4, 8, 16, 32] |
| ***Processor type*** | Transmeta Crusoe, Intel Xscale |

*Table 3-2. Experimental parameters*

It is shown that although $t_{RD}^{(k)}$ is a continuous variable, it cannot accept all values; therefore the slack time of tasks cannot be minimised. However, in the MMF-DVFS algorithm, the relation between those variables is:

$$f_{RD}^{(k)} t_{RD}^{(k)} = f_{min} t_1^{(k)} + f_{max} t_2^{(k)}$$

which is one equation with more than one variables $\left(t_1^{(k)}, t_2^{(k)}\right)$ and might have more eligible results; thus, appropriate values of these variables, with regard to the task conditions, can minimise the slack time and/or reduce energy consumption.

An overhead with MMF-DVFS is the transition time of switching from one frequency to another. An assumption that is almost always true is that the overhead of transition times is relatively much smaller than the execution times of tasks; therefore the transition times overhead can be neglected in calculations. In our experiments, tasks with $\left(T^{(k)}\right)$ at least 20 times larger than transition time are considered for the MMF-DVFS algorithm. Generally overhead of calculating MMF-DVFS for each task is fairly neglectable compare to the length of task.

*Table 3-3. The energy saving percentage of MMF-DVFS and other algorithms on 1800 random and real task graphs.*

| Experiment | Random tasks | Gauss-Jordan | LU-decomposition |
|---|---|---|---|
| *RDVFS* | 13.00% | 0.1% | 24.8% |
| *MMF-DVFS* | 13.50% | 0.11% | 25.5% |
| *Optimal Continuous Frequency* | 14.84% | 0.14% | 27.81% |

## 3.5 Summary and Remarks

Since most traditional static task scheduling algorithms in HPCS do not consider power management, the energy issue with task scheduling was addressed and the Maximum-Minimum-Frequency DVFS (MMF-DVFS) algorithm was presented. This algorithm adopted the DVFS technique, a recent advance in processor design, to reduce energy consumption.

While most existing DVFS-based approaches cover the idle time of scheduled tasks by using one frequency, MMF-DVFS uses a linear combination of the highest and lowest frequencies in the frequency range of the processors to process a task. After formulating an energy model in DVS-enabled processors, the MMF-DVFS algorithm was formulated as an optimisation problem of those frequencies. Simulation results of 1500 randomly generated task graphs and 300 real-world application task graphs showed that MMF-DVFS can improve energy saving by 0.7% compared to RDVFS algorithm.

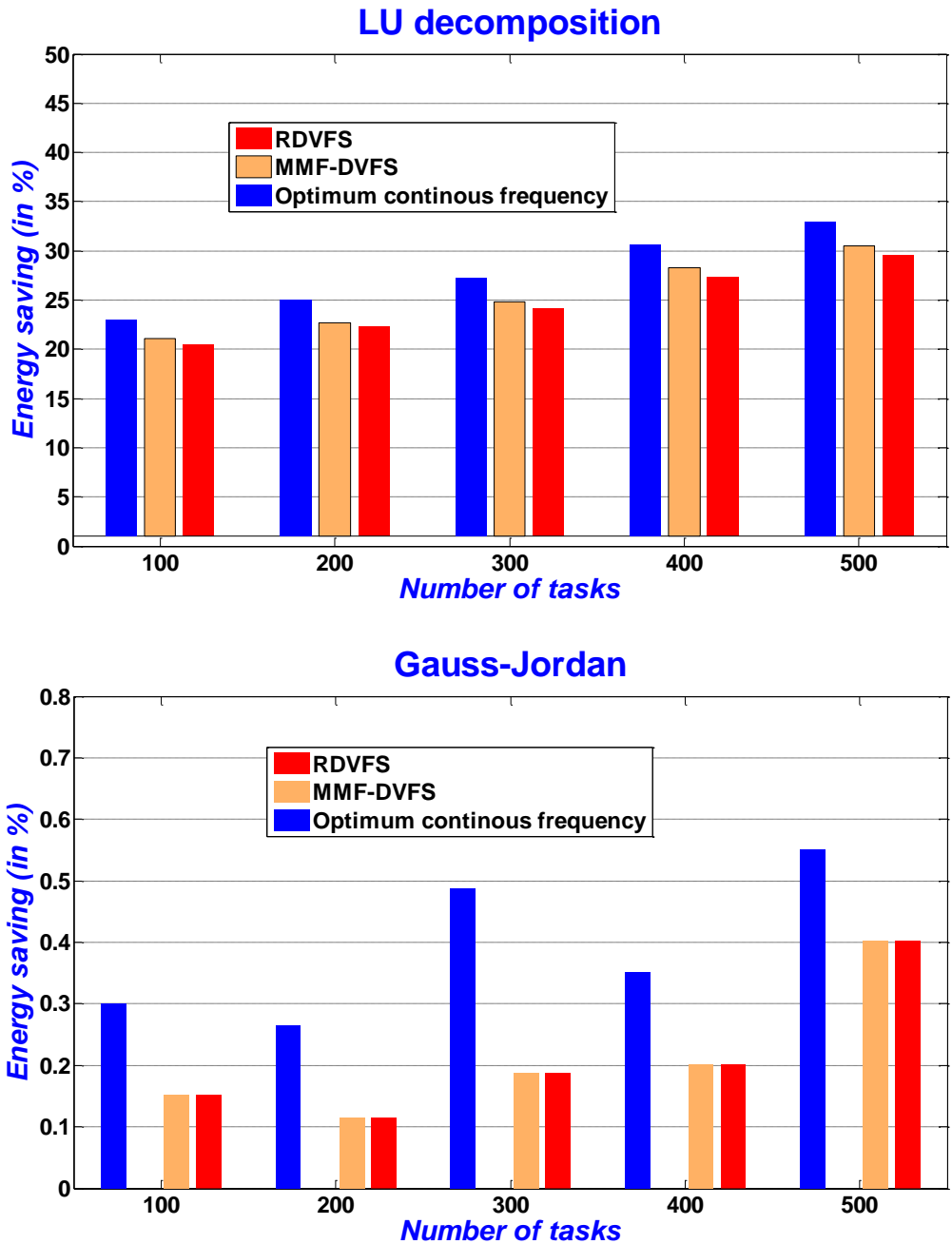

**Figure 3-6. The normalised energy consumption of MMF-DVFS and other algorithms on the number of tasks for two real-world applications: LU decomposition, and Gauss-Jordan.**

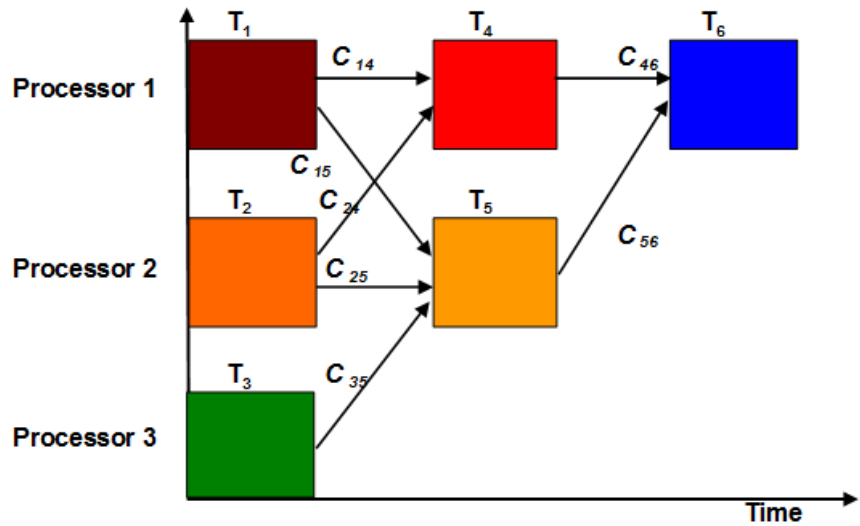

**(a)**

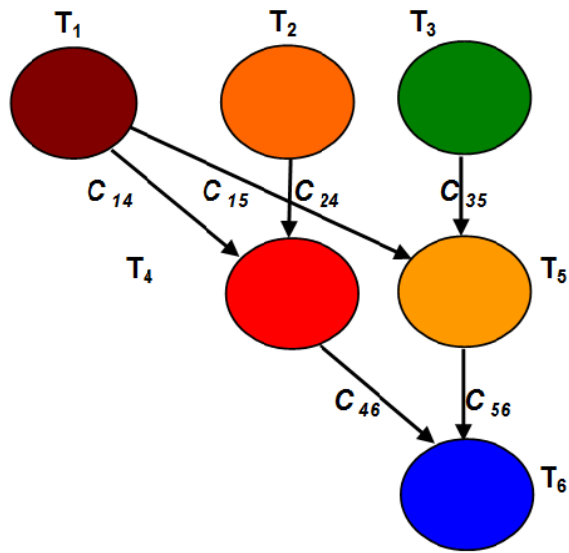

**(b)**

Figure 3-7. Gauss-Jordan task graph: (a) a sample scheduling of a three-level Gauss-Jordan task graph on three processors, (b) a Gauss-Jordan DAG for three levels. The communication cost$\left(C_{ij}\right)$ is equal to 10 time units for all *i* and *j*.

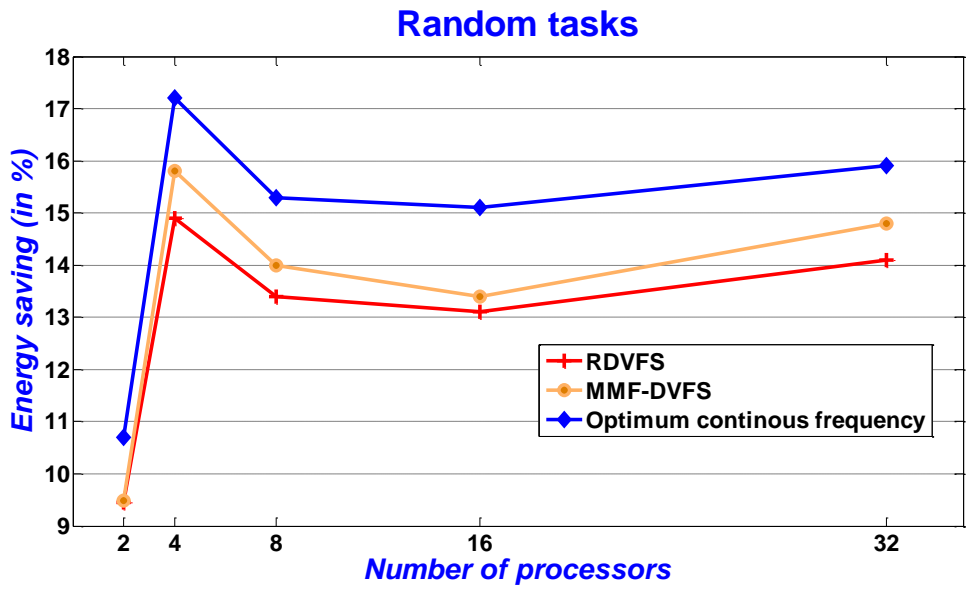

**(a)**

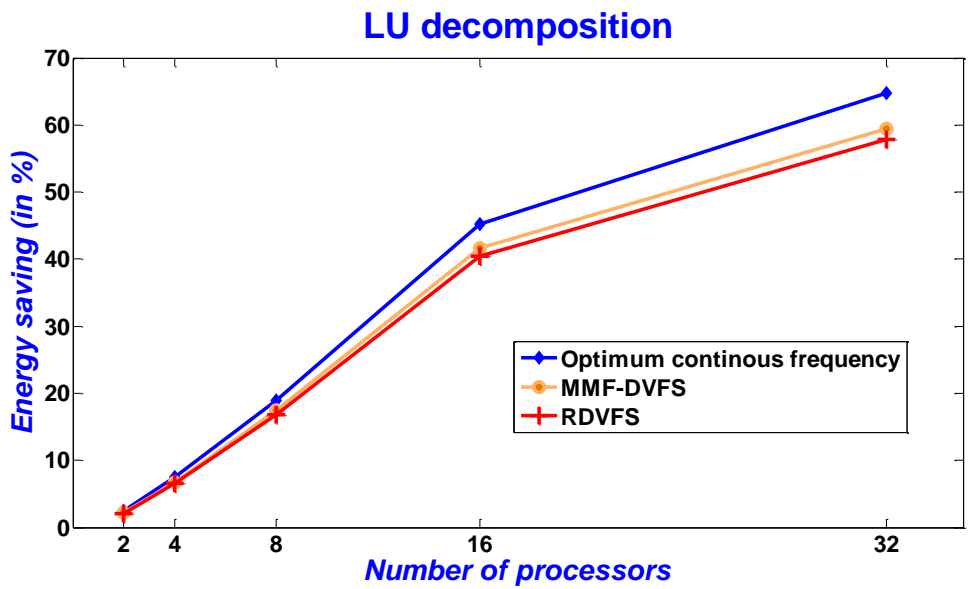

**(b)**

**Figure 3-8. The dependence of energy saving in MMF-DVFS and other algorithms on the number of processors: (a) 1500 randomly generated task graphs, (b) 300 LU decomposition task graphs.**

# Chapter 4. Multiple Frequency Selection in DVFS-Enabled Processors to Minimise Energy Consumption (second algorithm)

## 4.1 Introduction

In the previous chapter, the highest and lowest frequencies of processors were involved in slack time reclamation. In this chapter we investigate how the energy consumption of processors changes when a task is executed with a combination of all processors' frequencies.

## 4.2 Multiple-Frequency-Selection DVFS Algorithm (MFS-DVFS)

The key idea of MFS-DVFS is to execute tasks using a linear combination of all frequencies of DVFS-enabled processors so that their slack times are fully filled/covered. MFS-DVFS can be defined as finding the best combination of available frequencies $(f_1 < ... < f_N)$ to perform a predefined task with $K$ steps of computation within a predefined time $T$. Therefore, the power consumption minimisation of $k^{\text{th}}$-task $(A^{(k)})$ in MFS-DVFS algorithm is formulated in an optimisation form as follows:

$$\begin{cases} Min: E^{(k)} = \sum_{i=1}^{N} t_i^{(k)} \left( \alpha f_i^3 + \gamma \right) + P_{Idle} \left( T^{(k)} - \sum_{i=1}^{N} t_i^{(k)} \right) \\ subject\ to: \\ \qquad 1. \sum_{i=1}^{N} t_i^{(k)} f_i = K^{(k)} \\ \qquad 2. \sum_{i=1}^{N} t_i^{(k)} \leq T^{(k)} \\ \qquad 3. t_i^{(k)} \geq 0, \quad for\ i = 1,2,...,N \end{cases} \qquad (4-1)$$

where a simplified equation between power consumption and frequency is used (refer to Eqn. 3-4). The optimisation problem in Eqn. 4-1 represents the power consumption problem: how to choose $t_i^{(k)}$ so that the consumed energy of task $A^{(k)}$ minimises (Figure 4-1). For executing the task, the processor has to use the same number of tick clocks as constraint 1 in Eqn. 4-1. Applying theorems 1 and 2 in Chapter 3 simplifies the optimisation problem in Eqn. 4-1 to:

$$\begin{cases} Min\text{:}\, E^{(k)} = \sum_{i=1}^{N} t_i^{(k)} \left( \alpha f_i^3 + \gamma \right) \\ subject\ to\text{:} \\ \qquad 1.\, \sum_{i=1}^{N} t_i^{(k)} f_i = K^{(k)} \\ \qquad 2.\, \sum_{i=1}^{N} t_i^{(k)} \le T^{(k)} \\ \qquad 3.\, t_i^{(k)} \ge 0, \quad for\ i = 1,2,\dots,N \end{cases} \qquad (4-2)$$

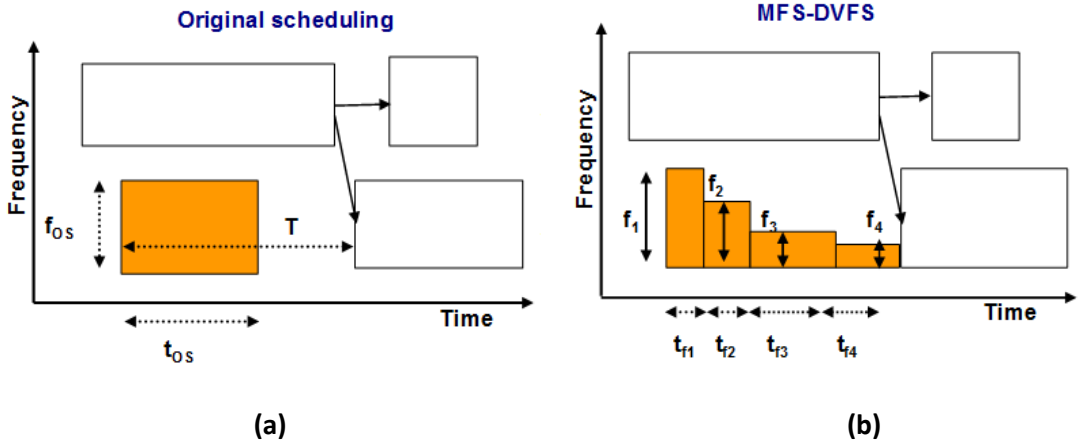

**Figure 4-1. Time representation of MMF-DVFS and other algorithms: (a) Original scheduling, and (b) MFS-DVFS algorithm.**

To find the best possible values of $t_i^{(k)}$, this optimisation algorithm must be applied to all tasks in the scheduling. There are cases in which MFS-DVFS cannot improve the power consumption, for example when a task reaches $f_1$ (the lowest frequency) in the RDVFS algorithm or it has no idle time. Therefore, to improve the speed of the MFS-DVFS algorithm, eligible tasks should be extracted before optimisation.

**Task eligibility:** to simplify the formulation, let us just consider four discrete values for frequencies (the real processors have normally 4–5 frequencies). In any case, the same procedure can be used for a higher number of frequencies. The problem in Eqn. 4-2 becomes:

**MFS-DVFS algorithm:** *linear combination of frequencies*

**Input:** *the scheduled tasks on a set of P processors*

1. **For** $k^{th}$-task $A^{(k)}$ *scheduled on processor* $P_j$

2. *Apply RDVFS algorithm on this task*

3. **if** $E_{RD}^{(k)} > 0$ *for this task* **then**

    - *This task is eligible for MFS-DVFS*

    - *Solve optimization problem in Eqn. 4-2 by linear programming*

    **else**

        *RDVFS is the optimal result*

4. **end if**

5. **end for**

6. **return** *(the voltages and frequencies of optimal execution of the task)*

**Figure 4-2. MFS-DVFS algorithm**

$$
\begin{cases}
Min: E^{(k)} = \sum_{i=1}^{4} t_i^{(k)} \left( \alpha f_i^3 + \gamma \right) \\
subject\ to: \\
\quad 1.\, t_1^{(k)} f_1 + t_2^{(k)} f_2 + t_3^{(k)} f_3 + t_4^{(k)} f_4 = K^{(k)} \\
\quad 2.\, t_1^{(k)} + t_2^{(k)} + t_3^{(k)} + t_4^{(k)} = T^{(k)} \\
\quad 3.\, t_i^{(k)} \geq 0, \quad for\ i = 1,2,3,4
\end{cases}
$$

Merging constraints 2 and 3 in Eqn. 4-2 results in:

$$
\begin{cases}
t_1^{(k)} = \dfrac{T^{(k)} f_2 - K^{(k)}}{f_2 - f_1} - t_3^{(k)} \dfrac{f_2 - f_3}{f_2 - f_1} - t_4^{(k)} \dfrac{f_2 - f_4}{f_2 - f_1} \\
t_2^{(k)} = \dfrac{-T^{(k)} f_1 + K^{(k)}}{f_2 - f_1} - t_3^{(k)} \dfrac{f_3 - f_1}{f_2 - f_1} - t_4^{(k)} \dfrac{f_4 - f_1}{f_2 - f_1}
\end{cases}
$$

Therefore, the power consumption function changes to

$$
E^{(k)} = a_0^{(k)} + a_1^{(k)} t_3^{(k)} + a_2^{(k)} t_4^{(k)} \qquad (4-3)
$$

where

$$\begin{cases} a_0^{(k)} = (\alpha f_1^3 + \gamma)\dfrac{T^{(k)}f_2 - K^{(k)}}{f_2 - f_1} + (\alpha f_2^3 + \gamma)\dfrac{-T^{(k)}f_1 + K^{(k)}}{f_2 - f_1} \\[2mm] a_1^{(k)} = (\alpha f_3^3 + \gamma) + (\alpha f_1^3 + \gamma)\dfrac{f_3 - f_2}{f_2 - f_1} - (\alpha f_2^3 + \gamma)\dfrac{f_3 - f_1}{f_2 - f_1} \\[2mm] a_2^{(k)} = (\alpha f_4^3 + \gamma) + (\alpha f_1^3 + \gamma)\dfrac{f_4 - f_2}{f_2 - f_1} - (\alpha f_2^3 + \gamma)\dfrac{f_4 - f_1}{f_2 - f_1} \end{cases} \qquad (4-4)$$

To guarantee less energy consumption using the MFS-DVFS algorithm, the following condition should be satisfied:

$$a_0^{(k)} + a_1^{(k)}t_3^{(k)} + a_2^{(k)}t_4^{(k)} < E_{RD}^{(k)} \qquad\qquad (4-5)$$

$a_0^{(k)} + a_1^{(k)}t_3^{(k)} + a_2^{(k)}t_4^{(k)}$ shows a 3-dimensional surface and the search region is where it satisfies the following three constraints: (1) $t_3^{(k)} \geq 0$, (2) $t_4^{(k)} \geq 0$ and (3) $E_{RD}^{(k)} > 0$. The first two constraints in Eqn. 4-5 are also considered by optimisation in Eqn. 4-2. The only one that specifies the search region is constraint 3. If a task satisfies this recent constraint, then it can be concluded that there is a valid search region for this task where MFS-DVFS gives a better result than RDVFS. Then linear programming explores this search region to find out the best suitable frequencies and their associated times. The detail of the MFS-DVFS algorithm has been shown in Figure 4-2.

## 4.3 Evaluation and Experimental Results

### 4.3.1 Simulation settings

Similar to the previous chapter, three list schedulers are used to distribute task graphs on five sets of 2, 4, 8, 16 and 32 processors. We evaluated the performance of algorithms with two sets of task graphs: randomly generated and real-word parallel applications. The two real-world applications used in our experiments were LU decomposition and Gauss-Jordan with DAGs extracted from [59]. We applied a large number of variations in the number of processors and tasks for each application in our simulations. The used processors are Transmeta Crusoe [10] and Intel Xscale [62], for which the voltage-frequency and also the model can be found in Table 3-1 and Figure 3-4.

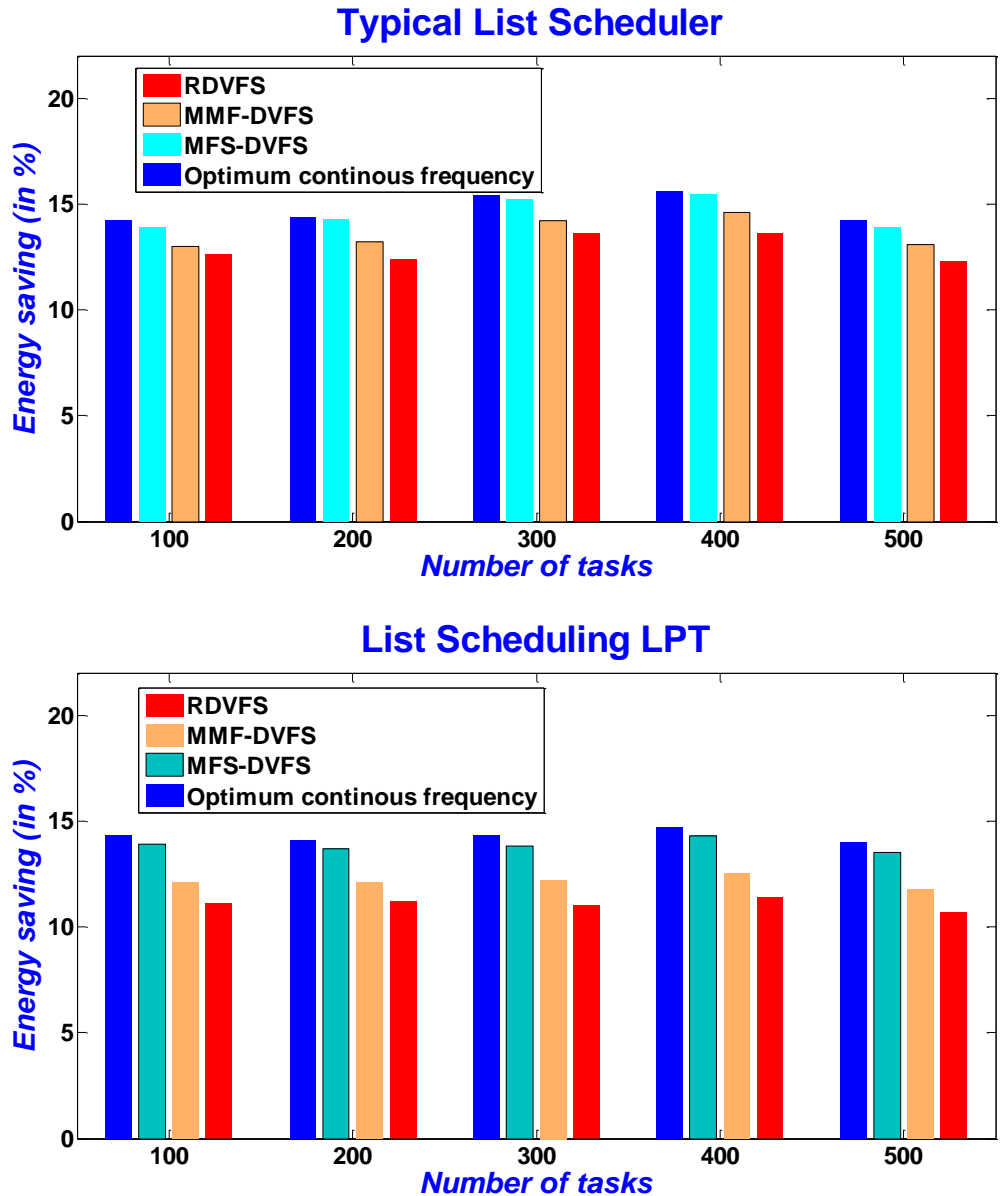

**Figure 4-3. The normalized energy consumption of MFS-DVFS on the number of tasks: The typical list scheduler, list scheduler with Longest Processing Time first (LPT), and list scheduler with Shortest Processing Time first (SPT).**

### 4.3.2 Results

The simulation results of normalised energy consumption for all DAGs (Figures 4-3 and 4-4) are shown in Table 4-1. This table clearly shows the superior performance of MFS-DVFS scheduling compared to the other approaches, including MMF-DVFS in the previous chapter, in all cases. This table also indicates that the energy saving of MFS-DVFS is closer to that of optimum continuous frequency than the other algorithms. Figure 4-4 shows that although the efficiency of all algorithms, including MFS-DVFS, in saving energy in LU decomposition is significant, these algorithms have a poorer performance on Gauss-Jordan tasks. This issue results from a lack of idle time among tasks of this application (Figure 3-7).

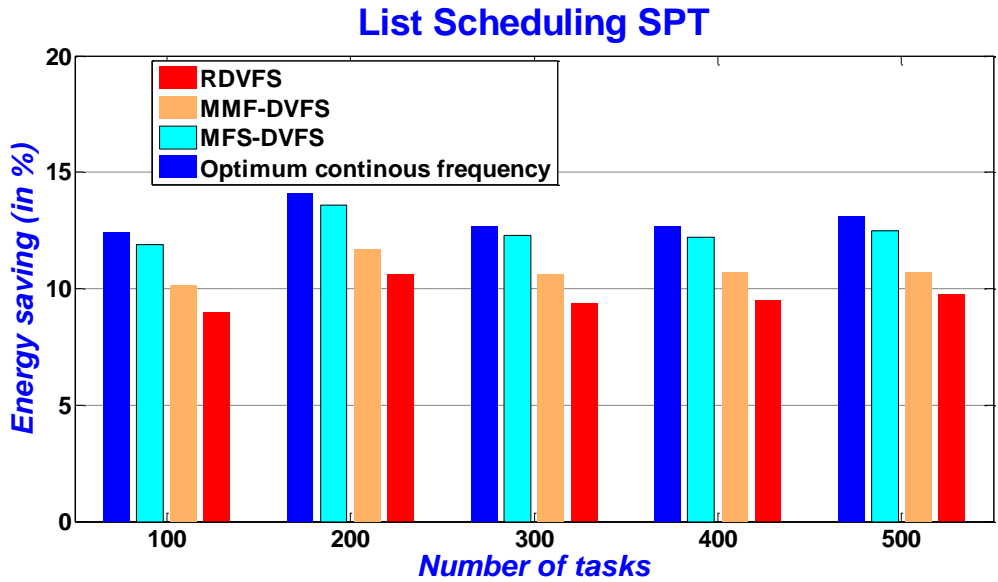

**Figure 4-3. (Continued)**

An interesting issue for further investigation is the relationship between energy consumption and number of processors in our experiments. Increasing the number of processors expedites the processing time and consequently reduces the makespan; however, as a drawback, it also increases the system slack time. Figure 4-5 addresses this issue and illustrates the percentage of overall energy saving of the system due to the number of processors for random and LU decomposition task graphs. The graphs in this figure reveal the fact that increasing the number of processors results in saving more energy.

*Table 4-1. The energy saving percentage of MFS-DVFS and other algorithms on 1800 random and real task graphs.*

| Experiment | Random tasks | Gauss-Jordan | LU-decomposition |
|---|---|---|---|
| ***RDVFS*** | 13.00% | 0.1% | 24.8% |
| ***MMF-DVFS*** | 13.50% | 0.11% | 25.5% |
| ***MVFS-DVFS*** | 14.40% | 0.11% | 27.0% |
| ***Optimal Continuous Frequency*** | 14.84% | 0.14% | 27.81% |

As with MMF-DVFS, an overhead with MFS-DVFS is the transition time of switching from one frequency to another. An almost-always true assumption is that the overhead of transition times is relatively much smaller than the execution times of tasks; therefore the transition times overhead can be neglected in calculations. In our experiments, a task with execution time ($T^{(k)}$) at least 20 times larger than transition time are considered for the MFS-DVFS algorithm.

## 4.4 Summary and Remarks

Following the MMF-DVFS algorithm in the previous chapter – using the highest and lowest frequencies of a DVFS-enabled processor – this chapter presented the Multiple-Frequency-Selection DVFS (MFS-DVFS) algorithm, which utilised a linear combination of all frequencies of the processors to reduce the energy consumption of scheduled tasks. First, the problem of MFS-DVFS was formulated as an optimisation problem. As optimisation for each task in DAG is time consuming, eligible tasks were mathematically extracted before optimisation. Simulation results of 1500 randomly generated task graphs and 300 real-world application task graphs showed the effectiveness of MFS-DVFS algorithm, with a 1.8% and 1.2% average energy saving compared to RDVFS and MMF-DVFS, respectively.

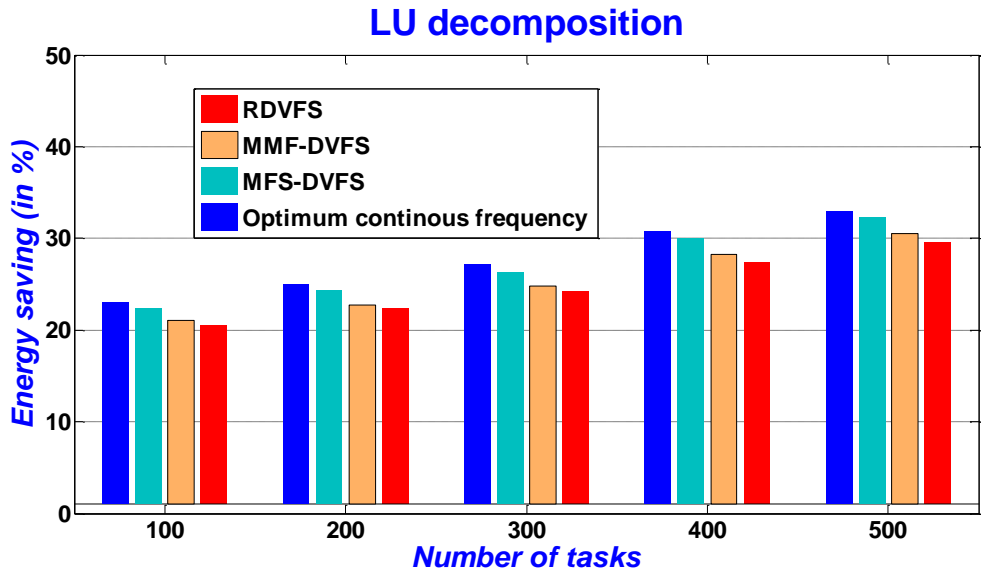

**(a)**

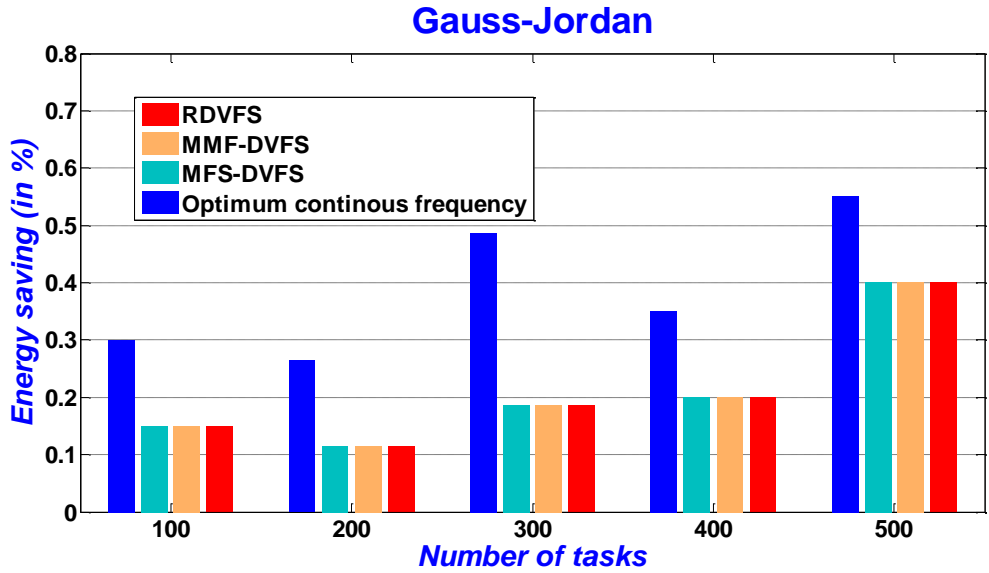

**(b)**

Figure 4-4. The normalized energy consumption of MFS-DVFS and other algorithms on the number of tasks for two real-world applications: (a) LU decomposition, (b) Gauss-Jordan.

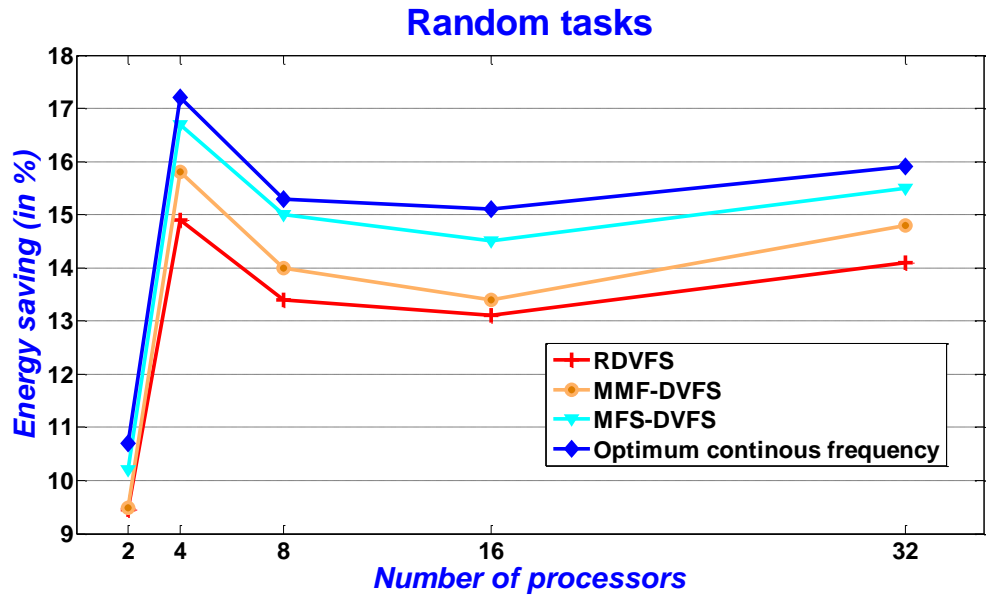

**(a)**

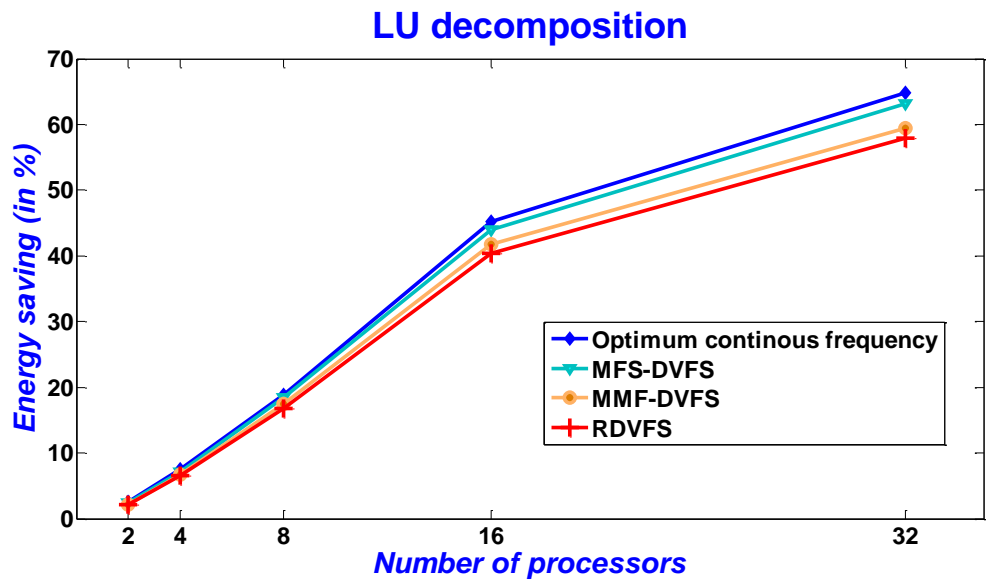

**(b)**

**Figure 4-5. The dependence of energy saving in MFS-DVFS and other algorithms on the number of processors: (a) 1500 randomly generated task graphs, (b) 300 LU decomposition task graphs.**

# Chapter 5. Observation of Optimal Frequency Selection in DVFS-based Energy Consumption Minimisation (third algorithm)

## 5.1. Introduction

Observations of simulation results in the previous chapter (Chapter 4) indicate that each task is executed by at most two frequencies of DVFS-enabled processors. Moreover, when a simplified version of frequency-power relation is used, these two frequencies are adjacent. This fact will be proven mathematically in this chapter for both the simplified and original frequency-power modelling of processors.

## 5.2. Problem Statement

Optimal energy consumption of the $k^{\text{th}}$-task can be defined as finding the best combination of available voltage-frequencies, $\{(f_1, v_1) < \cdots < (f_N, v_N)\}$, to perform a predefined task with $K$ clock ticks within a predefined time $T$. For the $k^{\text{th}}$-task, this optimal answer is defined as follows:

$$
\begin{cases}
Minimize: E^{(k)} = \displaystyle\sum_{i=1}^{N} t_i^{(k)} P_d(f_i, v_i) + P_l\left(T^{(k)} - \sum_{i=1}^{N} t_i^{(k)}\right) \\
subject\ to: \\
\qquad 1. \displaystyle\sum_{i=1}^{N} t_i^{(k)} f_i = K^{(k)} \\
\qquad 2. \displaystyle\sum_{i=1}^{N} t_i^{(k)} \leq T^{(k)} \\
\qquad 3. t_i^{(k)} \geq 0, \ \ for\ i = 1,2,\dots,N
\end{cases}
\qquad (5-1)
$$

Because our algorithm reclaims the slack time of each task independent from other tasks in DAG, the above formulation for the $k^{th}$-task is further simplified by replacing $t_i^{(k)}$, $T^{(k)}$ and $K^{(k)}$ with $t_i$, $T$ and $K$, respectively. Here, $t$ and $T$ are time values in milli-seconds and $K$ is an integer value.

## 5.3. Computing the Optimal Solution

To find the optimal solution for the problem defined by Eqn. 5-1, a simplified version of this problem is solved first, then generalised to find the solution for Eqn. 5-1. This simplified version uses only three frequencies $\big((f_1, v_1) < (f_2, v_2) < (f_3, v_3)\big)$ to perform a task in exact time (T) – as opposed to within – and is defined as follows:

$$\begin{cases} Minimize: \quad E = t_1 P_d(f_1, v_1) + t_2 P_d(f_2, v_3) + t_3 P_d(f_3, v_3) \\ subject\ to: \\ \quad 1.\ t_1 f_1 + t_2 f_2 + t_3 f_3 = K \\ \quad 2.\ t_1 + t_2 + t_3 = T \\ \quad 3.\ t_1 \geq 0, t_2 \geq 0, t_3 \geq 0 \end{cases} \qquad (5-2)$$

**Theorem 1:** The optimal solution for Eqn. 5-2 is obtained by at most two voltage-frequencies.

**Proof:** To prove this theorem, the general energy formulation using three voltage-frequencies is first computed, and then minimised.

From constraints 1 and 2:

$$\begin{rcases} t_1 = T - (t_2 + t_3) \\ K = t_1 f_1 + t_2 f_2 + t_3 f_3 = (T - t_2 - t_3)f_1 + t_2 f_2 + t_3 f_3 \end{rcases} \Rightarrow$$

$$\begin{cases} t_2 = \dfrac{(K - Tf_1) - t_3(f_3 - f_1)}{f_2 - f_1} \qquad (5-3) \\[2mm] t_1 = T - \left[\dfrac{(K - Tf_1) - t_3(f_3 - f_1)}{f_2 - f_1} + t_3\right] = \dfrac{(-K + Tf_2) - t_3(f_3 - f_2)}{f_2 - f_1} \qquad (5-4) \end{cases}$$

Based on constraint 3:

$$
\begin{cases}
t_1 \geq 0 \Rightarrow (Tf_2 - K) + t_3(f_3 - f_2) \geq 0 \Rightarrow t_3 \geq \dfrac{K - Tf_2}{f_3 - f_2} \\[3mm]
t_2 \geq 0 \Rightarrow (-Tf_1 + K) - t_3(f_3 - f_2) \geq 0 \Rightarrow t_3 \leq \dfrac{K - Tf_1}{f_3 - f_1} \\[3mm]
t_3 \geq 0
\end{cases}
$$

which results in:

$$
\begin{cases}
0 \leq t_3 \leq \dfrac{K - Tf_1}{f_3 - f_1} \\[3mm]
\dfrac{K - Tf_2}{f_3 - f_2} \leq t_3 \leq \dfrac{K - Tf_1}{f_3 - f_1}
\end{cases}
$$

By replacing $t_1$ and $t_2$ in the energy formulation based on $t_3$, the following equation for energy is obtained:

$$
E = t_3 \left( \frac{(f_3 - f_2)P_d(f_1, v_1) - (f_3 - f_1)P_d(f_2, v_2) + (f_3 - f_1)P_d(f_3, v_3)}{f_2 - f_1} \right)
$$

$$
+ \frac{(Tf_2 - K)P_d(f_1, v_1) + (K - Tf_1)P_d(f_2, v_2)}{f_2 - f_1}
$$

$$
= \alpha t_3 + \beta \qquad (5-5)
$$

This equation reveals that the energy consumption of a task can be represented as a linear function of $t_3$. Depending on the sign of $\alpha$, two scenarios might arise: (1) $\alpha < 0$; or (2) $\alpha > 0$.

Case 1: If $\alpha < 0$, then energy in Eqn. 5-5 is a strictly decreasing function of $t_3$. Therefore, it is minimised when $t_3$ is set to its highest possible value. Thus:

$$
\begin{cases}
t_3 = \dfrac{K - Tf_1}{f_3 - f_1} \\[3mm]
t_2 = 0 \\[3mm]
t_1 = T - t_2 - t_3 = T - \dfrac{K - Tf_1}{f_3 - f_1} = \dfrac{Tf_3 - K}{f_3 - f_1} \\[3mm]
E^*(f_1, f_3) = t_1 P_d(f_1, v_1) + t_3 P_d(f_3, v_3) = \dfrac{Tf_3 - K}{f_3 - f_1} P_d(f_1, v_1) + \dfrac{K - Tf_1}{f_3 - f_1} P_d(f_3, v_3)
\end{cases}
\qquad (5-6)
$$

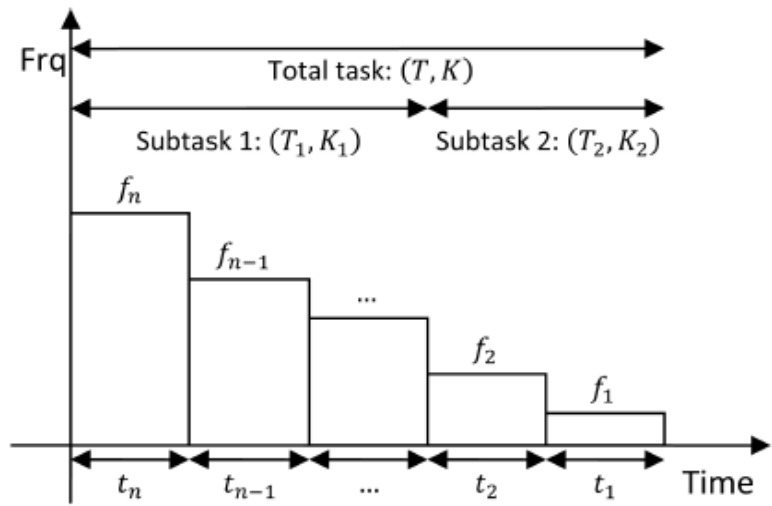

**Figure 5-1. Optimal answer for Eqn. 5-1 using multiple voltage-frequencies.**

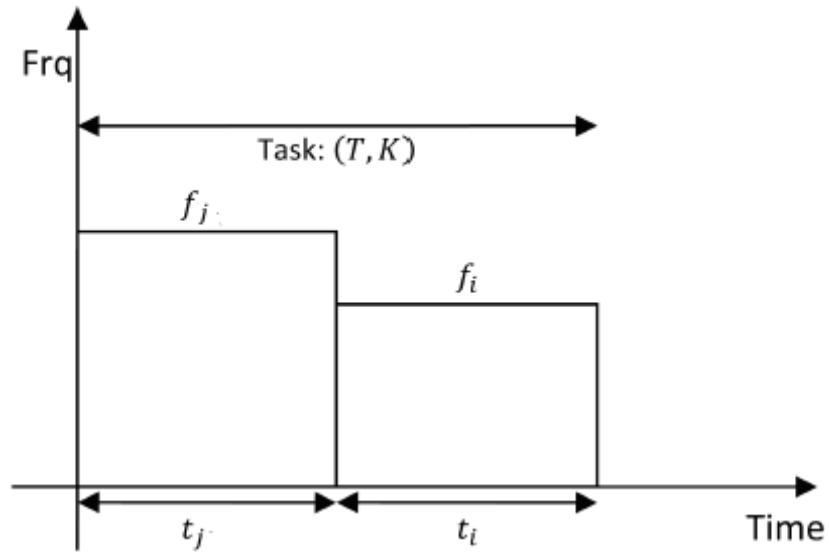

**Figure 5-2. Optimal answer for Eqn. 5-1 with two voltage–frequencies.**

Case 2: If $\alpha > 0$, then energy in Eqn. 5-5 is a strictly increasing function of $t_3$. Therefore, it is minimised when $t_3$ is set to its lowest possible value. Two minimal values might be set for $t_3$:

$t_3 = 0$

$$\Rightarrow \begin{cases} t_1 = T - t_2 - t_3 = T - \dfrac{K - Tf_1}{f_2 - f_1} = \dfrac{-K + Tf_2}{f_2 - f_1} \\[3mm] t_2 = \dfrac{(K - Tf_1) - t_3(f_3 - f_1)}{f_2 - f_1} = \dfrac{-Tf_1 + K}{f_2 - f_1} \\[3mm] E^*(f_1, f_2) = t_1 P_d(f_1, v_1) + t_2 P_d(f_2, v_2) = \dfrac{Tf_2 - K}{f_2 - f_1} P_d(f_1, v_1) + \dfrac{K - Tf_1}{f_2 - f_1} P_d(f_2, v_2) \end{cases} \quad (5-7)$$

or ,

$t_3 = \dfrac{K - Tf_2}{f_3 - f_2}$

$$\Rightarrow \begin{cases} t_1 = 0 \\[3mm] t_2 = \dfrac{(K - Tf_1) - \left(\dfrac{K - Tf_2}{f_3 - f_2}\right)(f_3 - f_1)}{f_2 - f_1} = \dfrac{Tf_3 - K}{f_3 - f_2} \\[3mm] E^*(f_2, f_3) = t_2 P_d(f_2, v_2) + t_3 P_d(f_3, v_3) = \dfrac{Tf_3 - K}{f_3 - f_2} P_d(f_2, v_2) + \dfrac{K - Tf_2}{f_3 - f_2} P_d(f_3, v_3) \end{cases} \quad (5-8)$$

Equations 5-6 to 5-8 show that, regardless of whether energy is a strictly decreasing or increasing function of $t_3$, always two voltage–frequencies provide the optimal energy consumption.

**Corollary 1:** If two voltage–frequencies $(f_j, v_j) > (f_i, v_i)$ are capable of performing a task, then their associated optimal energy consumption would be:

$$E^*(f_i, f_j) = \frac{Tf_j - K}{f_j - f_i} P_d(f_i, v_i) + \frac{K - Tf_i}{f_j - f_i} P_d(f_j, v_j) \quad (5-9)$$

**Proof:** direct observation from theorem 1.

**Corollary 2:** If two voltage–frequencies $(f_j, v_j) > (f_i, v_i)$ are capable of performing a task, then their associated execution times for optimal energy consumption would be:

$$\begin{cases} t_i = \dfrac{Tf_j - K}{f_j - f_i} \\[3mm] t_j = \dfrac{K - Tf_i}{f_j - f_i} \end{cases} \quad (5-10)$$

**Proof:** direct observation from theorem 1.

**Lemma 1 (Optimum continuous frequency):** If a processor is able to perform a task with a continuous range of voltage–frequencies, which is an unrealistic assumption, then the optimum energy to perform task $A$ is when the task's slack time (T) is fully utilised.

**Proof:** If $(f_j, v_j) > (f_i, v_i)$ are two voltage–frequencies to obtain optimal energy for a task, then Eqn. 5-2 can be rewritten as:

$$\begin{cases} Minimize: \quad E = t_i P_d(f_i, v_i) + t_j P_d(f_j, v_j) \\ subject\ to: \\ \quad 1.\ t_i f_i + t_j f_j = K \\ \quad 2.\ t_i + t_j = T \\ \quad 3.\ t_i \geq 0, t_j \geq 0 \end{cases}$$

By replacing $t_j$ with $t_j = T - t_i$, the energy formula would be:

$$E = t_i \left( P_d(f_i, v_i) - P_d(f_j, v_j) \right) + P_d(f_j, v_j) \times T \qquad (5-11)$$

Because $E$ in Eqn. 5-11 is a strictly decreasing function of $t_i$, it is minimised when $t_i = 0$. This implies that if a frequency can be chosen from a continuous spectrum, the energy is optimised using only one voltage-frequency. Further, this frequency would cover the whole slack time and could be calculated as follows:

$$t_i = 0 \Rightarrow \begin{cases} f_{ideal} = f_j = \dfrac{K}{T} \\ t_{ideal} = t_j = T \\ E_{Opt-Cont.} = T \times P_d(f_{ideal}, v_{ideal}) \end{cases} \qquad (5-12)$$

**Lemma 2:** If a processor's set of available voltage-frequencies is discrete, then two voltage-frequencies that would lead to the optimal energy consumption will be on both sides of $f_{ideal}$ in Eqn. 5-12.

**Proof:** Constraint 3 in Eqn. 5-2 implies that all time segments are greater or equal to zero. By applying this rule to the time values in Corollary 2 and with condition $(f_j, v_j) > (f_i, v_i)$, the following can be concluded:

$$\begin{cases} t_i \geq 0 \Rightarrow \dfrac{Tf_j - K}{f_j - f_i} \geq 0 \Rightarrow f_j \geq \dfrac{K}{T} \\[2mm] t_j \geq 0 \Rightarrow \dfrac{-Tf_i + K}{f_j - f_i} \geq 0 \Rightarrow f_i \leq \dfrac{K}{T} \end{cases} \Rightarrow f_i \leq \frac{K}{T} \leq f_j \qquad (5-13)$$

where $\frac{K}{T} = f_{ideal}$ by definition.

**Theorem 2:** Optimal answer for Eqn. 5-1 uses at most two voltage-frequencies.

**Proof (by contradiction):** To prove this theorem, we show that the optimal answer for Eqn. 5-1 cannot use more than two voltage-frequencies to minimise total energy consumption. If we assume that the optimal answer for Eqn. 5-1 utilises more than two voltage-frequencies, then its utilisation profile can be depicted as that in Figure 5-1 for $n \geq 3$. In this case, this task can be divided into two independent subtasks: (1) a subtask $(T_1, K_1)$ that uses three voltage-frequencies, e.g. $(f_n, v_n), (f_{n-1}, v_{n-1}), (f_{n-2}, v_{n-2})$; and (2) a subtask to cover the rest of calculations, i.e., $(f_{n-3}, v_{n-3}), \dots, (f_1, v_1)$.

$$\begin{cases} T = T_1 + T_2 = (t_n + t_{n-1} + t_{n-2}) + (t_{n-3} + \cdots + t_1) \\ K = K_1 + K_2 = (t_n f_n + t_{n-1} f_{n-1} + t_{n-2} f_{n-2}) + (t_{n-3} f_{n-3} + \cdots + t_1 f_1) \\ E = E_1 + E_2 = E_1\big((f_n, v_n), (f_{n-1}, v_{n-1}), (f_{n-2}, v_{n-2})\big) + E_2\big((f_{n-3}, v_{n-3}), \ldots, (f_1, v_1)\big) \end{cases}$$

Now, based on theorem 1, subtask $(T_1, K_1)$ can be performed with only two voltage-frequencies with less energy consumption, i.e.,

$$E_1\big((f_n, v_n), (f_{n-1}, v_{n-1}), (f_{n-2}, v_{n-2})\big) > \begin{cases} E^*\big((f_n, v_n), (f_{n-1}, v_{n-1})\big) \\ OR \\ E^*\big((f_{n-1}, v_{n-1}), (f_{n-2}, v_{n-2})\big) \\ OR \\ E^*\big((f_n, v_n), (f_{n-2}, v_{n-2})\big) \end{cases}$$

Thus:

$$E = E_1 + E_2 > E_2 + \begin{cases} E^*\big((f_n, v_n), (f_{n-1}, v_{n-1})\big) \\ OR \\ E^*\big((f_{n-1}, v_{n-1}), (f_{n-2}, v_{n-2})\big) \\ OR \\ E^*\big((f_n, v_n), (f_{n-2}, v_{n-2})\big) \end{cases}$$

This, in fact, contradicts the optimality of $E\big((f_n, v_n), \ldots, (f_1, v_1)\big)$; therefore, the optimal answer for Eqn. 5-1 cannot use more than two voltage-frequencies to minimise energy consumption (Figure 5-2).

Above, we proved that Eqn. 5-1 can only be minimised by using two voltage-frequencies. However, in all these formulas, constraint 2 of this problem was relaxed to use the maximum available time $T$ to find its optimal solution, although the optimiser is allowed to use less time than $T$. Therefore, in the following theorem we prove that using less time will always lead to more energy consumption. That is, the original assumption of replacing $t_1 + t_2 + \cdots + t_N \leq T$ with $t_1 + t_2 + \cdots + t_N = T$ was correct.

**Theorem 3:** In Eqn. 5-1, using less time will always result in consuming more energy.

**Proof:** To prove this theorem, a task is assumed to be executed with two voltage-frequencies $(f_i, v_i)$ and $(f_j, v_j)$ in times $T$ and $T_1 (< T)$. Based on corollary 1, the associated energy consumption for these two cases would be:

$$\begin{cases} T_1: & E(T_1) = \dfrac{T_1 f_j - K}{f_j - f_i} P_d(f_i, v_i) + \dfrac{K - T_1 f_i}{f_j - f_i} P_d(f_j, v_j) + P_l(T - T_1) \\[3mm] T: & E(T) = \dfrac{T f_j - K}{f_j - f_i} P_d(f_i, v_i) + \dfrac{K - T f_i}{f_j - f_i} P_d(f_j, v_j) \end{cases}$$

then,

$$E(T) - E(T_1) = f_j \frac{P_d(f_i, v_i)}{f_j - f_i}(T - T_1) + f_i \frac{P_d(f_j, v_j)}{f_j - f_i}(T - T_1) - P_l(T - T_1)$$

$$= \underbrace{\frac{\overbrace{T - T_1}^{>0}}{f_j - f_i}}_{>0} \Big( f_j \big( P_d(f_j, v_j) - P_l \big) - f_i \big( P_d(f_i, v_i) - P_l \big) \Big) \qquad (5-14)$$

As $\dfrac{f_j}{f_i} > 1$ and $\dfrac{P_d(f_j, v_j)}{P_d(f_i, v_i)} > 1$, then $\big[ f_j(P_d(f_i, v_i) - P_l) - f_i \big( P_d(f_j, v_j) - P_l \big) \big] < 0$; thus, $E(T) - E(T_1) < 0$. Therefore, the original assumption of replacing $t_1 + t_2 + \cdots + t_N \leq T$ with $t_1 + t_2 + \cdots + t_N = T$ is correct.

## 5.4. Computation of Optimal Energy Consumption for $k^{\text{th}}$-task

Based on corollary 2, the following post–processing scheduling algorithm is proposed to optimise the energy consumption of each task. For $k^{\text{th}}$-task, two voltage–frequencies $\big( (f_i, v_i), (f_j, v_j) \big)$ that satisfy constraints 1 and 2 from Eqn. 5-1 (capable of performing $k^{\text{th}}$-task in time $T$) are first obtained and then their associated deployment times ($t_i^{(k)}$ and $t_j^{(k)}$) are calculated as follows.

$$\begin{cases} t_i^{(k)} f_i^{(k)} + t_j^{(k)} f_j^{(k)} = K^{(k)} \\[2mm] t_i^{(k)} + t_i^{(k)} = T^{(k)} \end{cases} \Rightarrow \begin{cases} t_j^{(k)} = \dfrac{K^{(k)} - T^{(k)} f_i^{(k)}}{f_j^{(k)} - f_i^{(k)}} \\[3mm] t_i^{(k)} = \dfrac{T^{(k)} f_j^{(k)} - K^{(k)}}{f_j^{(k)} - f_i^{(k)}} \end{cases} \qquad (5-15)$$

Based on constraint 3 from Eqn. 5-1,

$$\begin{cases} t_j^{(k)} \geq 0 \\[2mm] t_i^{(k)} \geq 0 \end{cases} \Rightarrow \begin{cases} K^{(k)} - T^{(k)} f_i^{(k)} \geq 0 \\[2mm] T^{(k)} f_j^{(k)} - K^{(k)} \geq 0 \end{cases} \Rightarrow f_i^{(k)} \leq \frac{K^{(k)}}{T^{(k)}} \leq f_j^{(k)}$$

and the optimal energy is calculated as:

$$E^{(k)}\left(T^{(k)}, T^{(k)}\right)$$

$$= \frac{T^{(k)} f_j^{(k)} - K^{(k)}}{f_j^{(k)} - f_i^{(k)}} P_d\left(f_i^{(k)}, v_i^{(k)}\right)$$

$$+ \frac{T^{(k)} - T^{(k)} f_i^{(k)}}{f_j^{(k)} - f_i^{(k)}} P_d\left(f_j^{(k)}, v_j^{(k)}\right) \qquad (5-16)$$

The details of MVFS-DVFS algorithm has been described in Figure 5-3.

## 5.5. Simplified-Multiple Frequency Selection DVFS (SMFS-DVFS)

In most DVFS algorithms, it is assumed that processor energy consumption is a convex function of frequency (or voltage) as:

$$P_d(f, v) = \lambda f^3$$

The convex function relationship between power and voltage was used by Ishihara in [26] where CPU power is just a square function of voltage – not frequency. If the relationship between voltage and frequency in Eqn. 3-1 is assumed to be linear, then the Ishihara work will be similar to the SMFS-DVFS algorithm in this section. Generally, Eqn. 3-1 is an approximation of the real relation between voltage-frequency and power in CMOS circuits that may not be followed by a few current or future CPUs. This problem has been solved in the MVFS-DVFS algorithm in this chapter by considering a general form between power and voltage-frequency in CPUs as shown in Eqn. 3-2. The MVFS-DVFS algorithm claims that independent of the method of modelling the relationship between power and voltage-frequency, if Eqn. 3-2 is satisfied, two frequencies are always involved in the optimal energy consumption. In other words, the technique in [26] is a subset of the MVFS-DVFS technique described in this chapter. This simplification changes the problem statement in Eqn. 5-1 to:

$$\begin{cases} Minimize: E^{(k)} = \lambda \sum_{i=1}^{N} t_i^{(k)} f_i^3 + P_I \left(T^{(k)} - \sum_{i=1}^{N} t_i^{(k)}\right) \\ subject\ to: \\ \qquad 1. \sum_{i=1}^{N} t_i^{(k)} f_i = K^{(k)} \\ \qquad 2. \sum_{i=1}^{N} t_i^{(k)} \leq T^{(k)} \\ \qquad 3. t_i^{(k)} \geq 0, \quad for\ i = 1,2,\dots,N \end{cases} \qquad (5-17)$$

Here, to simplify the writing of the equations for the $k^{\text{th}}$-task, $t_i^{(k)}$, $T^{(k)}$, and $K^{(k)}$ are also replaced with $t_i$, $T$, and $K$, respectively. As this simplified problem is a case-study of the main problem, all the proved theorems and corollaries are still valid; therefore, (1) two frequencies $f_i$ and $f_j$ ($> f_i$) result in the optimal answer, and (2) these two frequencies are near $f_{ideal} = \frac{K}{T}$ or $f_i < f_{ideal} < f_j$. The following two theorems, exclusively proved for this case study, show that $f_i$ and $f_j$ must also be adjacent. In this case (cubic functions) the optimal result can be calculated as:

$$E^*(f_i, f_j) = \lambda K \frac{f_i^3 - f_j^3}{f_i - f_j} - \lambda T f_i f_j \frac{f_i^2 - f_j^2}{f_i - f_j} = \lambda \left[K\left(f_i^2 + f_i f_j + f_j^2\right) - T f_i f_j \left(f_i + f_j\right)\right] \qquad (5-18)$$

**Theorem 4:** if $f_i$ and $f_j$ are capable of performing a task and there exists $f_r$ such that $f_i < f_r < f_j$, then adding $f_r$ to the frequency pool will always reduce the total energy consumption.

**Proof:** to prove this theorem we need to prove that:

$$E(f_i, f_j) > \begin{cases} E^*(f_r, f_j) & , if \quad K - Tf_r \geq 0 \\ E^*(f_i, f_r) & , if \quad K - Tf_r < 0 \end{cases}$$

**Case 1: IF** $K - Tf_r \geq 0$ :

$$\begin{aligned} E(f_i, f_j) &- E^*(f_r, f_j) \\ &= \left[K(f_i^2 + f_i f_j + f_j^2) - Tf_i f_j(f_i + f_j)\right] \\ &- \left[K(f_r^2 + f_r f_j + f_j^2) - Tf_r f_j(f_r + f_j)\right] = \cdots \\ &= (f_r - f_i)(f_i + f_r + f_j)(Tf_j - K) \\ &= (f_r - f_i)(f_i + f_r + f_j)(f_j - f_r)t_r > 0 \end{aligned}$$

**Case 2: IF** $K - Tf_r < 0$:

$$\begin{aligned} E(f_i, f_j) &- E^*(f_i, f_r) \\ &= \left[K(f_i^2 + f_i f_j + f_j^2) - Tf_i f_j(f_i + f_j)\right] \\ &- \left[K(f_i^2 + f_r f_i + f_r^2) - Tf_r f_i(f_r + f_i)\right] = \cdots \\ &= (f_j - f_r)(f_i + f_r + f_j)(K - Tf_i) \\ &= (f_j - f_r)(f_i + f_r + f_j)(f_r - f_i)t_r > 0 \end{aligned}$$

Using the above mentioned theorems, the optimal answer for Eqn. 5-17 can now be calculated. Theorem 2 proves that regardless of the number of available frequencies for a processor, the optimal answer would use at most two frequencies, while theorem 5 proves that these two frequencies should be also adjacent.

**Theorem 5:** The two frequencies that minimise Eqn. 5-17 are adjacent.

**Proof (by contradiction):** Based on theorem 2, the optimal answer for Eqn. 5-17 can only be obtained by using two frequencies. Here, we prove that these two frequencies should also be adjacent. To prove that (by contradiction), we show that these two frequencies cannot be non-adjacent. If they are, as shown in Figure 5-2, theorem 4 suggests that adding any available frequency between these two frequencies will reduce the total energy consumption and yield another answer with less energy consumption. This, in fact, contradicts the optimality of the original non-adjacent frequency selection. Therefore, two frequencies that minimise Eqn. 5-17 should be adjacent.

So far, we proved that Eqn. 5-17 can only be minimised by using two adjacent frequencies. As the only adjacent frequencies before and after $f_{ideal} = \frac{K}{T}$ are $f_{RD}$ and $f_{RD-1}(< f_{RD})$, respectively, therefore the optimal energy is achieved by frequencies $(f_{RD}, f_{RD-1})$, where $f_{RD}$ is a frequency obtained from the RDVFS algorithm in [10] and is defined as the largest and closest frequency to $f_{ideal} = \frac{K}{T}$. For $k^{\text{th}}$-task, the associated times for $f_{RD}$ and $f_{RD-1}$ are calculated as follows:

$$\begin{cases} t_{RD-1} = \dfrac{Tf_{RD} - K}{f_{RD} - f_{RD-1}} \\ t_{RD} = \dfrac{K - Tf_{RD-1}}{f_{RD} - f_{RD-1}} \end{cases} \qquad (5-19)$$

The algorithm for SMFS-DVFS has been described at Figure 5-4. It is highly expected that the outcome of the SMFS-DVFS algorithm will be similar to the MFS-DVFS algorithm in the previous chapter.

### 5.5.1. An example

The following example shows how each of the algorithms uses a task's slack time to reduce its associated energy consumption. To simplify, it is assumed that the power consumption is a cubic function of frequency, as $P_d(f, v) = 1.367 \times 10^{-24} \times f^3$. Figure 3-1a shows the original scheduling of $k^{\text{th}}$-task $(A^{(k)})$ executed on a processor. Assuming $P_{idle} = 0$, the values of the parameters for this task are as shown in Table 5-1. Based on these parameters:

- By referring to Eqn. 5-11 and Eqn. 5-12, the optimum continuous frequency is calculated as $f_{ideal}^{(k)} = \frac{K^{(k)}}{T^{(k)}} = 53.84 \ MHz$, which is not a valid frequency in the processor frequency list. The energy corresponding to this frequency is $E_{Opt-Cont.}^{(k)} = 27.73 \ mW$ (Figure 3-1c). As we mentioned earlier, this energy is the optimum energy for this task.

| | |
|---|---|
| $f_{RD}^{(k)}$ | 60MHz |
| $f_{RD-1}^{(k)}$ | 50MHz |
| $t_{OS}^{(k)}$ | 70 msec |
| $T^{(k)}$ | 130 msec |
| $K^{(k)}$ | 7 million cycles |

*Table 5-1. Parameters in comparison example between MVFS-DVFS and other algorithms, i.e., RDVFS, MMF-DVFS, and optimum continuous frequency.*

- In the RDVFS algorithm, this processor executes the task with the closest available frequency to $f_{ideal}^{(k)}$. Referring to the previous table, this frequency $\left(f_{RD}^{(k)}\right)$ is $60MHz$. Referring to Eqn. 3-3, the energy calculated by this method is $E_{RD}^{(k)} = 34.25 \ mW$ (Figure 3-1b).

- SMFS–DVFS (the simplified version of our proposed method) attempts to find the optimal energy by a linear combination of all processor frequencies. We proved that for each task, always two neighbour frequencies produce the optimal energy. These two frequencies are $f_{RD}^{(k)}$ and $f_{RD-1}^{(k)}(< f_{RD}^{(k)})$, which are obtained from the RDVFS algorithm (Figure 5-3). The value of energy consumption of $k^{th}$-task in this example is calculated as $E_{SMFS-DVFS}^{(k)} = 28.43 \ mW$. It can be noted that SMFS–DVFS gives the closest energy to the optimum energy $\left(E_{Opt-Cont.}^{(k)}\right)$ compared with the RDVFS algorithm.

## 5.6. Summary and Remarks

Simulation results on MFS-DVFS described in the previous chapter showed that a task can reach its highest energy saving when it is executed with at most two frequencies. These two frequencies are on both sides of the optimum continuous frequency for that task. The problem becomes interesting when a simplified version of the frequency-power equation is used. In this case, these two frequencies are adjacent. This chapter proved these observations mathematically.

# Part II- Statistical Performance Prediction and Automatic Tuning of MapReduce Jobs

# Chapter 6. Background and Literature Review on Research in MapReduce

## 6.1. Introduction

Recently, businesses have started adopting MapReduce as a popular distributed computing framework for processing large-scaled data in both public and private clouds; e.g., many Internet endeavours have already deployed MapReduce architecture to analyse their core businesses by mining their produced data. Consequently, application developers stand to benefit from understanding performance trade-offs in MapReduce-style computations in order to better utilise their computational resources. After an overview of the MapReduce programming model, in this chapter we review current research areas and corresponding related works including performance monitoring, modelling/provisioning, and optimisation of Hadoop/MapReduce. Then we explain how the Hadoop/MapReduce cluster works and how one can profile performance in such clusters.

## 6.2. MapReduce Programming Model

MapReduce, introduced by Google in 2004 [63], is a framework for processing large quantities of data on distributed systems. Each computation on this framework consists of three major phases: Map, Shuffle, and Reduce as shown in Figure 6-1.

In the Map phase, after copying the input file to the MapReduce distributed file system (DFS) and splitting the file into smaller chunks (or Blocks), data inside the split files are converted into $< key, value >$ pairs (e.g., $key$ can be a line number and $value$ can be a word in an essay). These $< key, value >$ pairs are entered into Mappers, where the first part of the processing (based on map function) is applied to them. In fact, as Mappers in such a framework are designed independently, MapReduce applications are always naturally ready to parallelise. This parallelisation, however, can be bounded sometimes due to other issues such as the nature of the data source and/or the number of CPUs having access to the data simultaneously.

In the Shuffle phase, a network intensive job starts to collect intermediate produced $< key, value >$ pairs by Mappers and transfer them to Reducers. The outputs with the same intermediate $key$ obtained by applying map operations will be presented to the same Reducers. After this point, depending on the MapReduce configuration, a sort/combine stage may also be applied to expedite the whole process. Finally, in the Reduce phase, a reduce function is applied to the values of intermediate keys. The result is concurrently created and written in output files (typically one output file) in the file system.

The process of converting an algorithm into independent mappers and reducers causes the MapReduce model to be inefficient for algorithms with a sequential nature. In fact, MapReduce is designed for computing a massive amount of data (a.k.a. Big data) instead of making complicated computations on a small amount of data [64]. Due to its simple structure, MapReduce suffers from serious deficiencies, predominantly in scheduling, energy efficiency and resource allocation.

In distributed computing systems, MapReduce has been known as a Big data processing framework [64-66] indicating that performance (e.g., execution time, network load, CPU usage) is the most critical aspect of running an application on this framework. As a result, improving the performance of an application is crucial to customers to hire enough resources from MapReduce providers, as well as for these providers to schedule incoming jobs properly.

Among various parameters influencing the performance of a running application on a MapReduce cluster, in this thesis we will study the impact of configuration parameters, and in particular, the values for number of map tasks and number of reduce tasks. Generally, problems regarding the dependency between configuration parameters and performance of applications in a framework such as MapReduce can be grouped into four categories:

1) Which configuration parameters are involved in influencing the performance (such as resource utilisation, energy consumption, or execution time) for different types of applications? For MapReduce, the list of available parameters can be found in [67].

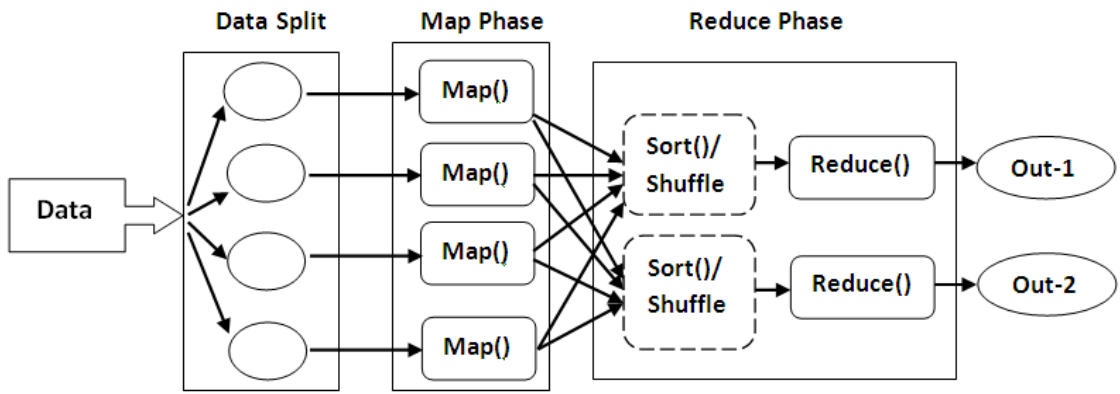

**Figure 6-1. MapReduce workflow**

2) Among these parameters, which ones are the most important? In other words, what are the most effective parameters influencing the performance? This question can be answered by P-value, t-statistics, linear/non-linear correlation analysis, and Principle Component Analysis (PCA); however, it is expected that the list of effective parameters will be application dependent.

3) As the relationship between these parameters and performance is not trivial, the next issue is how to statistically/mathematically model this relationship and thus calculate optimal values for these parameters. We have proposed to model the relationship between these two configuration parameters and total CPU tick clocks and execution time for MapReduce applications. However, in modeling I/O utilisation (Network, memory, disk, etc.) which is beyond the scope of this thesis, a linear model may not be applicable. In this case, it might be more appropriate to use non-linear modeling techniques like neural network and fuzzy logic. In the present study, we will study network load in MapReduce and model the relationship between network load and two main parameters in MapReduce using statistical regression. As addressed in the literature, one of the performance bottle-necks in MapReduce is its heavy load on the cluster network during the map, shuffle and reduce phases. The network load is of special concern with MapReduce as large amounts of traffic can be generated in the shuffle phase when the output of map tasks is transferred to reduce tasks. As each reduce task needs to read the output of all map tasks, a sudden explosion of network traffic can significantly deteriorate cluster/cloud performance. This is especially true when data has to traverse a greater number of network hops while going across *racks* of servers in a data centre [68]. Generally, the network of the MapReduce cluster is stressed during (1) the shuffle phase where each reducer contacts all other reducers – most probably on other machines in a cluster/cloud – to collect intermediate files; and (2) the reduce output phase where the final results of the whole job will be written to HDFS – usually with three replicas. Among them, the former is the most intensive period of network load and acts as a performance issue in most MapReduce applications. Therefore, from a performance perspective, it would be interesting and useful to analyse and provision the network load of a submitted application before its actual running.

4) Another issue is how to classify applications based on their performance. In other words, if a new unknown application arrives, how to find the existing application in the database that is most similar to it? The idea is that if two applications are classified into the same class, then they might have the same optimal values of configuration parameters; thus the configuration parameters of the new unknown application can be set without previous steps. Toward this issue, a statistical technique for matching between applications is proposed in this thesis.

The answers to the above-mentioned questions can be used to intelligently manage and schedule incoming jobs to a MapReduce cluster.

## 6.3. Statistical Machine Learning and Prediction of Performance in MapReduce

Machine learning (ML) is a promising discipline which designs a model of a system by using the historical data of the system. Typically, ML techniques have two stages: learning (or training) and prediction. In the former, a learner model can take advantage of examples (data) coming from the system to capture characteristics of interest of their unknown underlying probability distribution. Data can be seen as instances of the possible relationships between observed variables. A major focus of machine learning research is the design of algorithms that recognise complex patterns and make intelligent decisions based on input data. After learning the model, it is used to predict the value of new input data. Figure 6-2 shows our ML-based model for modelling and prediction of the performance of MapReduce. This framework can also be used with other ML-based applications with a few changes.

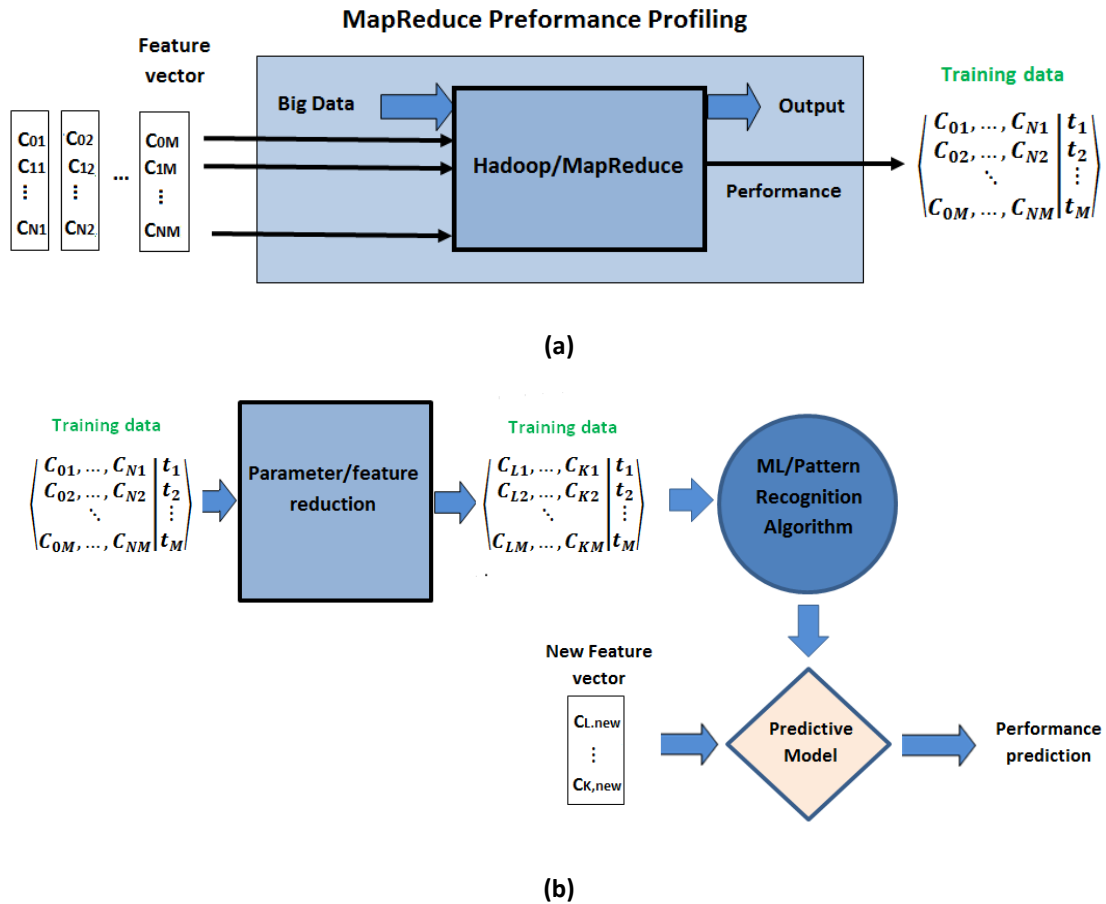

**Figure 6-2. Machine learning modelling and prediction of performance in MapReduce: (a) profiling and data gathering, (b) modelling and prediction.**

### 6.3.1. MapReduce performance profiling

The fuel of ML-applications is historical data. A robust environment is required to get meaningful data, which generally requires a few processing steps on historical data, including filtering, cleaning, and outlier removal. In the MapReduce framework, a set of jobs for an application is generated by giving different values for configuration parameters $< C_0, \dots, C_N >$ (in our work, two configuration parameters, i.e., number of map tasks and number of reduce tasks, are used). It is worth noting that in most ML techniques it is necessary that the number of jobs be much larger than the number of configuration parameters, i.e., $M \gg N$. Then, each job is run on the given MapReduce platform. While running each job, the resource utilisation (i.e., CPU usage, network usage) of the job is gathered – as training data – to build a trace for future deployments as

$$\begin{pmatrix} C_{01}, \ldots, C_{N1} & t_1 \\ C_{02}, \ldots, C_{N2} & t_2 \\ \ddots & \vdots \\ C_{0M}, \ldots, C_{NM} & t_M \end{pmatrix}$$

Where $t_i$ can be a real number (i.e., execution time, total CPU usage, total network load) or a vector with real number, such as CPU time series or network time series of the application on the given MapReduce cluster. Such data can be easily gathered in Linux through functions provided in the SysStat monitoring package [69] with almost no overhead. Within the system, the resource usage of the job, for each machine, is sampled from the time the mappers start (i.e., 'running job') to the time all reducers finish (i.e., 'job complete') with a time interval of one second (Figure 6-3, top left panel); for example, the top right panel in this figure indicates CPU utilisation samples, while the ones at the bottom of the figure are network load samples of the running job. Moreover, if the execution time of the job is of interest, it is calculated as the difference between 'running job' time and 'job complete' time.

### 6.3.2. Parameter/ feature dimension reduction

In ML, dimension reduction is the process of reducing the number of random variables under consideration. The complexity of ML modelling has a high dependency on the number of features/parameters; thus it is important to decrease the number of features (a.k.a. dimension reduction) to simplify the calculation. The dimension reduction can be achieved into two ways. In the first method, the input parameters/features are dependent. Therefore, Principle Component Analysis (PCA) and correlation analysis should be used to transform the feature vector to orthogonal space and keep most important parameters. However, there are two issues with using PCA. First, PCA and correlation analysis detect linear relation among features. Therefore, if two features are highly correlated in a non-linear way, PCA and correlation analysis may score low dependency (there are some kernel-PCAs which model non-linearity, but their efficiency depends on finding proper kernel function). Second, PCA, by definition, finds relationships between measurement features; this means that if two parameters are valued by a user – which is the case in our MapReduce performance modelling based on configuration parameters – there should not be any correlation between them, and therefore PCA cannot reduce dimension.

In the second method of achieving dimension reduction, the input parameters/features may be independent, but they may have influence on performance of system. In this case, correlation can be used to find the relationships between performance indicator and input parameters. This may fail due to the non-linear relationships between them. Other approaches use P-values and t-statistics, which both study standard deviation between performance indicator and input parameters. In P-value, a null hypothesis (for example, that 'number of map tasks' has no influence on network load or execution time) is tested based on historical data; as it does not rely on the linearity or non-linearity of the relationship, it should give a better dependency test; then the parameters/features with less influence on the performance indicator can be removed from the feature space. So, the training data changes to:

$$\begin{pmatrix} C_{L1}, \dots, C_{K1} \\ C_{L2}, \dots, C_{K2} \\ \ddots \\ C_{LM}, \dots, C_{LM} \end{pmatrix} \begin{vmatrix} t_1 \\ t_2 \\ \vdots \\ t_M \end{vmatrix}$$

Where $<C_L, \dots, C_K>$ is a subset of $<C_0, \dots, C_N>$. In the present study, as only two configuration parameters are studied, this step is neglected.

### 6.3.3. Machine learning technique

One of the major categories of ML algorithms is pattern recognition, which studies the assignment of a label/output to a given input value. Depending on the type of label output, on whether learning is supervised or unsupervised, and on whether the algorithm is statistical or non-statistical in nature, algorithms for pattern recognition can be divided into the following main groups:

- Regression algorithms (predicting real-valued labels): regression analysis is a statistical technique for predicting the relationships among variables. These variables can have discrete or continuous values. When output variables are continuous (i.e., accepts real-valued labels), a regression model predicts the behaviour of a system; when output variables are discrete, regression analysis is used for classification. More specifically, regression analysis studies how an output of a system (as a dependent variable) changes when an input (as an independent variable) varies, while other inputs are held fixed. From a statistical perspective, regression analysis tries to estimate the conditional expectation of the output variable (i.e., dependent variable) based on input variables (i.e., independent variables), which is the average value of the output variable when the input variables are fixed. In the case of performance provisioning of MapReduce applications (where output is a continuous variable) based on configuration parameters (inputs are discrete variables), regression is the main choice for modelling.

- Clustering algorithms (predicting categorical labels when categories are unknown): in ML, clustering (a.k.a. unsupervised learning) refers to the process of grouping unlabeled data into categories based on some measure of inherent similarity (e.g., the distance between instances, considered as vectors in a multi-dimensional vector space). The number of categories and also their structure are unknown and are obtained from the unlabeled data. Since the data given to the learner is unlabeled, there is no error or reward signal to evaluate a potential solution. There are many techniques in unsupervised learning that try to summarise and explain key features of input data. Three common approaches to unsupervised learning/clustering are k-means, mixture models, and hierarchical clustering.

  In MapReduce resource provisioning, clustering/unsupervised learning can be used to find groups or classes with a similar structure to MapReduce applications in term of performance (e.g., applications that are CPU/network intensive). These obtained classes are then used for classification of new applications by using classification algorithms.

- Classification algorithms (predicting categorical labels when categories are known): in ML and statistics, classification (a.k.a. supervised learning) is the process of assigning a new observation to a set of known classes/categories – these classes/categories may be identified by the user or obtained from clustering/unsupervised learning of historical observations. The first step in clustering/supervised learning is to analyse historical observations and identify a set of classes/categories; these classes are created based on a set of quantifiable properties, known as various explanatory variables, features, etc. These properties may be categorical (like 'spam', 'non-spam' for email), ordinal (like 'heavy', 'medium' or 'light' for a car), integer-valued (like the frequency of occurrences of a word in a text) or real-valued (like a measurement of car speed).

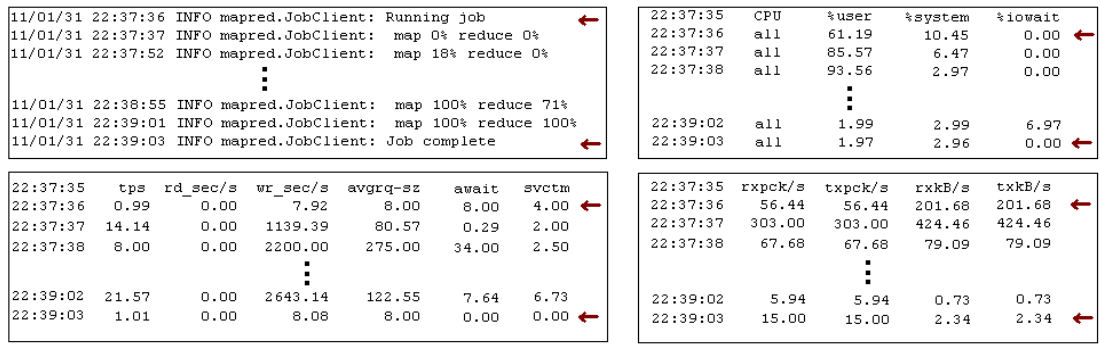

**Figure 6-3. The flow of the MapReduce job in Hadoop (left) and CPU usage time series extracted from actual system (right).**

An algorithm that implements classification is known as a classifier, which, based on key features, decides that a new observation belongs to a particular class/category.

In MapReduce resource provisioning, classification is a post-processing step after clustering to classify a new job/application into available known classes, where the classes group applications with similar performance in CPU usage, network usage, etc.

- Pattern matching and classification: generally, pattern matching searches for exact matching of an input and pre-existing patterns in a database (DB) and therefore is the opposite of pattern recognition; however, in the present study, we use uncertain pattern matching on historical data of different MapReduce applications as a pre-processing step to assign similar applications with the same label/category (which is a type of classification). The pattern matching algorithm in this work has two fundamental differences compared to common pattern matching algorithms: first, input and pre-existing patterns in DB have different lengths and thus their length must become the same; and second, points in both input and pattern in DB are uncertain and have normal distribution.

### 6.3.4. Predictive model and new inputs

The outcome of ML/pattern recognition is a predictive model which was trained from historical data. This predictive model is then used to estimate the output/label of new unseen input data (in our case, new values of configuration parameters).

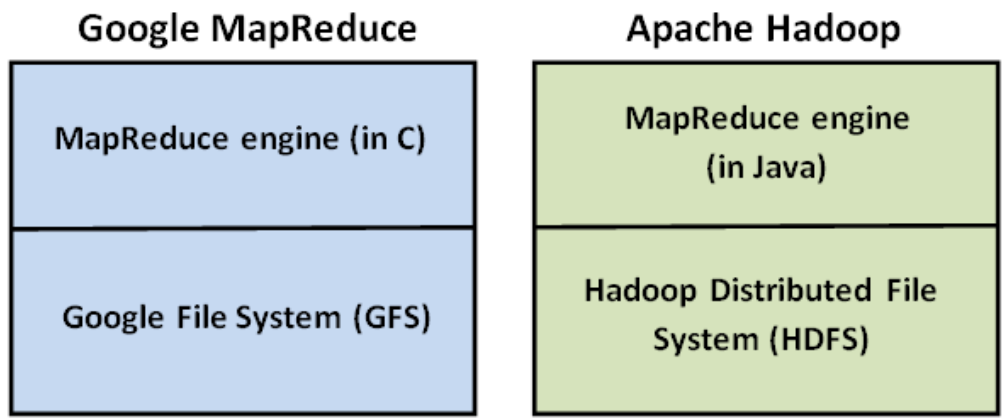

## 6.4. An Overview on Hadoop MapReduce Clusters

Hadoop is an open-source implementation of the MapReduce framework developed in Java by Apache [70]. The Hadoop core is divided into two fundamental layers: a MapReduce engine and HDFS. Hadoop uses Hadoop Distributed File System (HDFS) as its underlying layer rather than Google File System (GFS). The MapReduce engine is the computation engine running on top of HDFS as its data storage manager. As most of the techniques in the literature and also the algorithms in the present study were developed on a Hadoop cluster, in the three following sections we will review the details of the structure of the Hadoop cluster, Hadoop distributed file system (HDFS), and MapReduce engine. Figure 6-4 demonstrates the one to one correspondence of the distributed file system and MapReduce engine in two different implementations of MapReduce by Google and Apache.

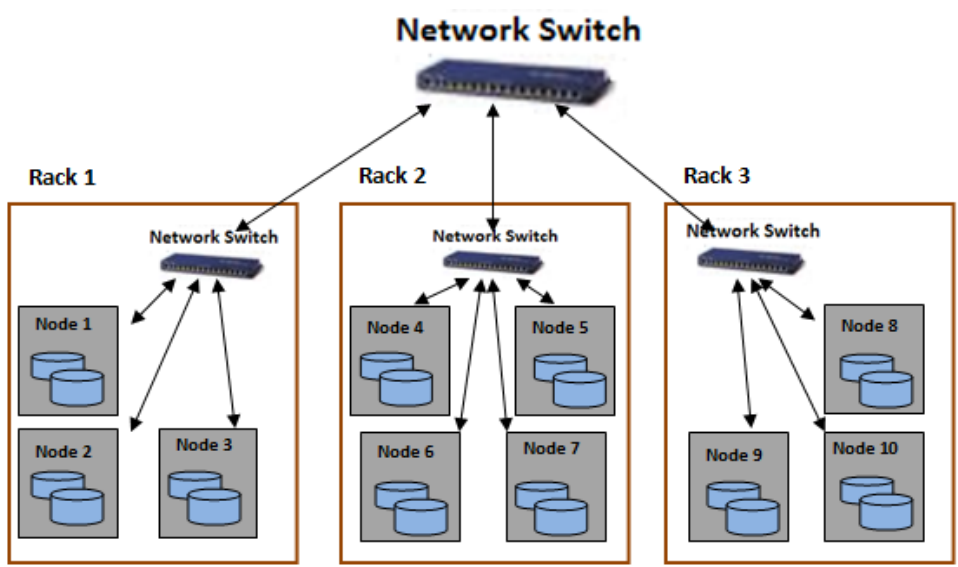

Figure 6-5. Hadoop cluster architecture

### 6.4.1 Hadoop cluster structure

Figure 6-5 shows how nodes in a Hadoop cluster are organised into a set of racks where all nodes in a rack are connected to a switch via high bandwidth links; then all racks' switches in the cluster are connected through another high bandwidth switch. Although the total bandwidth among nodes (a.k.a. inter-rack connection) in a rack is much higher than the total bandwidth among the racks in the cluster (a.k.a. intra-rack connection), bandwidth per node is much lower than bandwidth per rack (as inter-rack connection is shared by all nodes in a rack); this results in a higher risk of a bandwidth bottleneck among nodes of a rack compared to bandwidth problems among racks of a cluster; therefore, applications should be aware of network traffic within a rack and limit it whenever possible.

### 6.4.2 Hadoop Distributed File System

HDFS – a distributed file system in Hadoop very similar to Google File System (GFS) – organises files and stores their data on a distributed computing system. As shown in Figure 6-6, HDFS contains a single NameNode on a master node of a cluster, and a number of DataNodes as workers on other nodes. HDFS splits files (or data) into fixed-size data blocks (the size of data blocks is set by *'dfs.block.size'* parameter and is 64 MB by default) and distributes them among DataNodes; the placing of these data blocks is managed by the NameNode. To increase fault tolerance in Hadoop cluster, each data block is replicated three times (by default); it is also the responsibility of NameNode to track the place of each replica.

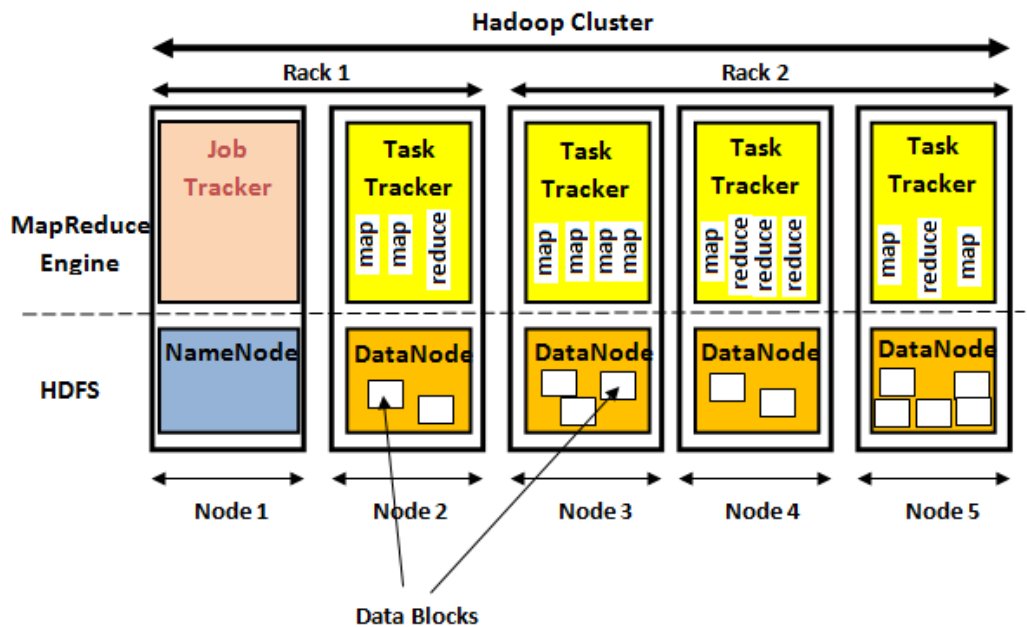

**Figure 6-6. HDFS Architecture [1]**

### 6.4.3 MapReduce engine

As shown in Figure 6-6, the MapReduce engine is the top layer of HDFS and is responsible for executing MapReduce applications. Similar to HDFS, this layer also has master/slave architecture: JobTracker lives on the master node (which is the home for NameNode in HDFS) and TaskTrackers lives on slave/worker nodes (which are the DataNodes locations in HDFS).

The workflow of processing a job in Hadoop MapReduce is as follows:

1) User loads data into HDFS, submits job to JobTracker, and gives value to configuration parameters.

2) The data is divided into several data blocks. Each data block is assigned to one map task. Therefore, the number of map tasks (or mappers) is forced by data blocks through the '*dfs.block.size*' configuration parameter.

3) Each map task is executed on an available map slot on worker nodes. Worker nodes (i.e., TaskTracker) execute several simultaneous map/reduce tasks depending on the number of simultaneous threads its hardware can support. For example, a 10 node cluster (i.e., 10 worker nodes) with quad-core CPUs has $10 \times 4 = 40$ slots.

4) The slots on the worker are configured with a fixed number of map slots and another fixed number of reduce slots via two configuration parameters:
   '*mapred.tasktracker.map.tasks.maximum*' for map slots and '*mapred.tasktracker.reduce.tasks.maximum*' for reduce slots, respectively; these two configuration parameters must be set in a *mapred-site.xml* file on each worker node in the Hadoop cluster before running an application.

5) The Hadoop scheduler is responsible for allocating map slots to map tasks. When there is no available map slot, pending map tasks must wait in the queue till some map slots become free. To reduce network traffic in the cluster, data blocks and their assigned map tasks should be located as close as possible.

6) The outputs of map tasks are shuffled through the network as inputs of reduce tasks. Although reduce tasks are started when map tasks are started, they only move the output of map tasks (i.e., intermediate records) until all map tasks are finished. A partitioning function moves the intermediate records with the same key to the same reduce tasks for processing; the number of reduce tasks is specified by the application via setting the '`mapred.reduce.tasks`' configuration parameter.

7) When all map tasks outputs are moved, the reduce phase starts. Allocation of available reduce slots to reduce tasks is another duty of the Hadoop scheduler. If there is a lack of available reduce slots, pending reduce tasks must wait in the queue till some reduce slots become free. The reduce function is called by reduce tasks to process the intermediate records and form the final output.

How to choose the right number of map/reduce slots and number of map/reduce tasks are still open research questions, but there are some rules of thumb:

- The slots can be divided in a way which integer-divides into the cores of a single machine: e.g., the ratio of number of map nodes to number of reduce nodes in a worker node with quad-core CPUs can be assigned as 3:1, 2:2, or 1:3. However, this rule of thumb does not work for all applications. For instance, if there is 4GB memory in a node with quad-core CPUs and four map/reduce slots, then each slot receives 1GB memory; if map tasks are memory intensive then this small amount of memory increases the execution time in map phase; therefore it is important to choose the number of map/reduce slots based on the type of application and its internal status.

- The number of map/reduce tasks is applied to application data in HDFS. The former splits input data into several blocks while the latter applies to the output of the map phase: e.g., with 10 TB of input data and 128 MB DFS block size, it will end up with 82 KB map tasks, unless '`mapred.map.tasks`' is even larger. Generally, the rule of thumb for the number of map tasks is around 10–100 map tasks per node, while this value for the number of reduce tasks is $(0.95 \text{ or } 1.75) \times \text{number of nodes} \times \text{'mapred.tasktracker.tasks.maximum'}$.

## 6.5. Related Works in MapReduce Performance Modelling and Optimisation

Early works on analyzing/improving MapReduce performance started around 2005; such as an approach introduced by Zaharia et al [71], which addresses the problem of improving the performance of Hadoop for heterogeneous environments. Their approach is based on the critical assumption in Hadoop for homogeneous cluster nodes – that tasks progress linearly. Hadoop utilises these assumptions to efficiently schedule tasks and (re)execute the stragglers. Their work introduces a new scheduling policy to overcome these assumptions. Besides their work, there are many other approaches to enhance or analyse the performance of different parts of MapReduce frameworks, particularly in scheduling [72], energy efficiency [73-75] and workload optimisation [76]. A statistics-driven workload modeling is introduced in [75] to effectively evaluate design decisions in scaling, configuration and scheduling. This framework is utilised to make appropriate suggestions to improve the energy efficiency of MapReduce applications. The authors in [77] propose a technique to model the total execution time of Hive queries, a higher level software for database interaction written for Hadoop. It uses Kernel Canonical Correlation Analysis to obtain the correlation between the performance feature vectors extracted from MapReduce job logs, and map time, reduce time, and total execution time. These features are acknowledged as critical characteristics for establishing any scheduling decisions.

A basic model for MapReduce computation utilisations is presented in [78, 79]. Here, at first, the map and reduce phases are independently modeled using dynamic linear programming; afterwards, these phases are combined to build a global optimal strategy for MapReduce scheduling and resource allocation. The authors in [80] describe an analytical model to predict the optimal execution time of MapReduce applications. This optimal execution time is calculated based on the required shortest time to move data; however, their evaluation shows that both MapReduce and Hadoop implementations are not as efficient as estimated. In [81], a model based on queuing theory is proposed for MapReduce applications. Assuming homogenous workload with many instances of the same job, the model tries to predict waiting time for map and reduce tasks. Starfish [82] is a self-tuning model for 'big data' analysis which splits tasks into several stages and models each stage with a different model considering all resources including CPU, disks and network. In [83], an evaluation of MapReduce performance was studied through a simulation approach. MPPref has been introduced [84] as a MapReduce simulator to facilitate exploration of MapReduce design space. As suggested by the authors, this simulator can be used to systematically understand the performance of MapReduce systems and realise the effect of tweaking the configuration parameters.

As reported by the authors in [85], Hadoop – as well-known implementation of MapReduce – fails to guarantee high performance in the shared MapReduce computation environment; this comes from the Hadoop assumption that all cluster nodes are dedicated to a single user. To solve this problem, the authors in this paper present two optimisation schemes, prefetching and pre-shuffling, and implement them in a plugin. MapReduce online [86] is a modified version of Hadoop that supports online aggregation and continuous queries, and allows pipelining data between tasks and jobs. This extension of the MapReduce programming model can reduce completion time of batch jobs and also improve resource utilisation. In [87], the problem of poor performance of running MapReduce on volunteer computing system (e.g., desktop computers through frameworks like Condor) with high rate of resource unavailability is studied. Then, MOON framework as an extension of Hadoop is proposed to run reliable MapReduce services on hybrid resource architectures. The scheduling algorithm in this framework detects different types of both MapReduce data and nodes in the MapReduce environment, and thus effectively places both tasks and data on both volatile and dedicated nodes.

Outlier detection in MapReduce environments is reported in [88] through introducing Mantri monitoring system. The outliers, which come from disk failures, bandwidth rate variation and congestion in network, and run-time confliction for hardware resources, alter job completion; therefore, the Mantri tries to fix these outliers by restarting them, performing network-aware replacement of tasks and protecting outputs of valuable tasks. The same authors report the Scarlett system in [89] to improve data replication in a MapReduce cluster by making more copies of popular data across the cluster. This system analyses data, predicts popular data, and replicates them with minimal interference to running jobs. The authors in [90] present Orchestra to analyse the network traffic pattern in MapReduce and Dryad, and find that transferring a massive amount of data between their internal states (e.g., shuffle phase in MapReduce) takes a significant portion of job completion time — depending on the nature of the job and the amount of intermediate data produced in these frameworks, these transfers may take more than 50% of job completion time. Therefore, a set of algorithms are proposed: first, to improve the transfer times of broadcasting and shuffling; and second, to allow scheduling policies at the transfer level (e.g., prioritising a transfer over other transfers). A distributed memory cache service is introduced in [91] through PACMan framework for both MapReduce and Dryad; this framework improves access latency and reduces load on disks by caching popular data blocks (i.e., data blocks that have been accessed multiple times) and loading them for future access directly from memory. The original Hadoop uses a first-come-first-served policy to schedule tasks, which causes large jobs to block processing of small jobs. Therefore in [71], a fair scheduler is proposed to tackle this problem.

A specific area in cluster computing systems including MapReduce and Dryad is workload consolidation. Workload consolidation studies how different types of workloads (e.g., CPU intensive, I/O intensive and so on) can share cluster resources to reach different performance goals (e.g., consumed energy, throughput, and response time). A similar work to workload consolidation is the purpose of the task scheduler proposed in [92], which was designed to predict the performance of concurrent MapReduce workloads and adjust their resources so that job response times are minimised. Delay scheduling proposed in [93] addresses the job latency problem on Hadoop clusters at Facebook and focuses on studying the trade-off between fairness in the scheduler and data locality in Hadoop applications. Quincy [94], a platform-specific scheduler implemented on a Dryad distributed execution engine, is a fair-share scheduler also addressing the same problem. The researchers in [95] address the problem of performance unpredictability and variance in EC2 cloud for MapReduce applications, and discover that unpredictability is greatly related to poor workload consolidation. There are also other studies where consolidation is used to optimise power and energy. Energy-aware workload consolidation in [96] is an attempt to conserve energy for disk-/CPU-intensive applications in cloud computing environments; the approach, however, lacks accurate workload characterisation. In [97], a novel runtime framework is proposed to dynamically consolidate instances from different workloads into a single GPU workload. They also propose GPU performance and power models for effective workload consolidation on GPUs. Joulemeter [98], which was initially designed as a tool for power usage measurement of virtual machines, aims to consolidate multiple workloads on fewer servers for improving resource utilisation and power costs.

Several studies analyse the effect of a last level processor cache (LLC) on workload consolidation. For example, research in [99] studies the behaviour of consolidated workloads, particularly on sharing caches across a variety of configurations. In [100], the authors also study shared resource monitoring to understand resource usage and ways to improve overall throughput as well as the quality of service of a datacentre. Further, an analytical model has been proposed to predict the effect of cache contention on the performance of consolidated workloads [101].

# Chapter 7. A Study on Using Uncertain Time Series Matching Algorithms in MapReduce Applications (fourth algorithm)

## 7.1. Introduction

In this chapter, the CPU utilisation time patterns of several MapReduce applications are studied. After extracting running patterns of several applications, the patterns along with their statistical information are saved in a reference database to be later used to tweak system parameters to efficiently execute future unknown applications. To achieve this goal, the CPU utilisation patterns of new applications and their statistical information are compared with the already known applications in the reference database to find/predict their most probable execution patterns. Because of different pattern lengths, Dynamic Time Warping (DTW) is utilised for such comparison; a statistical analysis is then applied to DTWs' outcomes to select the most suitable candidates in the reference database. Furthermore, a new algorithm is proposed to classify applications according to similar CPU utilisation patterns. Finally, the dependency between minimum distance/maximum similarity of applications and their scalability (in both input size and number of virtual nodes) is studied.

This part of our work is inspired by another discipline (speaker recognition) in which the similarity of objects is also very important. In speaker recognition (or signature verification) applications, it has been already shown that if two voices (or signatures) are significantly similar – based on the same set of parameters as well as their combinations – then they are most likely produced by a unique person [102]. Inspired by this proven fact, our proposed technique in this chapter uses the same logic for pattern feature extraction and matching, an area which is widely used in pattern recognition, sequence matching in bio-informatics and machine vision. Here, we extract the CPU utilisation pattern of unknown/new MapReduce applications using a small amount of data (not the whole data) and compare it with already known patterns in a reference database to find similarity between the applications. As a result, the optimal values of configuration parameters (the number of map tasks, and number of reduce tasks) for unknown/new applications can be set based on the already calculated optimal values for similar known applications in the database. This work has parallels with the work in [103] as well as our previous approach where matching algorithms (specifically DTW) were also used to find the relationship between historical data of applications and the new application on both Cloud and MapReduce platforms. Despite their many interesting findings, both of the aforementioned works suffer from a simplistic assumption that applications are run in noise-free environments, an assumption that sometimes contradicts reality. In fact, due to the dynamic nature of clusters/grids/clouds, identical experiments – i.e., running an application with the same set of data and configuration parameters – may result in different outcomes; e.g., different CPU time series. Our new approach is to address this issue by assuming uncertainty in the collected CPU time series of applications, and then to design a more sophisticated pattern-matching algorithm to consider such uncertainty in discovering relationships among different applications.

## 7.2. Theoretical Background

Pattern matching is a well-known approach used to transform a time series pattern into a mathematical space. Such transformation is essential to extract the most suitable running features of an application before comparing it with reference applications in a database to find its counterpart. Such approaches have two general phases: (1) the profiling phase, and (2) the matching phase. In the profiling phase, time series patterns of several applications are extracted. Several mathematical operations are then applied on these patterns (including magnitude normalisation), and results are stored in a database to be used as a reference for the matching phase. In the matching phase, the same procedure is repeated for an unknown/new application first, and then the time series of this application is compared with those stored in the database – using a pattern matching algorithm – to find the most similar ones.

### 7.2.1. Uncertain time series

A time series $\varphi_c(.)$ is called a certain time series when its data values are fixed/certain: $\varphi_c(.) = [\varphi_c[1], ..., \varphi_c[N]]$ where $\varphi_c[i]$ is the value of time series at time $i$. A time series $\varphi_u(.) = [\varphi_u[1], \ldots, \varphi_u[N]]$ is called **un**certain when there is uncertainty in its data values [104] and can be formulated as:

$$\varphi_u[i] = \varphi_c[i] + e_\varphi[i]$$

where $e_\varphi[i]$ is value of error/uncertainty in $i^{th}$ data point. Due to uncertainty, the value of each point is considered as an independent random variable with statistical mean ($\mu_\varphi[i]$) and standard deviation($\sigma_\varphi[i]$). These values are calculated during analysis of the time series in the profiling phase.

In a MapReduce application, the length of the CPU utilisation time series as well as the values in each point of the time series may change for even several identical executional environments of an application – i.e., the same input data file size, number of map tasks, and number of reduce tasks – thus, the CPU utilisation time series are considered 'uncertain' and become suitable for statistical similarity measurements.

### 7.2.2. Pattern matching

Similarity measurement algorithms have been frequently used in pattern matching, classification and sequence alignment in bio-informatics. The measurement of similarity between two (normalised) uncertain time series means to find a function:$SIM(\varphi_u, \phi_u)$ where $\varphi_u(.)$ and $\phi_u(.)$ are two time series without the same length. This function is typically designed as $0 \leq SIM(\varphi_u, \phi_u) \leq 1$, where greater values means higher similarities. In this case, $SIM(\varphi_u, \phi_u) = 1$ should be obtained for identical series only, and, $SIM(\varphi_u, \phi_u) = 0$ should reflect no similarity at all. To this end, 'similarity distance' is defined as a specific distance between two uncertain time series to reflect the level of their similarity.

### 7.2.2.1. Dynamic Time Warping (DTW)

DTW is among the common techniques to calculate distance/similarity between two certain time series of different lengths. This approach cannot be used to find similarity between two uncertain time series as it usually results in unacceptable outcomes. DTW uses a nonlinear search to map corresponding samples of each series to find such similarity. The following recursive operation shows how such distance/similarity between two certain time series $\varphi_c(.) = [\varphi_c[1], \ldots, \varphi_c[N]]$ and $\chi_c(.) = [\chi_c[1], \ldots, \chi_c[M]]$ $(N \geq M)$ is computed:

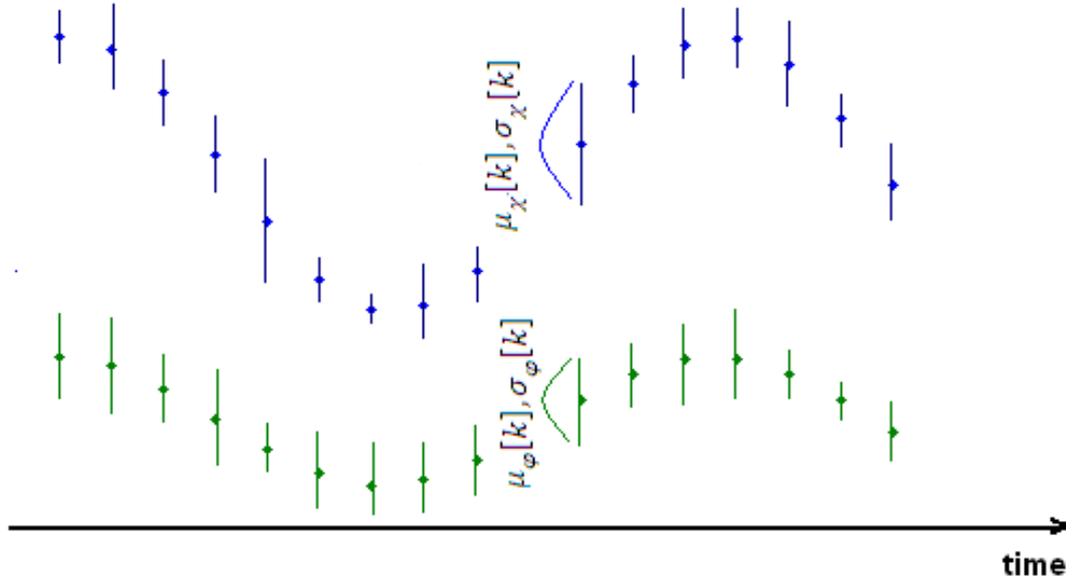

Figure 7-1. The distance between two uncertain time series and normal distribution of uncertainty in the $k^{th}$ points.

$$D(i,j) = \begin{cases} D(i,j-1) \\ D(i-1,j) \\ D(i-1,j-1) \end{cases} + d(\varphi_c[i], \chi_c[j]) \qquad (7-1)$$

where $d(.,.)$ is the Euclidean distance between corresponding points in each series, and

$$d(\varphi_c[i], \chi_c[j]) = \|CPU(\varphi_c[i]) - CPU(\chi_c[j])\|$$

where $CPU(\varphi_c[i])$ is the value of CPU utilisation at time $i$ in $\varphi_c$.

The result of these formulations is the $D[\varphi_c, \chi_c]$ matrix, in which each element, $D(i,j)$, reflects the minimum distance between $[\varphi_c[1], \chi_c[1]]$ and $[\varphi_c[i], \chi_c[j]]$. As a result, $D(N,M)$ reflects the distance/similarity between $\varphi_c$ and $\chi_c$. In this case, $\varphi'_c$ and $\chi'_c$ with equal length can always be made from $\varphi_c$ and $\chi_c$, respectively, so that $\chi'_c[i]$ is aligned with $\varphi'_c[i]$. $\varphi'_c$ and $\chi'_c$ are always made from $\varphi_c$ and $\chi_c$, respectively, by repeating some of their elements – based on $D[\varphi_c, \chi_c]$.

As mentioned earlier, because DTW cannot be directly used for uncertain time series, we only use it to produce temporary time series $(\varphi'_c, \chi'_c)$ with the same lengths. This procedure is then followed by applying DTW on the certain time series $(\varphi_c = mean(\varphi_u), \chi_c = mean(\chi_u))$ parts of the two uncertain time series $(\varphi_u, \chi_u)$ as follows:

$$\varphi_u[i] = \mathbb{N}(\underbrace{mean(\varphi_u[i])}_{\varphi_c[i]}, \underbrace{var(\varphi_u[i])}_{e_\varphi[i]}), 1 \leq i \leq N \text{ and}$$

$$\chi_u[j] = \mathbb{N}(\underbrace{mean(\chi_u[j])}_{\chi_c[j]}, \underbrace{var(\chi_u[j])}_{e_\chi[j]}), 1 \leq j \leq M,$$

where $N \geq M$. Then, to calculate the Euclidian distance between two uncertain but same length time series:

$$\varphi'_u[i] = \mathbb{N}(\underbrace{mean(\varphi'_u[i])}_{\varphi'_c[i]}, \underbrace{var(\varphi'_u[i])}_{e_{\varphi'}[j]}), 1 \leq j \leq R \text{ and}$$

$$\chi'_u[j] = \mathbb{N}(\underbrace{mean(\chi'_u[j])}_{\chi'_c[j]}, \underbrace{var(\chi'_u[j])}_{e_{\chi'}[j]}), 1 \leq j \leq R$$

where $R \geq N, M$. As a result:

$$[\varphi'_c, \chi'_c] = DTW(\varphi_c, \chi_c) \qquad\qquad (7-2)$$

It is worth noting that DTW in this chapter is only utilised to provide same length data series for $\varphi_u$ and $\chi_u$, and not to provide their actual similarity, because DTW does not affect the statistical mean and variance of points in the two uncertain time series. Thus, if DTW maps $\varphi'_c[i]$ to $\varphi_c[j]$, then

$$\begin{cases} mean(\varphi'_c[i]) = mean(\varphi_c[j]) \\ var(\varphi'_c[i]) = var(\varphi_c[j]) \end{cases}$$

### 7.2.2.2. Similarity measurement

Despite many methods in which the distance between two time series is calculated by summing up distances from points aligned by the DTW procedure, here, the square of Euclidian distance is used to calculate such similarity. To be more specific, two certain time series $\varphi'_c$ and $\chi'_c$ are similar when the square Euclidian distance between them is less than a distance threshold $(r)$:

$$DST(\varphi'_c, \chi'_c) = \sum_{i=1}^{N}(\varphi'_c[i] - \chi'_c[i])^2 \leq r$$

For uncertain time series $\varphi'_u$ and $\chi'_u$, where the problem is not straightforward as before, similarity is defined as [104]:

$$\Pr\left(DST(\varphi'_u, \chi'_u) = \sum_{i=1}^{N} D^2[i] \le r\right) \ge \tau \qquad (7-3)$$

where $D[i]$ is a random variable equal to $\varphi_u[i] - \chi'_u[i]$. Therefore, two uncertain time series are assumed to be similar when probability of their Euclidian distance is more than a pre-defined threshold ($0 < \tau \le 1$).

Because $\varphi'_u[i]$ and $\chi'_u[i]$ are independent random variables (Figure 7-1), both $D[i]$ and $DST(\varphi'_c, \chi'_c)$ are also independent random variables. Therefore, if $<\mu_{\varphi'}[i], \sigma_{\varphi'}[i]>$ and $<\mu_{\chi'}[i], \sigma_{\chi'}[i]>$ are $<$statistical mean, standard derivation$>$ of $\varphi'_u[i]$ and $\chi'_u[i]$, respectively, according to [104], $DST(\varphi'_u, \chi'_u)$ has the following normal distribution:

$$DST(\varphi'_u, \chi'_u) \sim \mathbb{N}\left(\sum_{i=1}^{N} E(D^2[i]), \sum_{i=1}^{N} Var(D^2[i])\right) \qquad (7-4)$$

where

$$\sum_{i=1}^{N} E(D^2[i]) = \sum_{i=1}^{N}\left[\mu_{\varphi'}^2[i] + \sigma_{\varphi'}^2[i] - 2\mu_{\varphi'}[i]\mu_{\chi'}[i]\right] \qquad (7-5)$$

$$+ \sum_{i=1}^{N}\left[\mu_{\chi'}^2[i] + \sigma_{\chi'}^2[i]\right]$$

and

$$\sum_{i=1}^{N} Var(D^2[i]) = 4\sum_{i=1}^{N}\left[\left(\sigma_{\varphi'}^2[i] + \sigma_{\chi'}^2[i]\right)\left(\mu_{\varphi'}[i] - \mu_{\chi'}[i]\right)^2\right] \qquad (7-6)$$

The standard normal distribution function of $DST(\varphi'_u, \chi'_u)$ can be calculated as:

$$DST_{norm}(\varphi'_u, \chi'_u) \sim \mathbb{N}(0,1) = \frac{DST(\varphi'_u, \chi'_u) - \sum_{i=1}^{N} E(D^2[i])}{\sqrt{\sum_{i=1}^{N} Var(D^2[i])}} \qquad (7-7)$$

Thus the problem in Eqn. 7-3 can be rewritten as:

$$\Pr\left(DST_{norm}(\varphi'_u, \chi'_u) \le r_{norm}(\varphi'_u, \chi'_u)\right) \ge \tau \qquad (7-8)$$

<u>Definition 1</u>: $r_{boundry,norm}$ is a minimum distance bound value that finds the lower bound for the standard normal probability in Eqn. 7-8; that is [104]:

$$\Pr\left(DST_{norm}(\varphi'_u, \chi'_u) \leq r_{boundry,norm}\right) = \tau \qquad (7-9)$$

where $r_{boundry,norm} = \sqrt{2} \times \mathrm{erf}^{-1}(2\tau - 1)$ for standard normal distribution and $\mathrm{erf}(.)$ is an error function obtained from statistics tables [105] when working on $DST(\varphi'_u, \chi'_u)$ instead of $DST_{norm}(\varphi'_u, \chi'_u)$:

$$r_{boundry} = \frac{r^2_{boundry,norm} - \sum_{i=1}^{N} E(D^2[i])}{\sqrt{\sum_{i=1}^{N} Var(D^2[i])}} \qquad (7-10)$$

<u>Definition 2</u>:

Two uncertain time series $\varphi'_u$ and $\chi'_u$ are similar with probability more than $\tau$ [104]:

$$\Pr(DST(\varphi'_u, \chi'_u) \leq r) \geq \tau \qquad (7-11)$$

when

$$r \geq r_{boundry} \qquad (7-12)$$

Here, $r_{boundry}$ defines the minimum distance between two uncertain series with probability $\tau$; that is:

$$\Pr\left(DST(\varphi'_u, \chi'_u) = r_{boundry}\right) = \tau \qquad (7-13)$$

Based on Eqns. 7-1 to 7-13, from this point forward, uncertain time series are used only; thus, $\varphi'$, $\chi'$, $\mu_{\varphi'}$, $\sigma_{\varphi'}$, $\mu_{\chi'}$ and $\sigma_{\chi'}$ are used instead of $\varphi'_u$, $\chi'_u$, $\mu_{\varphi'_u}$, $\sigma_{\varphi'_u}$, $\mu_{\chi'_u}$ and $\sigma_{\chi'_u}$, respectively.

---

**Profiling phase**

    **1. For $i^{th}$ application in database $(\varphi_i)$:**

    **2.  For $r^{th}$ small input data $(Size_r)$:**

    **3.    For $j^{th}$ set of configuration parameters values $(M_j, R_j)$:**

    **4.      Counter=1**

    **5.      Do**

    **6.       Run application with the $j^{th}$ set of parameters on a small input data$(Size_r)$**

    **7.       Capture CPU utilization Time Series with XenAPI $(\varphi_{i,j})$**

    **8.      Counter = Counter + 1**

    **9.    While (Counter <= 10)**

    **10.   For $k^{th}$ point in $\varphi_{i,j}$**

    **11.    Calculate $< \mu_{\varphi_i}[k],\ \sigma_{\varphi_i}[k] >$**

    **12.    $\mu_{\varphi_i} = \left[\ \mu_{\varphi_i}[1], \ldots,\ \mu_{\varphi_i}[N]\ \right]$**

    **13.    $\sigma_{\varphi_i} = \left[\ \sigma_{\varphi_i}[1], \ldots,\ \sigma_{\varphi_i}[N]\ \right]$**

    **14.   End**

    **15.   Save $\left\{ \varphi_i, (M_j, R_j), Size_r, \mu_{\varphi_i},\ \sigma_{\varphi_i} \right\}$ in Reference database**

    **16.   End**

    **17.  End**

    **18. End**

---

**Figure 7-2. Algorithms for profiling and pattern matching phases.**

## 7.3. Pattern Matching in MapReduce Applications

This section describes the uncertain time series matching technique used to find the distance/similarity between uncertain CPU utilisation time series of different MapReduce applications. The technique consists of two phases: profiling and matching.

### 7.3.1. Profiling phase

In the profiling phase, the CPU utilisation time series of several MapReduce applications in the database, along with their statistical information, is extracted. For each application, a set of experiments is generated with several small input data and two main MapReduce configuration parameters (number of map tasks and number of reduce tasks) on a given platform. Figure 7-2 shows the profiling algorithm. While running each experiment, the CPU utilisation time series of the experiment in each virtual node of a cloud is gathered to build a trace to be later used as the training data – this statistic can be gathered easily in virtual node (running on Linux) with the XenAPI monitoring package. Within the system, the CPU usage of the experiment is sampled in a native system from starting mappers until finishing reducers with a time interval of one second. If virtual nodes are homogeneous, the CPU time series of nodes for an application are assumed to be approximately similar. Thus, the final CPU time series of an application is computed by averaging CPU utilisation values at each point. Because of the temporal changes, several identical experiments – i.e., same input data and configuration parameters – may result in different values in each point of the extracted CPU utilisation time series. Therefore, each experiment is repeated ten times and then the statistical $< mean, variance >$ of each point in the time series is extracted. It is worth noting that the completion times of these experiments were insignificantly different from each other, and thus their differences could safely be ignored. Upon completion of ten experiments, the time series with its related set of configuration parameters values as well as its normalised statistical features is stored in the reference database. This procedure was repeated for all applications.

### 7.3.2. Matching phase

In the matching phase, the profiling is also performed for the newly submitted application, followed by several steps to find its distance/similarity with already known applications. As shown in Figure 7-2, the matching phase consists of two stages: statistical information extraction and candidate selection. In the statistical information extraction stage, the CPU utilisation time series of a new unknown application ($\chi$) is captured by XenAPI; then statistical $< mean, variance >$ at each point $< \mu_\chi[k], \sigma_\chi[k] >$ of the time series is extracted under the same input data sizes and configuration parameters. Here, because the length and magnitude of the new application time series might be different from those in the reference database ($\varphi_i$), they are first normalised and then DTW is used to make them of equal length. The result is two new uncertain time series for the new application and an application in database ($\varphi_i'$ and $\chi'$) to be

**Matching phase**

*For a new unknown application* $(\chi)$*:*

1. *For $i^{th}$ application in database* $(\varphi_i)$*:*
2. *For $r^{th}$ small input data* $(Size_r)$*:*
3. *For $j^{th}$ set of configuration parameters values* $(M_j, R_j)$*:*
4. *Counter=1*
5. *Do*
6. *Run $\chi$ with the $j^{th}$ set of parameters values parameters on a small input data($Size_r$)*
7. *Capture CPU utilization Time Series with XenAPI* $(\chi_j)$
8. *Counter = Counter + 1*
9. *While (Counter <= 10)*
10. *Calculate $\mu_{\varphi_i^{\square}}$ and $\mu_\chi$ under $j^{th}$ set of parameters values*
11. *$[\mu_{\varphi_i'}, \mu_{\chi_i'}] = DTW(\mu_{\varphi_i^{\square}}, \mu_\chi)$ : align mean times series of $\chi_j$ to mean time series of $\varphi_i^{\square}$ and form new mean time series $\varphi_i'$ and $\chi_j'$*
12. *For $k^{th}$ point in both new uncertain time series $\varphi_i'$ and $\chi_i'$*
13. *Calculate $< \mu_{\chi_j'}[k], \sigma_{\chi_j'}[k] >$ and $< \mu_{\varphi_i'}[k], \sigma_{\varphi_i'}[k] >$*
14. *$\mu_{\chi_j'} = \left[ \mu_{\chi_j'}[1], \dots, \mu_{\chi_j'}[R] \right]$ and $\sigma_{\chi_j'} = \left[ \sigma_{\chi_j'}[1], \dots, \sigma_{\chi_j'}[R] \right]$*
15. *$\mu_{\varphi_i'} = \left[ \mu_{\varphi_i'}[1], \dots, \mu_{\varphi_i'}[R] \right]$ and $\sigma_{\varphi_i'} = \left[ \sigma_{\varphi_i'}[1], \dots, \sigma_{\varphi_i'}[R] \right]$*
16. *End*
17. *Form $\left\{ \chi_j', (M_j, R_j), Size_r, \mu_{\chi_j'}, \sigma_{\chi_j'} \right\}$*
18. *End*
19. *End*
20. *End*

21. Set pre-defined Probability threshold $(\tau = 0.95)$
22. *For $r^{th}$ small input data* $(Size_r)$*:*
23. *For $j^{th}$ set of configuration parameters values* $(M_j, R_j)$*:*
24. *For $i^{th}$ application in database* $(\varphi_i)$*:*
25. *Calculate joint mean and variance of distance between $\varphi_i'$ and $\chi_i'$ From Eqns. 7-5, and 7-6*
26. *Calculate $r_{boundry}$ from Eqn. 7-10*
27. *$< \varphi_i, r_{boundry} >$ is added to candidature pool of $\chi_j$*
28. *End*
29. *End*
30. *End*

*In candidature pool of $\chi_j$, the application with lowest $r_{boundry}$ is chosen as the highest similar application to $\chi_j$*

**Figure 7-2.** *(Continued)*

later analyzed by extracting their statistical information at each point $<\mu_{\varphi'}[k], \sigma_{\varphi'}[k] >$ and $< \mu_{\chi'}[k], \sigma_{\chi'}[k] >$.

In the candidate selection stage, the mathematical analysis described in Section 7.2.2 is applied to calculate the similarity between the twisted version of the normalised uncertain time series in the database $(\varphi_i')$ as well as the new unknown application $(\chi')$. Consequently, based on Eqn. 7-13 the time series in the database which gives the minimum $r_{boundry}$ for predefined Euclidian distance probability $(\tau)$ are chosen as the most similar applications to the new application in the candidature pool. Raising the value of probability threshold $(\tau)$ will reduce the number of applications in the candidate pool, and consequently, increases the similarity selection accuracy.

## 7.4. Evaluation and Experimental Results

Four widely known/used applications (three text processing and one sorting) were deployed and implemented to evaluate the effectiveness of our method in this study. Figure 7-3 shows the structure of the private cloud used to conduct the experiments; it has the following specifications:

- Physical H/W: includes five servers, each server was a dual-core Intel Genuine 3.00 GHz with 4 GB memory, 1 GB cache and 250 GB of shared iSCSI network drive.

- Xen Cloud Platform (XCP) is used for virtualisation and has been used on top of the physical H/W. The Xen-API [106] provides functionality in high level languages like Java, C# and Python to manage virtual machines inside XCP, measure their detailed performance statistics as well as live-migrate them in a private cloud environment.

Debian images are used to provide Hadoop nodes (version 0.20.2) on the servers (the replication factor for data replicas in HDFS has been set to one); each virtualised debian was set to use 1 CPU, 1 GB RAM, and, 50 GB of disk. The number of virtual nodes were 5, 10, 15, 20, or, 25. To collect runtime CPU utilisation of these nodes, XenAPI is utilised on an Intel(R) Core i7 (four cores, eight logical processors, and 16 GB of RAM) desktop PC to monitor/extract the CPU utilisation time series of applications. Performance statistics for each experiment were collected from the

| | Exim MainLog Parsing | | | |
|---|---|---|---|---|
| | | *S-1* | *S-2* | *S-3* | *S-4* |
| **WordCount** | *S-1* | **24044** | 117017 | 94472 | 228071 |
| | *S-2* | 80648 | *64063* | 58351 | 138222 |
| | *S-3* | 79431 | **63232** | **56114** | 104255 |
| | *S-4* | 147014 | 83655 | 81434 | **70427** |

| | Exim MainLog Parsing | | | |
|---|---|---|---|---|
| | | *S-1* | *S-2* | *S-3* | *S-4* |
| **Terasort** | *S-1* | **27400** | *65102* | 65606 | 132799 |
| | *S-2* | 155038 | *67293* | 68455 | **69927** |
| | *S-3* | 123668 | 76859 | **51876** | 76589 |
| | *S-4* | 166234 | 77829 | 81751 | *74693* |

| | WordCount | | | |
|---|---|---|---|---|
| | | *S-1* | *S-2* | *S-3* | *S-4* |
| **Distributed Grip** | *S-1* | **21529** | 105309 | 90012 | 199451 |
| | *S-2* | 79965 | **62890** | 68553 | 122279 |
| | *S-3* | 77549 | 62949 | **54309** | 101280 |
| | *S-4* | 142703 | 83089 | 72987 | **70198** |

Table 7-1. A sample of the minimum distance ($r_{boundry}$) between the used applications for $\tau = 0.95$ for processing 5 G of input data on 10 virtual nodes.

'*running job*' stage to the '*job completion*' stage with a sampling time interval of one second. All CPU usages samples were then combined to form the CPU utilisation time series of an experiment. For each application, $8 \times 8 \times 5 (= 320)$ experiments were executed, where the number of map tasks and reduce tasks were 4,8,12,16,20,24, 28, or, 32; the size of input data was 5 GB, 10 GB, 15 GB or 20 GB; and each experiment was run ten times to collect its statistical information. Benchmark applications were WordCount, TeraSort, Distributed Grep and Exim Mainlog parsing. These benchmarks were chosen because they roughly represent a variety of MapReduce applications, and they are also used as valid MapReduce benchmarks in other approaches.

- WordCount [107, 108] is one of the most well-known applications used to analyse the performance of MapReduce applications, mainly by leading researchers in industry (e.g., Intel [109] and IBM [110]) and academics (e.g., MIT [111], and UC-Berkeley [71]). WordCount reads data from a text file and counts the frequency of each word; it writes its results into another text file in which each line contains a word along with its number of occurrences. During runtime, each mapper in WordCount picks a line as input and breaks it into pairs of $< key, value >$ where key is a 'word' in the line and its value is set to '1' – i.e., as $< word, 1 >$. In the reduce stage, each reducer counts the values of pairs with the same key and returns the frequency (the number occurrence) of each word.

- TeraSort [76] is a sorting algorithm used as a standard benchmark in international TeraByte sort competition [65, 112] as well as by many researchers in IBM [84, 113], Intel [109], INRIA [114] and UC-Berkeley [115]. In the reduce phase, all keys with $Sample[i-1] \leq key \leq Sample[i]$ are sent to the $i$-th reducer to guarantee that outputs of the $i$-th reduce are always less than those of the ($i$+1)-th reducer.

- Distributed Grep is used by researchers in IBM [110] and UC-Berkeley [115] to scan large-sized text files for occurrences of a particular expression. Here, mappers count the number of times the expression appears, while reducers sum up these counts and output the final result.

- Exim MainLog parsing [116] is a message transfer agent (MTA) for logging information of sent/received emails on Unix systems. This information is saved in exim_mainlog files and usually produces extremely large-sized files in mail-servers. To organise such a massive amount of information, a MapReduce application is used to parse the data – in an exim_mainlog file – into individual transactions, each separated and arranged by a unique transaction ID.

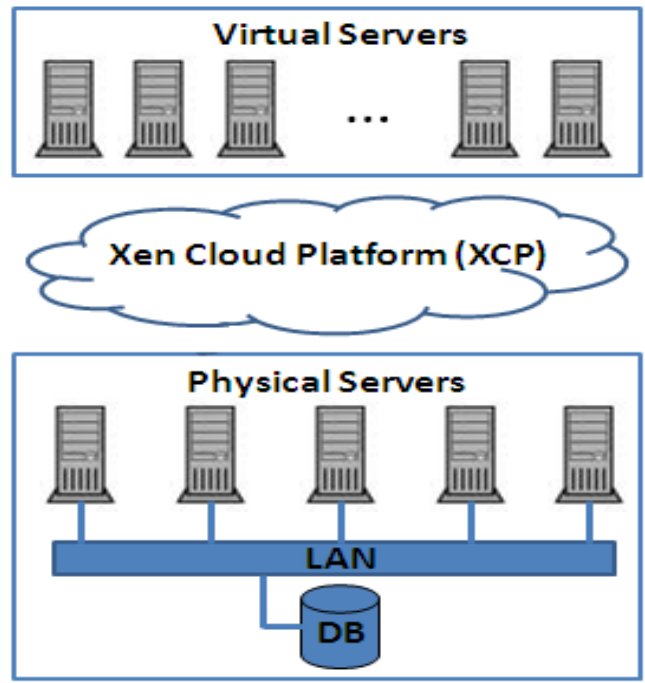

**Figure 7-3. Overall architecture of the private cloud for application pattern matching.**

Each of the aforementioned applications is executed on several small-sized input data files with a combination of difference configuration parameters (number of map and reduce tasks) to form its related CPU utilisation time series.

### 7.4.1. Application similarity

Table 7-1 and Figure 7-4 indicate the minimum distance ($r_{boundry}$) between CPU utilisation patterns of these four applications in our experiments (for 5 G of input data on 10 virtual nodes) for Euclidian distance probability of 95% ($\tau = 0.95$). In Table 7-1, the lowest and second lowest minimum distance between two applications are indicated with bold and bold-italic, respectively. Here $\{S-1, ..., S-8\}$ is the set of number of mappers and reducers used in the experiments. Figure 7-4 shows a diagonal line which represents the minimum possible distance between the two applications when they are run with the same values of the configuration parameters, and the points show the position of the calculated minimum distance between two applications, which are on or close to the diagonal line. The results in Table 7-1 and Figure 7-4 show that the diagonal numbers of these tables are always either the lowest or the penultimate lowest of all numbers, demonstrating that two computationally similar applications always have the minimum distance between them when run with similar configuration parameters. Based on this observation,

another candidate selection algorithm is proposed in which such similarities are taken into account for selecting the best set of running parameters for one application based on its similar counterparts. This new approach is detailed in Figure 7-5 and replaces the first attempt in Figure 7-2.

Based on observations, it is hypothesised that if two applications are considered 'computationally similar' for short data files, they will be fairly 'similar' for large data sizes. If this hypothesis is correct, the optimal number of map and reduce tasks for running a new unknown MapReduce application may be found through first categorising it based on its CPU utilisation patterns, and then estimating its optimal running parameters based on similar applications in the same category/class.

As an example, assume there are N applications in the database $\varphi = \{\varphi_1, \dots, \varphi_N\}$ along with their optimal configuration parameters. For a new unknown application $\chi$, this application is executed for the same set of input data and parameters – used to collect optimal values for $\{\varphi_1, \dots, \varphi_N\}$ – then the running parameters of $\chi$ can be chosen based on the most similar application in the database. The results shown in Table 7-1 indicate that WordCount, Exim and Distributed Grep can be categorised in the same CPU time series class, while Terasort forms another class.

### 7.4.2. Auto-similarity of applications

To further investigate the hypothesis, the auto similarity of an application is also studied. Here, it is expected that the diagonal numbers from calculating the auto-similarity of all applications must be significantly larger than all other off-diagonal numbers. Table 7-2 shows the results and confirms the hypothesis. This table in fact proves that only similar configurations of parameters can produce comparatively small Euclidean distances between different experiments.

### 7.4.3. $\tau$ and minimum distance ($r_{boundry}$) relation

One of the parameters influencing the minimum distance between the CPU utilisation time series of applications ($r_{boundry}$) is the value of Euclidian distance probability($\tau$). Euclidian distance probability greatly depends on the level of similarity between two applications. As expected, increasing $\tau$ always results in raising $r_{boundry}$. This observation is also well justified from a mathematical point of view as shown in Eqns. 7–9 and 7–10. Based on these equations, greater values of ($\tau$) should result in greater values of $erf^{-1}(2\tau - 1)$, and consequently, greater values of the minimum distance ($r_{boundry}$) as well.

## Distributed Grep vs. WordCount

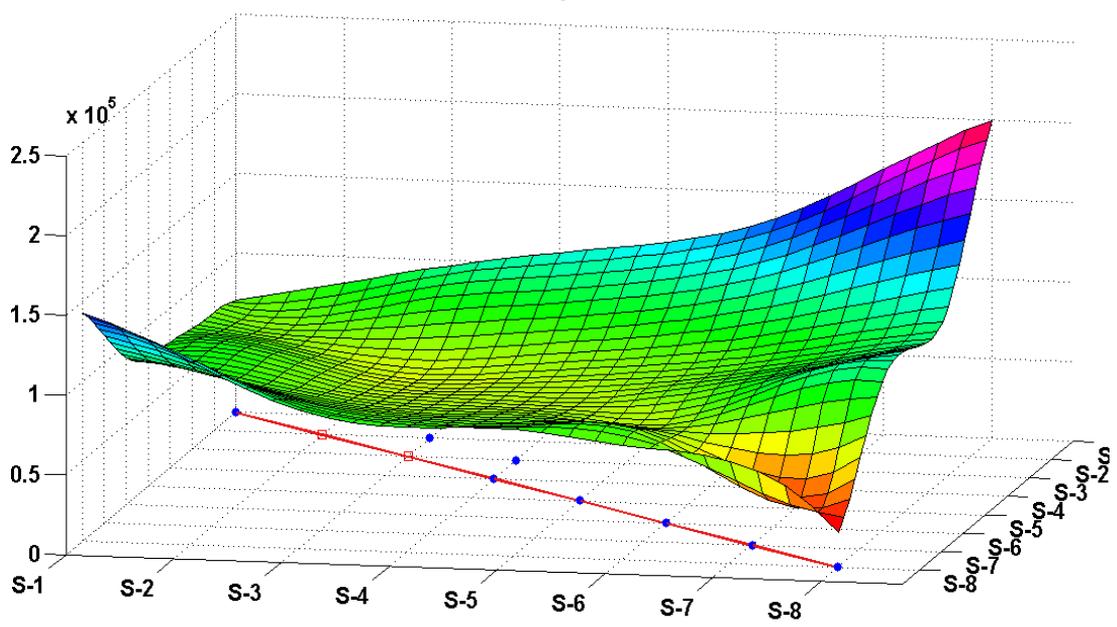

## TeraSort vs. WordCount

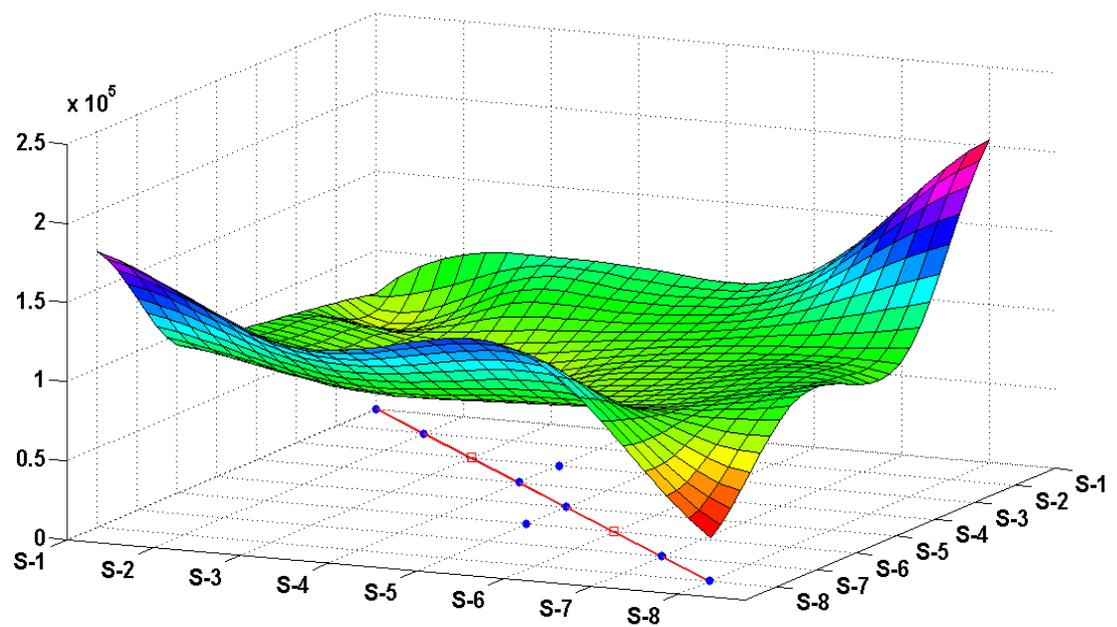

Figure 7-4. The minimum distance ($r_{boundry}$) between the four applications for $\tau = 0.95$ in processing 5 G of input data on 10 virtual nodes. The lowest minimum distance indicates the highest similarity. The y-axis shows the value of minimum distance between applications. The straight lines show the diagonal line and the points show the position of the minimum distance between two applications, which are almost all on the diagonal line.

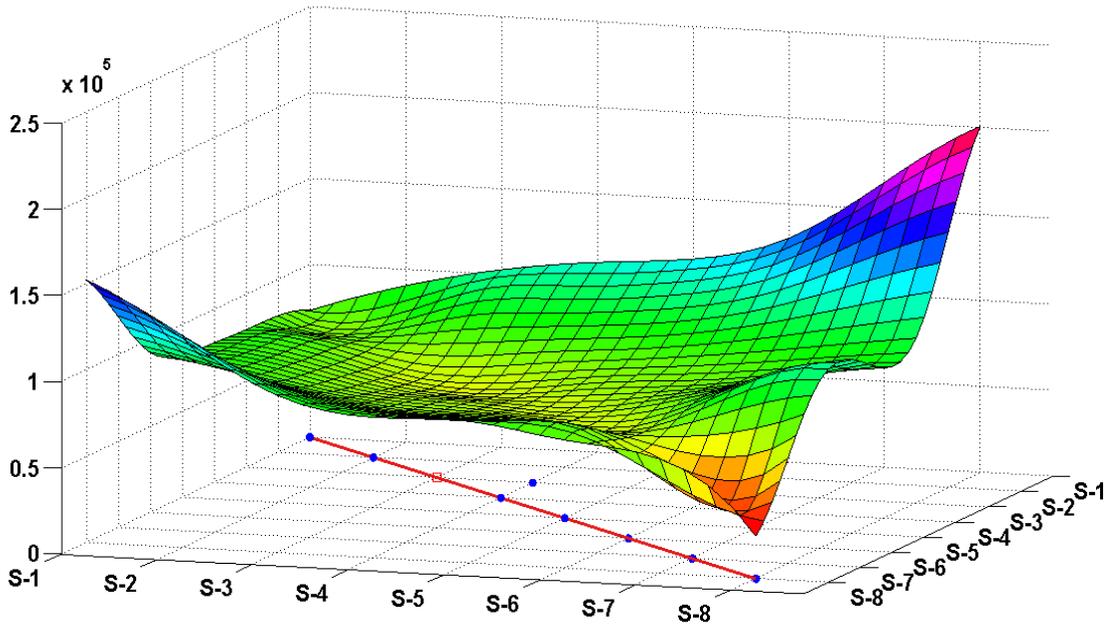

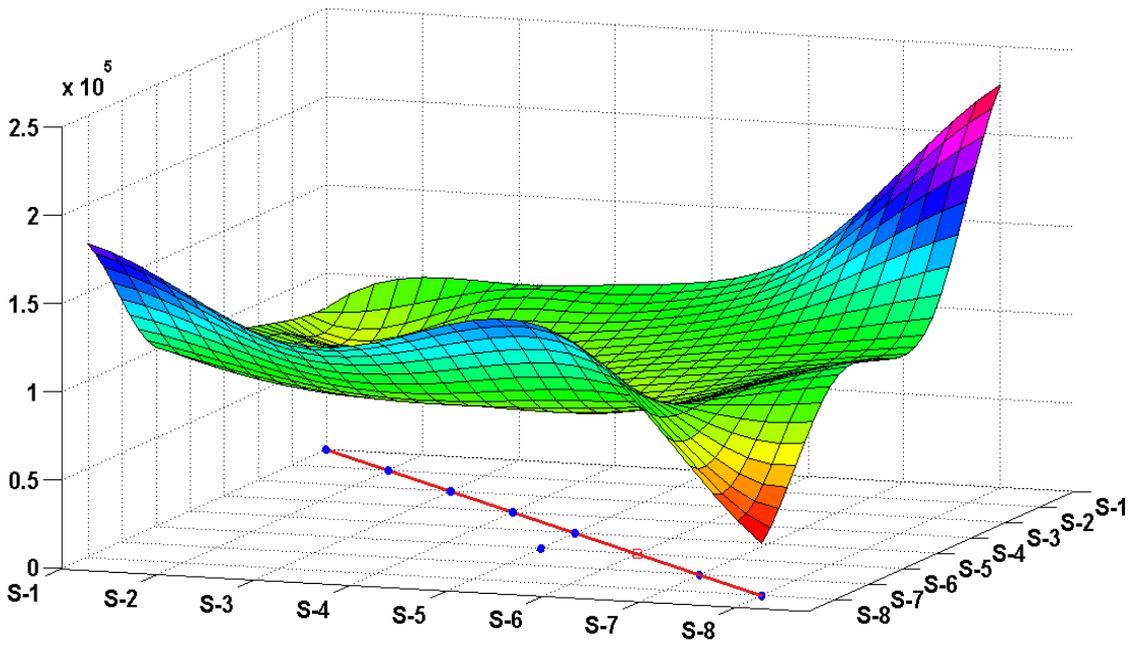

Figure 7-4 (*Continued*)

## TeraSort vs. Exim Mainlog Parsing

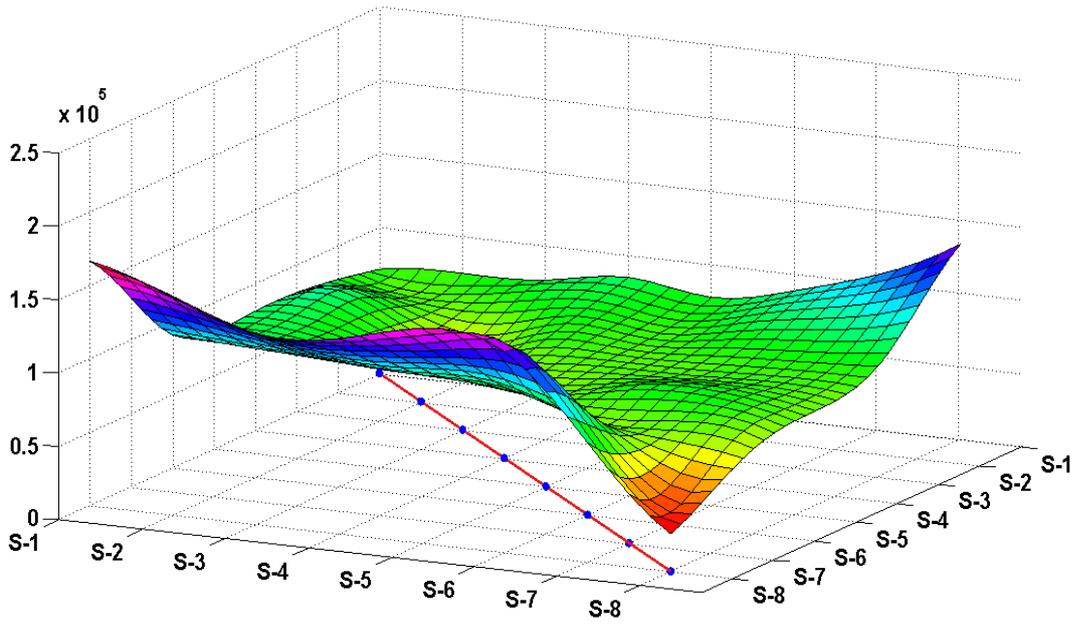

## WordCount vs. Exim MainLog Parsing

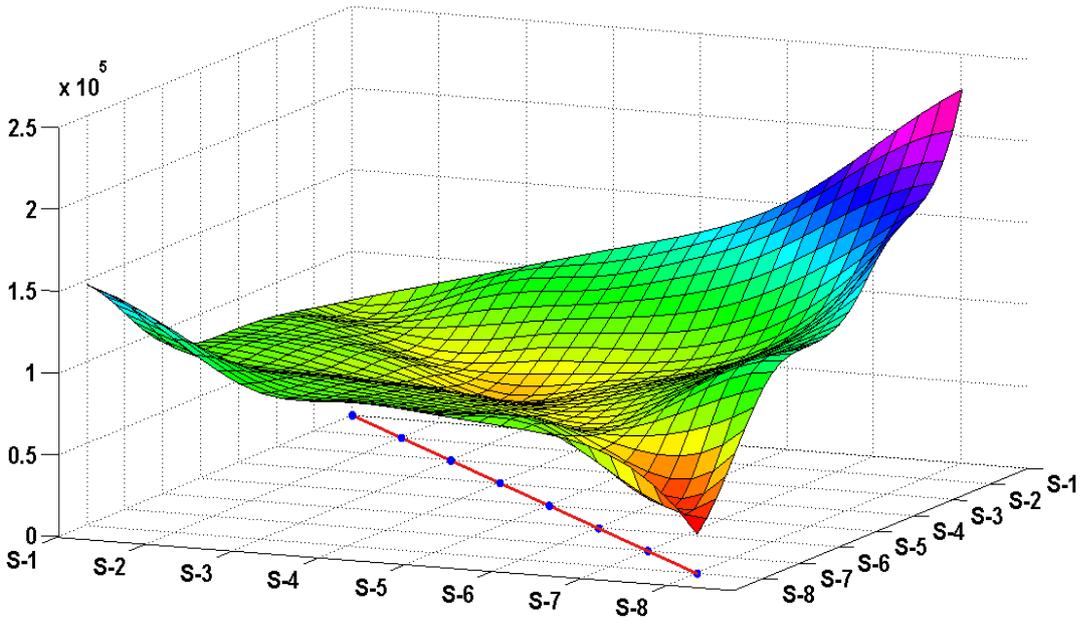

**Figure 7-4 (*Continued*)**

### 7.4.3.1. Scalability for the size of the input data

Upon finding the minimum distance between two applications for sets of parameters in Table 7-1 and Figure 7-4, the relationship between the scalability of input size and the distances is investigated for $\tau = 0.95$ on 10 virtual nodes; the results have been shown in Figure 7-6. Here, the input file size is 5 G, 10 G, 15 G, or, 20 G. This figure indicates that increasing the size of the input data file always results in a relatively greater minimum distance as well. This observation can be justified/explained by considering the fact that larger data files always need more time to be executed. Therefore, the length of CPU time series will also increase, and consequently more points will be compared.

### 7.4.3.2. Scalability for the number of virtual nodes

Figure 7-7 shows the relationship between the scalability of the number of nodes (5 to 25) in an experiment, and the minimum distances for $\tau = 0.95$ on a 5 G data file; this figure also shows that this relationship is much more complicated than the previous case. Further investigation and analysis, however, reveals the following facts: (1) When the number of nodes is increased – for the same data – because the input data are mapped onto more nodes, it results in shorter lengths of the CPU time series. Furthermore, because nodes are assumed to be identical, they are expected to produce very similar CPU time series as well. However, results show that more nodes always results in dispersed time series with more variance at each data point. This in turn causes greater levels of uncertainty, less accuracy in a system, and greater minimum distances between experiments. (2) Increasing the number of nodes expedites execution of an application and thus results in decreasing the number of points in its CPU time series. Therefore, as mentioned above in relation to scalability for the size of input data, this also leads to lower minimum distance values for experiments.

Comparison of Figures 7-6 and 7-7 shows that increasing either the size of data or the number of virtual nodes will result in greater minimum distance values. Nevertheless, as discussed, the increasing slope for scalability of virtual nodes is less than that of the size of the input data; thus, scaling the number of virtual nodes requires a lower similarity threshold value compared with scaling the size of the input data. It should be also noted that although scaling the experiments in either size of data or number of nodes increases the minimum distance, the hypothesis about applications' similarities still holds. It means that scaling (in both cases) does not affect similarity among applications, mainly because the order of lowest or the second lowest minimum distances among applications rarely changes when executed with a different number of map and reduce tasks, size of data file, and/or number of virtual nodes.

| WordCount | | | | | |
|---|---|---|---|---|---|
| | | *S-1* | *S-2* | *S-3* | *S-4* |
| **WordCount** | *S-1* | **1079** | 116850 | 89382 | 216474 |
| | *S-2* | 116850 | **2293** | 61541 | 119652 |
| | *S-3* | 89382 | 61541 | **2524** | 118166 |
| | *S-4* | 216474 | 119652 | 118166 | **2980** |

| Exim | | | | | |
|---|---|---|---|---|---|
| | | *S-1* | *S-2* | *S-3* | *S-4* |
| **Exim** | *S-1* | **325** | 63003 | 69015 | 138062 |
| | *S-2* | 63003 | **987** | 42078 | 71601 |
| | *S-3* | 69015 | 42078 | **1136** | 65632 |
| | *S-4* | 138062 | 71601 | 65632 | **1706** |

| Terasort | | | | | |
|---|---|---|---|---|---|
| | | *S-1* | *S-2* | *S-3* | *S-4* |
| **Terasort** | *S-1* | **361** | 135619 | 105092 | 150639 |
| | *S-2* | 135619 | **1046** | 60107 | 74998 |
| | *S-3* | 105092 | 60107 | **1102** | 69114 |
| | *S-4* | 150639 | 74998 | 69114 | **1588** |

*Table 7-2. The minimum distance ($r_{boundry}$) between each application with itself for $\tau = 0.95$, input data size=5G on 10 virtual nodes.*

|  | **Distributed Grip** | | | |
|---|---|---|---|---|
| **Distributed Grip** | | *S-1* | *S-2* | *S-3* | *S-4* |
| | *S-1* | **1895** | 106728 | 79264 | 2124544 |
| | *S-2* | 111743 | **3609** | 72784 | 104756 |
| | *S-3* | 79536 | 66530 | **4706** | 996546 |
| | *S-4* | 207634 | 98648 | 102746 | **6098** |

*Table 7-2. (Continued)*

### 7.4.3.3. The cost of profiling and modeling

During the experiments, execution time of these algorithms is also carefully logged. Table 7-3a reflects the average execution time as well as the total time for only the profiling phase of a whole set of experiments for each application on 5 G of input data; each application is executed 8 (possible number of mappers) times 8 (possible number of reducers) times 10 (repeating the whole experiment), i.e., 640 times in total. Table 7-3b shows the required time for the pattern-matching phase, i.e., comparing the CPU time series of an application on 5 G of input data to others with the same input data size in MATLAB [117]. Table 7-3b also shows that matching TeraSort with others always takes more time; an educated guess to explain this could be related to the nature of TeraSort, which is vastly different to the other applications.

## 7.5. Summary and Remarks

In this chapter, a new statistical approach was presented to find similarity among uncertain CPU utilisation time series of CPU intensive MapReduce applications on a private cloud platform. Through two phases of the approach (profiling and pattern-matching), the execution behaviour of known applications was used to estimate the behaviour of unknown ones. Profiling was performed through sampling, while the novel combination of DTW and Euclidean distance was used for pattern matching. Four applications (WordCount, Distributed grep, Exim MainLog parsing and TeraSort) were deployed on a private cloud to experimentally evaluate the performance of the approach. The results were very promising and showed how the CPU utilisation patterns of known applications were related and therefore could be used to estimate those of the unknown ones. Moreover, the scalability of the approach was studied against the size of the input data and the number of nodes in the private cloud environment.

*"Candidate Selection"*

21. Set pre-defined Probability threshold ($\tau = 0.95$)

*31. For $r^{th}$ small input data ($Size_r$):*

22.   *For $j^{th}$ set of configuration parameters values $S_j = (M_j, R_j)$:*

23.     *For $i^{th}$ application in database ($\varphi_i$):*

19.       *Run $\varphi_i$ for a small input data($Size_r$)*

24.       *Calculate joint mean and variance of distance between $\varphi_i'$ and $\chi_j'$ from Eqns. 7-5 and 7-6*

25.       *Calculate $r_{boundry}$ from Eqn. 7-10*

26.       *$< Size_r, \varphi_i,\ r_{boundry,i} >$ is added to candidature pool of $\chi_j$*

27.     *End*

28.   *End*

*"Applications similarity"*

*In candidature pool of $\chi_j$,*

*30. For $r^{th}$ small input data ($Size_r$):*

31.   *For $j^{th}$ set of configuration parameters values:*

32.     *Pick candidate with lowest $r_{boundry}$, e.g. $\varphi_k$*

33.     *Add k to the $Class_\chi$ ($Class_\chi$ keeps the index of similar applications to $\chi$)*

34.   *End*

*35. End*

*Hypothesis: The application with the highest index frequency in $Class_\chi$ is the most similar application to $\chi$ for all sets of configuration parameters. Therefore they belong to the same category regard to their CPU utilization patterns.*

**Figure 7-5. Candidate selection with application pattern matching algorithm based on similarity hypothesis**

|  | Average time for one experiment | Total time of 8*8*10 experiments per application |
|---|---|---|
| **WordCount** | ~19 min | ~202 h and 40 min |
| **Exim MainLog** | ~6.5 min | ~69 h and 20 min |
| **Terasort** | ~9 min | ~96 h |
| **Distributed Grep** | ~12.2 min | ~130 h and 8 min |

**(a)**

|  | WordCount | Exim MainLog | Terasort | Dist. Grep |
|---|---|---|---|---|
| **WordCount** | --- | 44 sec | 184 sec | 56 sec |
| **Exim MainLog** | 44 sec | --- | 96 sec | 64 sec |
| **Terasort** | 184 sec | 96 sec | --- | 93 sec |
| **Dist. Grep** | 56 sec | 64 sec | 93 sec | --- |

**(b)**

*Table 7-3. Time required for (a) profiling and (b) pattern matching for executing the proposed algorithm for all experiments on 5G of data and on 10 virtual nodes.*

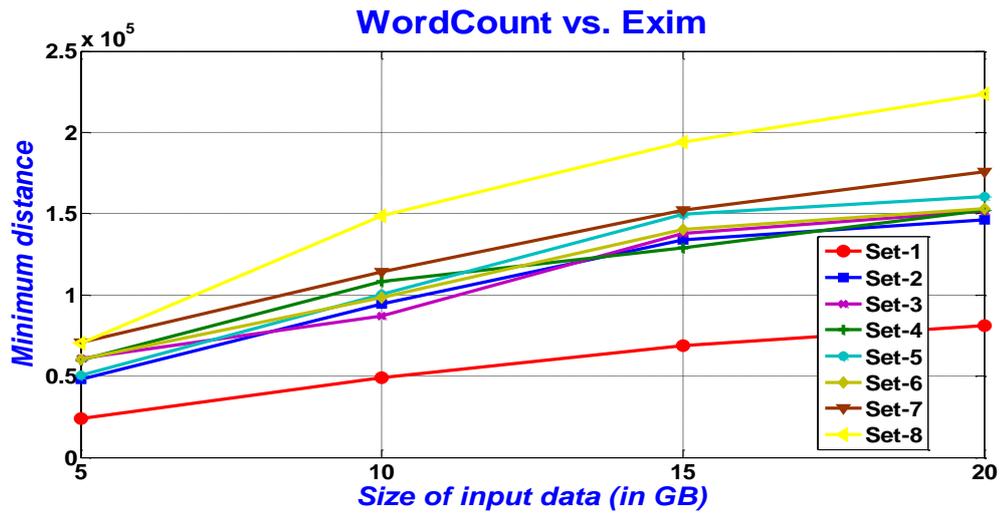

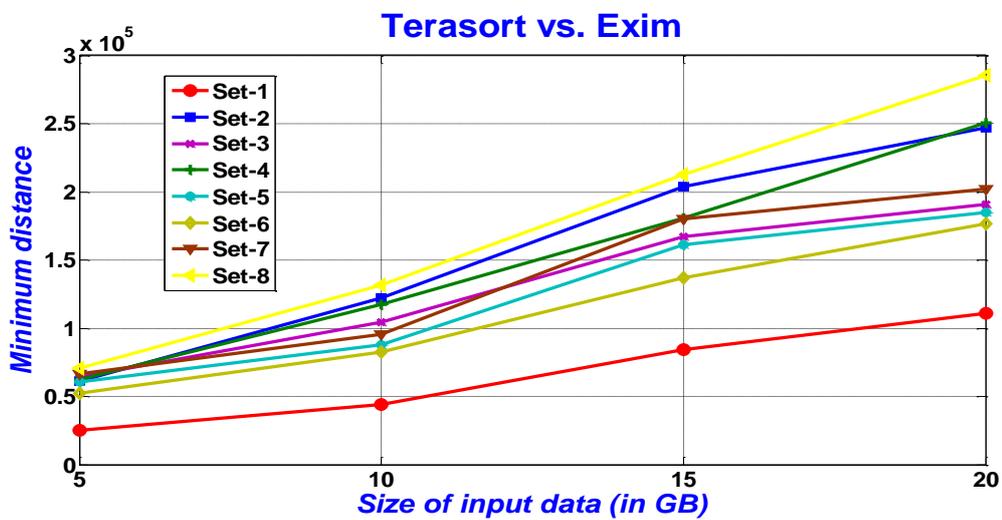

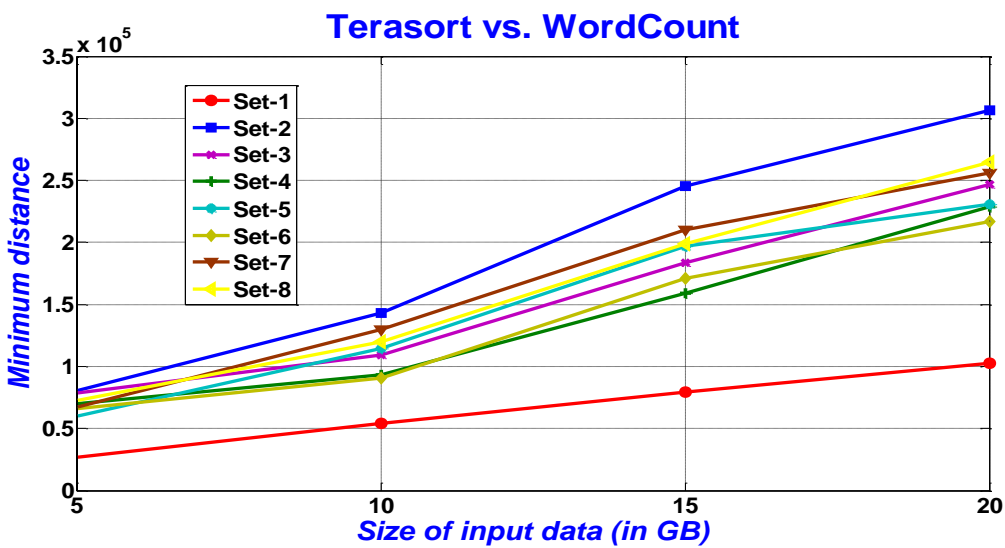

**Figure 7-6. The dependence of the minimum distance/maximum similarity (scalability in input file size) between applications on the size of input file for $\tau = 0.95$, and input data size of 5 G, 10 G, 15 G and 20 G on 10 virtual nodes. The y-axis shows the value of minimum distance between applications.**

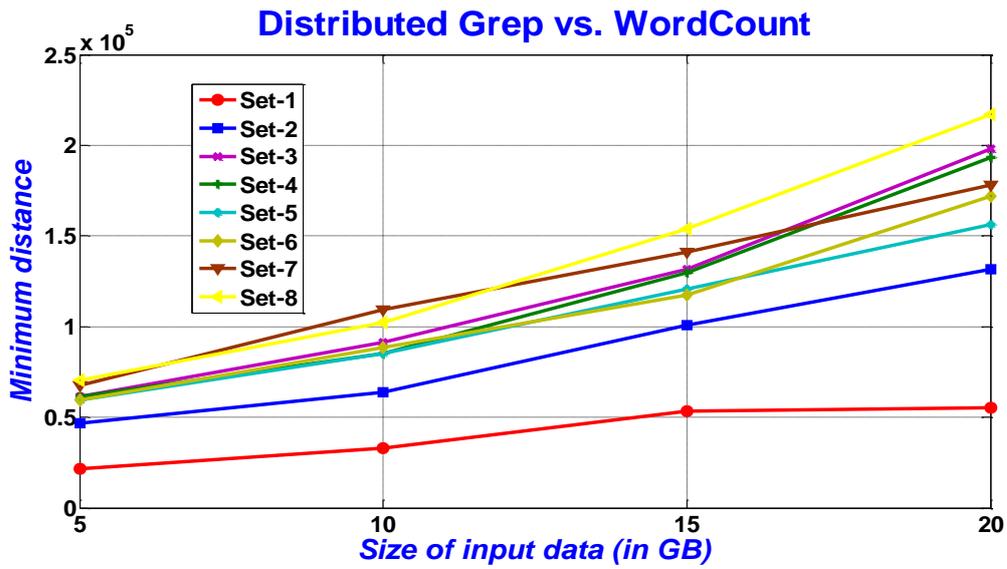

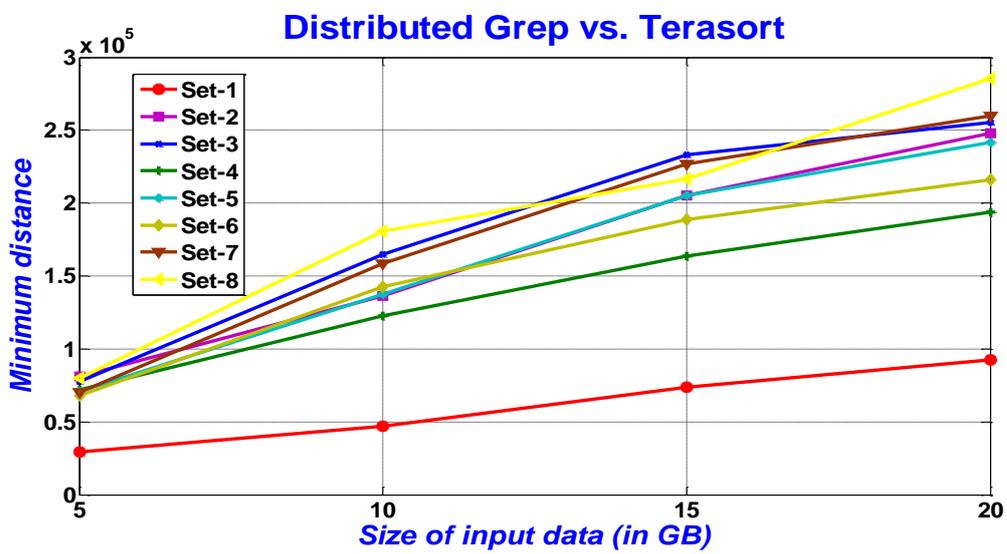

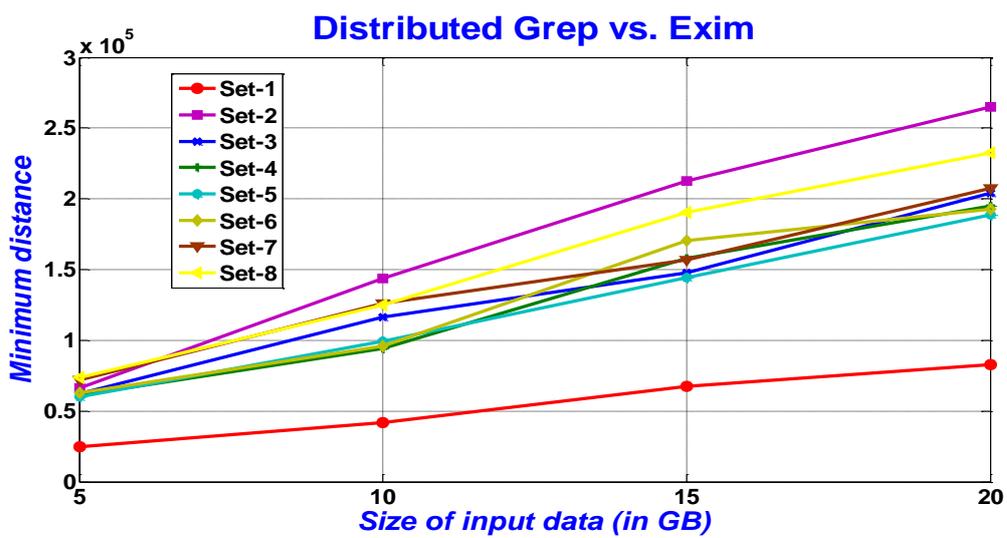

**Figure 7-6. (*Continued*)**

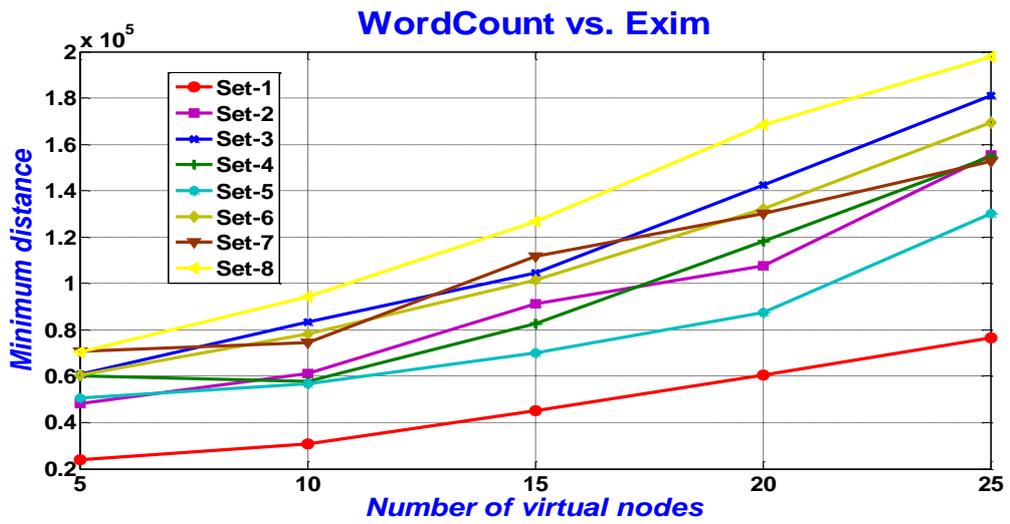

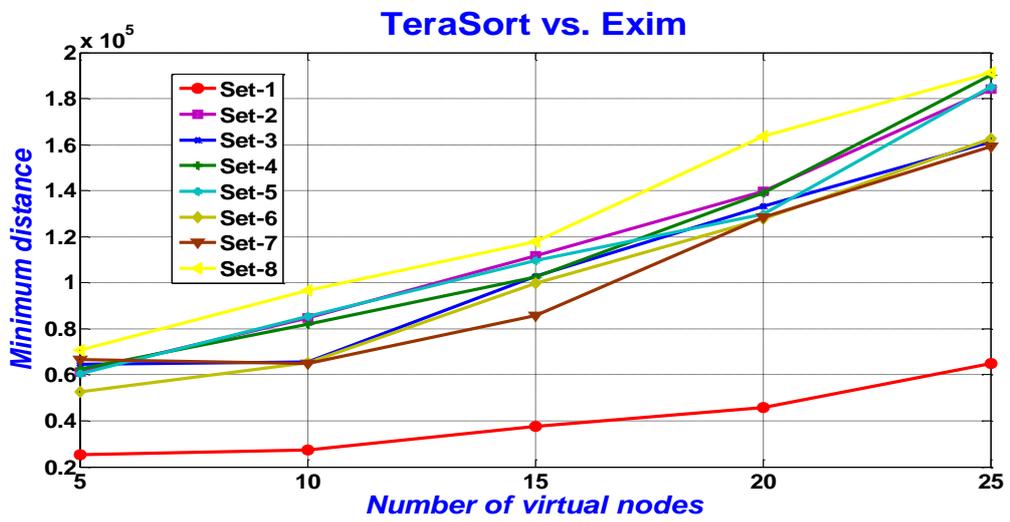

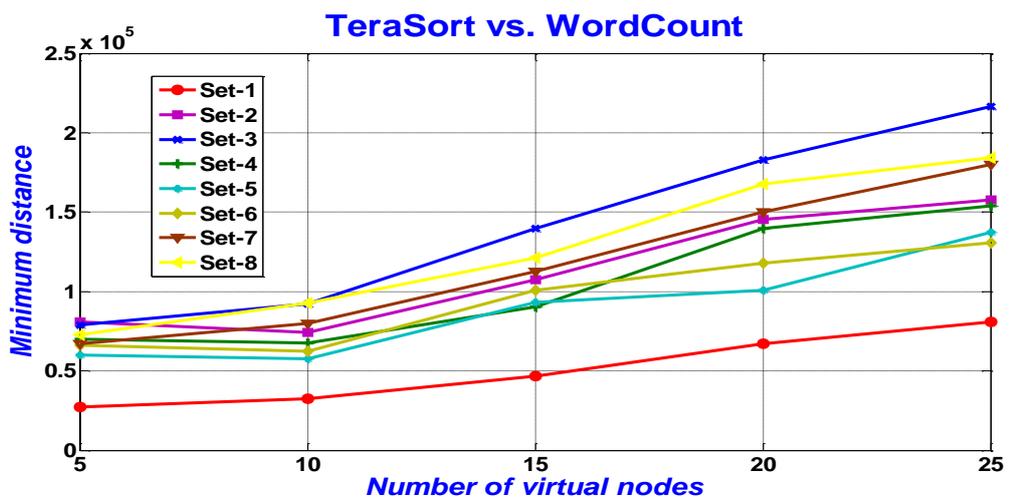

**Figure 7-7. Dependence of the minimum distance/maximum similarity (scalability in number of nodes) between applications on the number of virtual nodes for $\tau = 0.95$ and 5, 10, 15, and 20 virtual nodes, with 5 G of input data.**

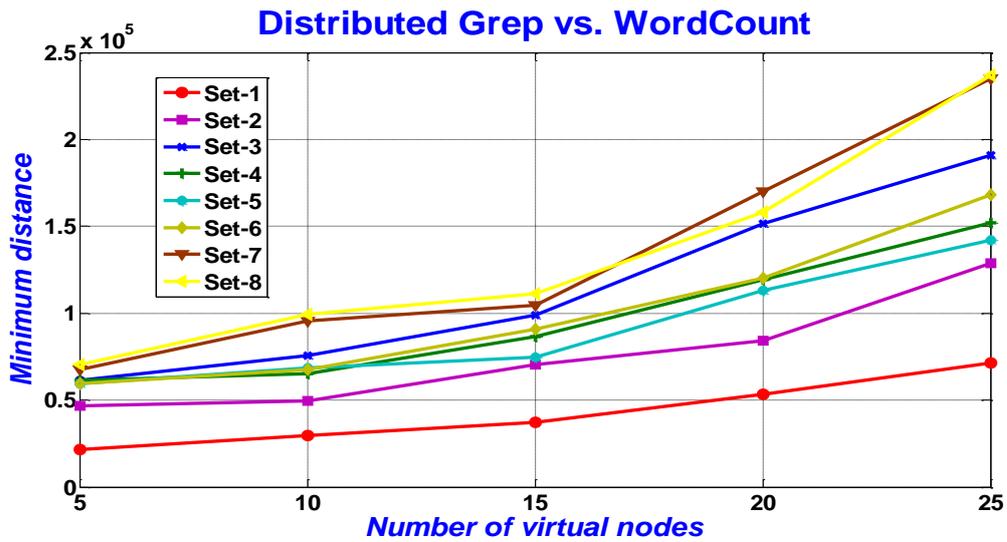

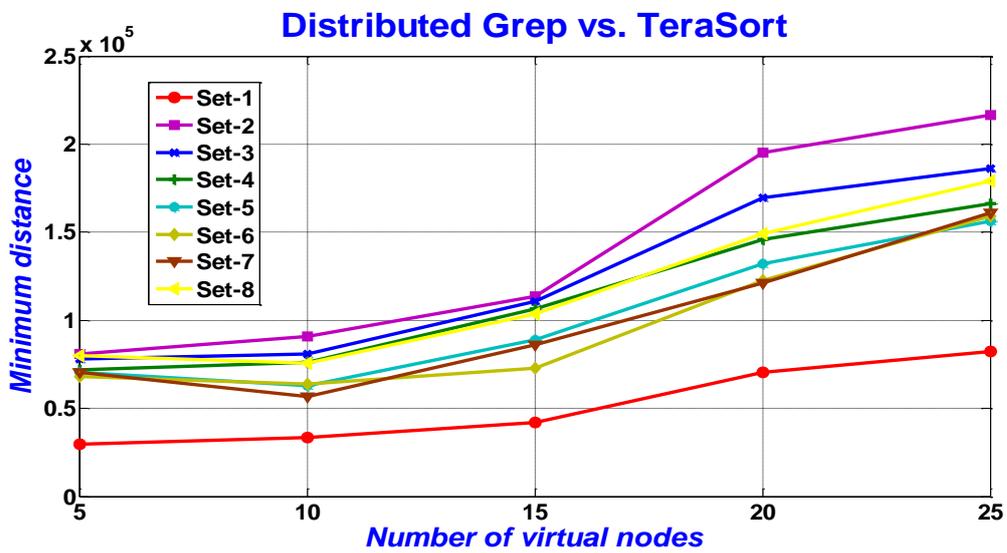

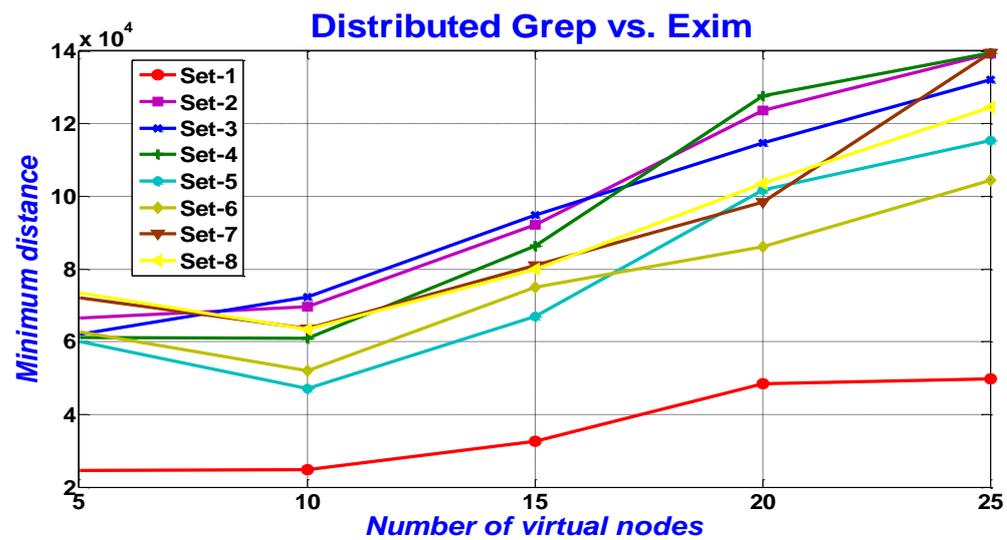

**Figure 7-7. (*Continued*)**

# Chapter 8. Network Load Provisioning of MapReduce Applications (fifth algorithm)

## 8.1. Introduction

This chapter attempts to study and model the network load of MapReduce applications in their shuffle phase. For a 5-node private MapReduce cluster, three well-known applications (i.e., WordCount, Exim MainLog Parsing, and TeraSort) run iteratively with different values of two configuration parameters (number of map tasks, and number of reduce tasks) on fixed-size input data, and network load in the shuffle phase of these applications is gathered. Then for each application, a model is constructed by applying polynomial multi-linear regression on the set of configuration parameter values (as input) and obtained network loads of the application (as output).

This modeling, however, works under some assumptions. First, both the degree of complexity of an application and proper model selection influence the accuracy of modeling, resulting in less accuracy for high complex applications. Second, even though the modeling is valid for applications on different platforms, different MapReduce/Hadoop clusters should result in different values for model parameters.

## 8.2. Model Generation and Evaluation

### 8.2.1. Profiling

For each application, several experiments are carried out with different values of the number of map tasks and the number of reduce tasks on a given cluster. After running each experiment, the network load in the shuffling phase of an application is extracted (using SysStat API[69]) as training data for future use by the model. Due to the temporal changes, it is expected that several runs of an experiment – with the same configuration parameters – may result in slightly different network loads.

Therefore, the average of several runnings of an experiment is considered as the network load.

## 8.2.2. Model generation

This section explains how to model the relation between the configuration parameters and network load of an application in MapReduce. The problem of modelling based on multivariate linear regression involves choosing suitable coefficients of the modelling such that the model's output well approximates a real system's response.

Consider three degree linear algebraic equations for $M$ number of experiments of an application for $N$ effective configuration parameters $(M \gg N)$:

$$\begin{cases} Net_{load}^{(1)} = \alpha_0 + \alpha_{11}p_1^{(1)} + \alpha_{12}\left(p_1^{(1)}\right)^2 + \alpha_{13}\left(p_1^{(1)}\right)^3 + \\ \qquad ... + \alpha_{N1}p_N^{(1)} + \alpha_{N2}\left(p_N^{(1)}\right)^2 + \alpha_{N3}\left(p_N^{(1)}\right)^3 \\ Net_{load}^{(2)} = \alpha_0 + \alpha_{11}p_1^{(2)} + \alpha_{12}\left(p_1^{(2)}\right)^2 + \alpha_{13}\left(p_1^{(2)}\right)^3 + \\ \qquad ... + \alpha_{N1}p_N^{(2)} + \alpha_{N2}\left(p_N^{(2)}\right)^2 + \alpha_{N3}\left(p_N^{(2)}\right)^3 \\ \qquad\qquad\qquad \vdots \\ Net_{load}^{(M)} = \alpha_0 + \alpha_{11}p_1^{(M)} + \alpha_{12}\left(p_1^{(M)}\right)^2 + \alpha_{13}\left(p_1^{(M)}\right)^3 + \\ \qquad ... + \alpha_{N1}p_N^{(M)} + \alpha_{N2}\left(p_N^{(M)}\right)^2 + \alpha_{N3}\left(p_N^{(M)}\right)^3 \end{cases} \qquad (8-1)$$

where $Net_{load}^{(k)}$ is the value of network load during the shuffle phase of an application in the $k^{\text{th}}$ experiment and $\left(p_1^{(k)}, p_2^{(k)}, ..., p_N^{(k)}\right)$ are the values of N configuration parameters for the same experiment, respectively. With matrix P as:

$$\boldsymbol{P} = \begin{bmatrix} 1, p_1^{(1)}, \left(p_1^{(1)}\right)^2, \left(p_1^{(1)}\right)^3, ..., p_N^{(1)}, \left(p_N^{(1)}\right)^2, \left(p_N^{(1)}\right)^3 \\ 1, p_1^{(2)}, \left(p_1^{(2)}\right)^2, \left(p_1^{(2)}\right)^3, ..., p_N^{(2)}, \left(p_N^{(2)}\right)^2, \left(p_N^{(2)}\right)^3 \\ \vdots \\ 1, p_1^{(M)}, \left(p_1^{(M)}\right)^2, \left(p_1^{(M)}\right)^3, ..., p_N^{(M)}, \left(p_N^{(M)}\right)^2, \left(p_N^{(M)}\right)^3 \end{bmatrix} \qquad (8-2)$$

Eqn. 8-1 can be rewritten in the matrix format as:

$$\underbrace{\begin{bmatrix} Net_{load}{}^{(1)} \\ Net_{load}{}^{(2)} \\ \vdots \\ Net_{load}{}^{(M)} \end{bmatrix}}_{Net_{load}} = \boldsymbol{P} \underbrace{\begin{bmatrix} \alpha_0 \\ \alpha_{11} \\ \alpha_{12} \\ \alpha_{13} \\ \vdots \\ \alpha_{N1} \\ \alpha_{N2} \\ \alpha_{N3} \end{bmatrix}}_{A} \qquad (8-3)$$

Using the above formulation, the approximation problem is converted to estimating the values of model parameters, i.e., $\widehat{\alpha_0}, \widehat{\alpha_{11}}, \widehat{\alpha_{12}}, \widehat{\alpha_{13}} \ldots, \widehat{\alpha_{N1}}, \widehat{\alpha_{N2}}, \widehat{\alpha_{N3}}$, to optimise a cost function between the approximation and real values of the network load. Then, an approximated total network usage $\left(\widehat{Net_{load}{}^{(*)}}\right)$ of the application for a new unseen experiment is predicted as:

$$\widehat{Net_{load}{}^{(*)}} = \widehat{\alpha_0} + \widehat{\alpha_{11}}\, p_1^{(*)} + \widehat{\alpha_{12}}\left(p_1^{(*)}\right)^2 + \widehat{\alpha_{13}}\left(p_1^{(*)}\right)^3 + \cdots + \widehat{\alpha_{N1}}\, p_N^{(*)}$$
$$+ \widehat{\alpha_{N2}}\left(p_N^{(*)}\right)^2$$
$$+ \widehat{\alpha_{N3}}\left(p_N^{(*)}\right)^3 \qquad (8-4)$$

It can be mathematically proved that the model parameters can be calculated by minimising the least square error between real and approximated values as:

$$A = (\boldsymbol{P^T P})^{-1} \boldsymbol{P^T} Net_{load} \qquad (8-5)$$

## 8.3. Evaluation Criteria

The accuracy of the fitted models, generated from regression, is evaluated based on a number of metrics [118]: Mean Absolute Percentage Error (MAPE), PRED(25), Root Mean Squared Error (RMSE) and R2 Prediction Accuracy . These metrics are described in the following subsections.

### 8.3.1. Mean Absolute Percentage Error (MAPE)

The Mean Absolute Percentage Error [118] for a prediction model is given by the following formula:

$$MAPE = \frac{\sum_{i=1}^{M} \dfrac{\left| Net_{load}{}^{(i)} - \widehat{Net_{load}{}^{(i)}} \right|}{Net_{load}{}^{(i)}}}{M}$$

| | RMSD | MAPE | $R^2$ prediction accuracy | PRED |
|---|---|---|---|---|
| **WordCount** | 0.24 | 1.78 | 0.93 | .93 |
| **Exim MainLog parsing** | 0.29 | 2.63 | 0.91 | .96 |
| **TeraSort** | 0.31 | 7.61 | 0.80 | 0.82 |

*Table 8-1. The prediction evaluation*

where $Net_{load}{}^{(i)}$ is the actual output of the application, $\widehat{Net_{load}}{}^{(i)}$ is the predicted output and $M$ is the number of observations in the dataset for which the prediction is made. A lower value of MAPE implies a better fit of the prediction model; i.e., indicating superior prediction accuracy.

### 8.3.2. PRED(25)

The measure PRED(25) [118] is defined as the percentage of observations whose prediction accuracy falls within 25% of the actual value. A more formal definition of PRED(25) is as follows:

$$PRED(25) = \frac{\#\ of\ observations\ with\ relative\ error\ less\ than\ 25\%}{\#\ of\ total\ observations}$$

It is intuitive that a PRED(25) value closer to 1.0 indicates a better fit of the prediction model.

### 8.3.3. Root Mean Square Error (RMSE)

The metric Root Mean Square Error (RMSE) [118] is defined by the following formula:

$$RMSE = \sqrt{\frac{\sum_{i=1}^{M}(Net_{load}{}^{(i)} - \widehat{Net_{load}}{}^{(i)})^2}{M}}$$

A smaller RMSE value indicates a more effective prediction scheme.

### 8.3.4. $R^2$ Prediction Accuracy

The $R^2$ Prediction Accuracy [118] is a measure of the goodness-of-fit of the prediction model. The formula of $R^2$ Prediction Accuracy is:

$$R^2 = 1 - \frac{\sum_{i=1}^{M}(Net_{load}^{(i)} - \widehat{Net_{load}}^{(i)})^2}{\sum_{i=1}^{M}(\widehat{Net_{load}}^{(i)} - \sum_{r=1}^{M}\frac{Net_{load}^{(r)}}{M})}$$

Note that the $R^2$ value falls within the range [0, 1]. This metric is commonly applied to Linear Regression models. In fact, $R^2$ Prediction Accuracy determines how the fitted model approximates the real data points. A $R^2$ prediction accuracy of 1.0 indicates that the forecasting model is a perfect fit.

## 8.4. Experimental Setting and Results

### 8.4.1. Experimental setting

Three applications are used to evaluate the effectiveness of the proposed method in this chapter. The method has been implemented and evaluated on a 5-node physical MapReduce platform running Hadoop version 0.20.2 – Apache implementation of MapReduce developed in Java [70]. The hardware specification of the nodes in our 5-node MapReduce platform is:

- Master/node-0 and node-1: Dell with one processor: 2.9 GHz, 32-bit, 1 GB memory, 30 GB Disk, and 512 KB cache.

- Node-2, node-3, and node-4: Dell with one processor: 2.5 GHz, 32-bit, 512 MB memory, 60 GB Disk, and 254 KB cache.

These nodes were connected via LAN network links. In the training phase of the modelling, 64 *s*ets of experiments are conducted where the number of map/reduce tasks are a value in [4,8,12,16,20,24,28,32]; the size of input data is fixed to 12 GB. To overcome temporal changes, each experiment is repeated ten times. Then in the prediction phase, the accuracy of the application model is evaluated with 30 new/unseen experiments on the same input data size and random number of map/reduce tasks – as an integer value – in a range of [4 … 32] .

The benchmark applications were WordCount, TeraSort, and Exim Mainlog parsing. These benchmarks were described in the previous chapter (section 7.4).

### 8.4.2. Results

To test the accuracy of an application's network load model, it is used to predict the network load of several experiments on different applications with a random number of map/reduce tasks. All experiments are executed on a 5-node cluster, and simultaneously their real network load is gathered to determine the prediction error. Figure 8-1 shows the prediction accuracies and MAPE prediction errors of these applications, comparing actual values of network load with their predicted values. Table 8-1 addresses the RMSD, MAPE, and R2 prediction accuracy, and PRED(25) of the predictions for these applications. From this table, it can be seen that MAPE for WordCount and Exim MainLog parsing is in a reasonable margin (1.78 and 2.63, respectively) whereas it is slightly high for TeraSort (7.61). This implies that three-degree polynomial regression performs well for WordCount and Exim, while it almost fails to correctly model the network load of TeraSort; therefore a better model must be used for this application. Again, the RMSDs of both WordCount and Exim MainLog are smaller than that of TeraSort, and experientially prove the applicability of this model for these applications over TeraSort. This finding is also supported by the PRED(25) results, as these two applications have a higher prediction accuracy than TeraSort.

## 8.5. Summary and Remarks

This chapter utilised polynomial regression to model the dependency between two major MapReduce configuration parameters (number of map tasks and number of reducer tasks) and network load during the shuffle phase of MapReduce applications with fixed-size input data. After extracting the network load of several experiments of an application with different values for the numbers of map/ reduce tasks, multivariate regression was used to model the relationship between the extracted network load and the used values for these two configuration parameters. Evaluation results on three applications on a 5-node MapReduce cluster showed that the modelling technique can effectively predict the network load of these applications with root mean squared error of less than 8%.

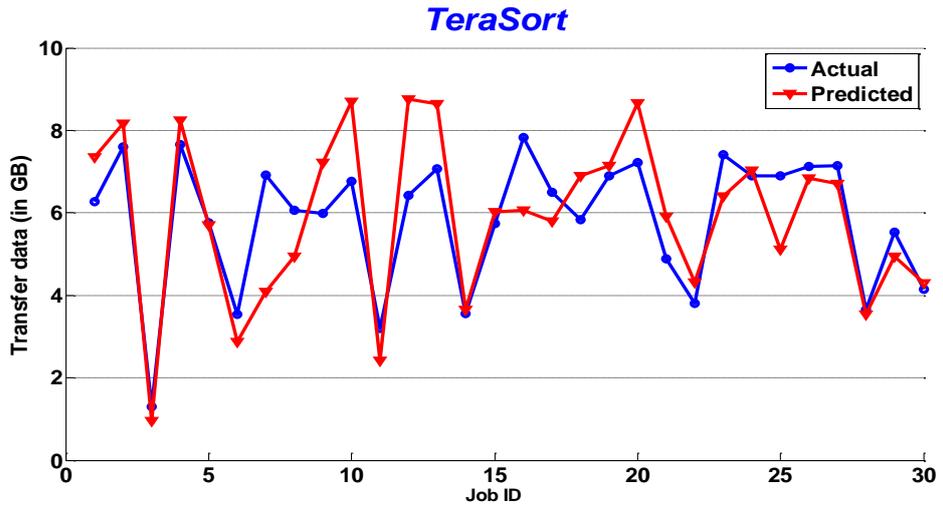

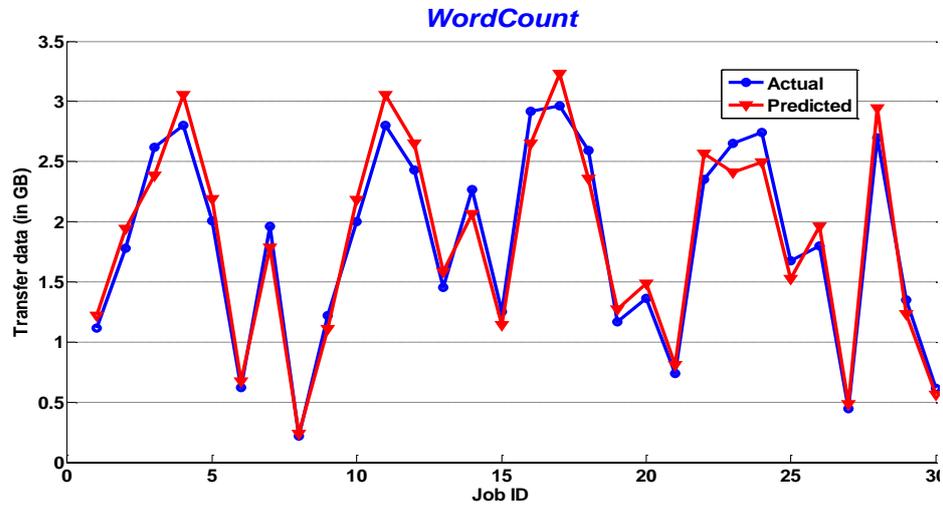

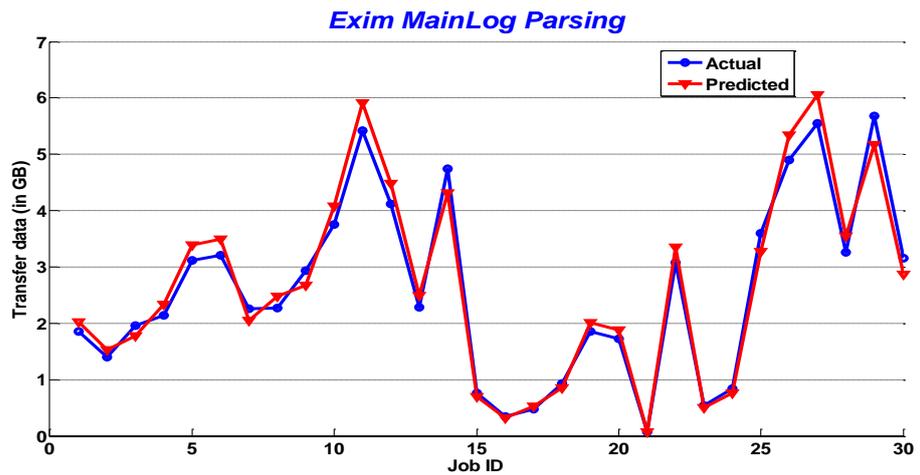

Figure 8-1. The actual and predicted total network load for benchmark applications. The X-axis is the index of new/unseen experiments.

# Chapter 9. Conclusion and future works

This thesis has recounted the journey of my PhD study on the application of algorithmic tools: (1) on energy efficient task scheduling in DVFS-enabled processors of distributed/cloud systems; and (2) performance provisioning and modelling in MapReduce, a famous distributed/network-based computing service on cloud platforms (e.g., Amazon Elastic MapReduce). In summary, the present thesis consists of two parts. In the first part, energy efficient task scheduling in DVFS-enabled processors was studied, and an in-depth survey of the existing methods was given. Through proposing and testing two new algorithms, we observed that a task reaches its optimal energy when it is executed by at most two available frequencies in the DVFS-enabled processor; this observation was proved mathematically. As a by-product, this proof addresses a bottleneck problem of research on energy efficient task scheduling in DVFS.

As addressed by researchers at UC-Berkeley [6], the most energy-efficient way of scheduling a computation is to put all hardware into the highest-performance state and race to complet as quickly as possible. Then drop the hardware to low power modes; therefore, n the second part of this thesis, we proposed two algorithms for performance provisioning of MapReduce jobs; this is based on the fact that smart scheduling of jobs on distributed/cloud computing platforms necessitates analysis of those jobs' resource requirements (i.e., CPU usage, Network usage, etc.). In the first algorithm, statistical pattern matching analysis was used to find similar CPU utilisation time patterns between two applications. Based on pattern matching concepts, it was hypothesised that if two applications are considered 'computationally similar' for short data files, they will be fairly 'similar' for large data sizes too. This assumption can be used to find the optimal number of map and reduce tasks for running a new unknown MapReduce job through first categorising it based on its CPU utilisation patterns, and then estimating its optimal running parameters based on similar applications in the same category/class. As MapReduce is a parametric distributed computing model, which means a few configuration parameters should be properly set by the user, in the second algorithm, we studied the correlation/dependency between two main configuration parameters (i.e., number of map/reduce tasks) and MapReduce resource usage, and then modelled this dependency with statistical regression.

Since this thesis encompasses two areas of research, we discuss the future research trends in both of these areas. First, this work employs DVFS to reclaim slack times of scheduled tasks. Generally, there are two ways to combine scheduling and DVFS: (1) independent slack reclamation, and (2) integrated scheduling generation. The former was addressed in this work, so our next work is to focus on the latter by combining DVFS and task scheduling. The existing methods in the literature that are based on this combination have a major limitation: almost all of them choose one frequency for each task to integrate DVFS and scheduling, while we would like to apply our multiple frequency selection technique and scheduling to achieve more energy saving.

Second, resource modeling and provisioning of jobs on the MapReduce platform – a similar procedure can be used for other distributed/cloud computing platforms – is relatively a new area of research. In addition to CPU and network usage provisioning, studied in the present work, the prediction of execution time of a job is also an important issue; such prediction can help the VM scheduler to accept/reject new jobs or intelligently re-schedule both running and new jobs with a guarantee of all jobs' Service Level Agreements (SLA). A basic analysis of the execution time of such jobs was reported in our Technical Report. The results from evaluation of our prediction models for CPU usage, network usage and execution time showed that there is a high correlation between these metrics and two MapReduce configuration parameters (i.e., number of map/reduce tasks). Our hypothesis is that strong dependency may exist between these metrics and most of the other configuration parameters. Therefore, it becomes more interesting and valuable to cover more configuration parameters to model execution time and other resources. Moreover, we observed that statistical regression cannot reach the necessary accuracy for some applications, probably due to over-fitting and under-fitting issues of regression-based models, and the nonlinear behavioural of such applications; thus, our next move is to deploy techniques like combination of k-nearest neighbor and regression or fuzzy logic to improve the prediction accuracy. Finally, as clearly addressed in [65], several runnings of a job with identical configuration parameters may result in different execution times – also known as uncertain execution time. Although there are a few recent methodologies to estimate the execution time of MapReduce jobs [119, 120], to the best of our knowledge, there is no practice to study the aforementioned uncertainty. Therefore, another future work could be to study and model such uncertainty based on configuration parameters using Gaussian Process Regression. Regard to energy efficiency in MapReduce/Hadoop, our plan is (1) to do further analysis and synthetic workloads to explore under what type of application characteristics would the better or worse results as far as energy savings, and (2) to correlate and extend performance modeling and predictions for MapReduce to include power.